\documentclass[12pt,amsmath,amssymb,showkeys,prb]{revtex4}
\usepackage{xcolor}
\usepackage{setspace}
\usepackage{graphicx}

\newcommand{\citen}[1]{%
  \begingroup
    \romannumeral-`\x 
    \setcitestyle{numbers}%
    \cite{#1}%
  \endgroup
}

\usepackage{amssymb}
\usepackage{amsmath}
\usepackage{bbold}

\usepackage{mathrsfs}
\usepackage{calligra}
\usepackage{bm}

\usepackage{xcolor}

\DeclareSymbolFont{matha}{OML}{txmi}{m}{it}
\DeclareMathSymbol{\varw}{\mathord}{matha}{119}   
\DeclareMathSymbol{\varpsi}{\mathord}{matha}{32}   

\newcommand{\np}{{n_{\rm p}}}
\newcommand{\nw}{{n_{\rm w}}}

\newcommand{\muwr}{ \mu } 

\newcommand{\Vol}{\Omega}

\newcommand{\TT}{\mathscr{T}}
\newcommand{\UU}{\mathscr{U}}
\newcommand{\HH}{\mathscr{H}}

\newcommand{\DD}{\mathscr{D}} 

\newcommand{\bra}[1]{\left\langle #1 \right|}
\newcommand{\ket}[1]{\left| #1 \right\rangle}

\newcommand{\lb}{l_{\rm B}}

\newcommand{\Z}{{\cal Z}} 

\newcommand{\kk}{{\bf{k}}} 
\newcommand{\kzero}{{\bf 0}} 
\newcommand{\R}{{\bf R}} 
\newcommand{\rr}{{\bf r}} 

\newcommand{\Qp}{{\cal Q}_{\rm p}} 
\newcommand{\Qw}{{\cal Q}_{\rm w}} 
\newcommand{\Qn}{{\cal Q}_{\rm n}} 

\newcommand{\kB}{k_{\rm B}} 

\newcommand{\ii}{{\rm i}}
\newcommand{\dint}{{\rm d}}
\newcommand{\dipole}{{ \bf{m} }}
\newcommand{\charge}{ c}
\newcommand{\RPAM}{{ \hat{\Delta}} }

\newcommand{\gmm}[1]{{ g^{\rm mm}_{#1} }}
\newcommand{\gmc}[1]{ { g^{\rm mc}_{#1}}}
\newcommand{\gcc}[1]{{ g^{\rm cc}_{#1}}}

\begin{document}

$\null$
\hfill {April 2, 2021}
\vskip 0.3in

\begin{center}
{\Large\bf A Simple Explicit-Solvent Model of Polyampholyte}\\ 

\vskip 0.3cm

{\Large\bf Phase Behaviors and its Ramifications for}\\ 

\vskip 0.3cm

{\Large\bf Dielectric Effects in Biomolecular Condensates}\\

\vskip .5in
{\bf Jonas W{\footnotesize{\bf{ESS\'EN}}}},$^{1,\dagger}$
{\bf Tanmoy P{\footnotesize{\bf{AL}}}},$^{1,\dagger}$
{\bf Suman D{\footnotesize{\bf{AS}}}},$^{1,\dagger}$
{\bf Yi-Hsuan L{\footnotesize{\bf{IN}}}},$^{1,2}$
 and\\
{\bf Hue Sun C{\footnotesize{\bf{HAN}}}}$^{1,*}$

$\null$

$^1$Department of Biochemistry,
University of Toronto, Toronto, Ontario M5S 1A8, Canada;\\
$^2$Molecular Medicine, Hospital for Sick Children, Toronto, 
Ontario M5G 0A4, Canada\\

\vskip 1.3cm

%

\end{center}

\vskip 1.3cm

\noindent
$\dagger$ Contributed equally to this work.
\\

\noindent
$*$Corresponding author\\
{\phantom{$^\dagger$}}
E-mail: chan@arrhenius.med.utoronto.ca;
Tel: (416)978-2697; Fax: (416)978-8548\\
{\phantom{$^\dagger$}}
Mailing address:\\
{\phantom{$^\dagger$}}
Department of Biochemistry, University of Toronto,
Medical Sciences Building -- 5th Fl.,\\
{\phantom{$^\dagger$}}
1 King's College Circle, Toronto, Ontario M5S 1A8, Canada.\\

\vfill\eject

\noindent
{\large\bf Abstract}\\

\noindent
Biomolecular condensates such as membraneless organelles, underpinned by
liquid-liquid phase separation (LLPS), are important for physiological
function, with electrostatics---among other interaction types---being a
prominent force in their assembly. Charge interactions of intrinsically
disordered proteins (IDPs) and other biomolecules are sensitive to the aqueous
dielectric environment. Because the relative permittivity of protein is
significantly lower than that of water, the interior of an IDP condensate is a
relatively low-dielectric regime, which, aside from its possible functional
effects on client molecules, should facilitate stronger electrostatic
interactions among the scaffold IDPs. To gain insight into this LLPS-induced
dielectric heterogeneity, addressing in particular whether a low-dielectric
condensed phase entails more favorable LLPS than that posited by assuming IDP
electrostatic interactions are uniformly modulated by the higher dielectric
constant of the pure solvent, we consider a simplified multiple-chain model of
polyampholytes immersed in explicit solvents that are either polarizable or
possess a permanent dipole. Notably, simulated phase behaviors of these systems
exhibit only minor to moderate 
differences from those obtained using implicit-solvent
models with a uniform relative permittivity equals to that of pure solvent.
Buttressed by theoretical treatments developed here using random phase
approximation and polymer field-theoretic simulations, these observations
indicate a partial compensation of effects between favorable solvent-mediated 
interactions among the polyampholytes in the condensed phase and favorable 
polyampholyte-solvent interactions in the dilute phase, often netting
only a minor enhancement of overall LLPS propensity from the very dielectric
heterogeneity that arises from the LLPS itself. Further ramifications of this
principle are discussed.

\vfill\eject

\noindent
{\large\bf INTRODUCTION}\\

Biomolecular condensates,\cite{rosen2017}---physico-chemically based in
large measure upon liquid-liquid  phase  separation 
(LLPS)\cite{CellBiol,NatPhys,linPRL,tanja2019} underpinned
by multivalent interactions among intrinsically
disordered proteins (IDPs), proteins containing intrinsically disordered
regions (IDRs), folded domains, folded proteins, and nucleic 
acids\cite{brangwynne2009,Rosen12,McKnight12,Nott15,tanja2015,parker2015}---are
now widely recognized as a means for living organisms, eukaryotes as
well as prokaryotes, to functionally compartmentalize their internal 
space.\cite{Michnick2016,cliff2017,Monika2018Rev,weber2020} 
Among their multifaceted physiological 
roles,\cite{rosen2017,chong2016,babu2018,mingjie2020,roland2020} 
the droplet-like condensates, especially the intracellular varieties
referred to as membraneless organelles, provide specialized, tunable
micro-environments, distinct from their surroundings, that facilitate certain
biochemical processes.\cite{shorter2019,mekhail2020} One of the many possible
mechanisms by which they serve this task is by preferential partitioning of
certain reactants into the condensates. Examples include retention of selected
types of small nuclear ribonucleoproteins in Cajal bodies, a function likely
plays a critical role in pre-mRNA splicing,\cite{Schaffert_etal} partitioning
of single-stranded but not double-stranded DNA into condensates of RNA helicase 
Ddx4 that mimic nuage membraneless organelles---thus serving probably as a
molecular filter for functional nucleic acid processing,\cite{Nott15,Nott16}
and concentration of transcription regulators in TAZ (tafazzin) nuclear
condensates to promote gene expression.\cite{Lu_etal2020}

LLPS of IDPs/IDRs is known to underlie an 
increasing number of biomolecular condensates,\cite{VeronikaRev2016}
as exemplified by well-studied condensates of N-terminal IDR domains of, 
respectively, the DEAD-box RNA helicase 4 (Ddx4),\cite{Nott15}
the RNA-binding protein fused in sacroma (FUS),\cite{Fawzi2017} and 
the P-granule DEAD-box RNA helicase LAF-1.\cite{cliff2015}
Compared to amino acid sequences encoding for globular proteins,
IDPs/IDRs are depleted of hydrophobic but enriched with polar and
charged residues.\cite{BabuRev2014} Consistent with these trends,
electrostatics is one of the major driving forces for biomolecular 
LLPS,\cite{Nott15,linPRL,linJML,HXZhouRev2018} together with cation-$\pi$ 
and $\pi$-$\pi$ interactions which can be equally or sometimes even more 
important,\cite{robert,moleculargrammar,julieRev,TanjaScience2020,SumanPNAS} 
and hydrophobic 
interactions which can be important\cite{keeley2018} 
for systems that exhibit a lower critical solution 
temperature\cite{biochemrev,roland18,jeetainACS} rather than an upper 
critical solution temperature.\cite{Roland2019} 
As far as electrostatics is concerned,\cite{Brocca2020,obermeyer2020} 
experiments on a charge scrambled 
mutant of Ddx4,\cite{Nott15} phosphorylated variants of FUS,\cite{Fawzi2017} 
phosphorylated variants of an IDR of the fragile X mental retardation protein 
(FMRP),\cite{BrianTsang2019} and charge shuffled variants of 
LAF-1\cite{Schuster2020} indicate that effects of charged interactions 
on LLPS and complexation behaviors depend not only on 
an IDP/IDR's net charge but also its sequence charge 
pattern.\cite{rohit2013,kings2015,ZhaoetalJCP2015,linPRL,lin2017,jacob2017,Alan}
Recently, some of these and related sequence-dependent effects on LLPS 
have been modeled successfully and rationalized by polymer theory using 
random phase approximation (RPA),\cite{linPRL,linJML,njp2017}
restricted primitive models,\cite{singperry2017}
Kuhn length renormalization,\cite{kings2020}
coarse-grained explicit-chain molecular 
dynamics (MD)\cite{SumanPNAS,Schuster2020,dignon18,suman1,jeetainPNAS,suman2,koby2020}
and Monte Carlo\cite{stefan2019} sampling, and 
field-theoretic simulations.\cite{joanElife,joanJPCL,joanPNAS,joanJCP,Pal2021}
\\


In these recent theoretical and computational studies of electrostatic 
effects in LLPS, the relative permittivity, $\epsilon_{\rm r}$, for the 
solvent-mediated intra- and interchain electrostatic interactions among 
charged IDPs are often taken to be uniform and equal to that of bulk water 
($\epsilon_{\rm r}\approx 80$),\cite{jeetainACS,Schuster2020,singperry2017,dignon18,suman1,jeetainPNAS,koby2020,joanElife}
or left unspecified as a constant $\epsilon_{\rm r}$ that is subsumed in
the Bjerrum length or a reduced 
temperature,\cite{linJML,lin2017,njp2017,kings2020,suman2,joanJPCL,joanPNAS,joanJCP}
or as a fitting parameter for matching theory to experiment, yielding constant 
$\epsilon_{\rm r}$ values in the range 
of $\approx 30$--$60$, which are smaller than but still within the same order
of magnitude of that of bulk water.\cite{Nott15,linPRL,linJML}
A common feature of these formulations is that while the value of 
$\epsilon_{\rm r}$ may differ from model to model, $\epsilon_{\rm r}$ is 
always assumed to be a position-independent constant for a given model
irrespective of the local IDP density. However, 
proteins\cite{honig,bertrand0,bertrand,vanGunsteren2001,warshel2001,plotkinPCCP,kings2017} 
and nucleic acids\cite{DNA}
have substantially lower relative permittivities than bulk 
water. Inside the droplet, the density of water is reduced
because of the higher concentrations of protein and nucleic acids.
Moreover, the water molecules that remain, being likely 
near protein surfaces, generally entail a smaller relative 
permittivity than that of bulk water due to
restrictions on their rotational freedom.\cite{Geim} 
Taken together, these considerations lead one to expect that the effective 
permittivity\cite{bragg,jackson,markel2016,uversky2016,njp2017}
inside a condensed-phase IDP droplet is lower than that of the dilute
phase.\cite{mann2020} Such a different dielectric environment 
could affect the behaviors of proteins and 
nucleic acids\cite{bloomfield1995,feig2014} 
as well as the partitioning and sequestration of a variety of ``client'' 
molecules in the condensed droplet\cite{spruijt2019}
and thus contribute to the maintenance of biomolecular condensates
and the functional biophysical and biochemical processes within
them.\cite{deniz2018,uversky2018,uversky2018_1,Prasad2019}

The significantly different dielectric environments expected for
the dilute phase on one hand and the condensed phase on the other
raise basic questions regarding the aforementioned common
practice of using a single, constant $\epsilon_{\rm r}$---rather than 
a variable $\epsilon_{\rm r}$ that depends on local IDP density---for
the electrostatic interactions in LLPS theories and computational models.
One obvious concern is that because the effective $\epsilon_{\rm r}$
inside the condensed phase is substantially lower than that of water,
the LLPS-driving electrostatic interactions in the condensed phase should 
be modulated by this lower value of $\epsilon_{\rm r}$ and therefore much
stronger than that posited by assuming a uniform $\epsilon_{\rm r}$ of
bulk water. This consideration suggests that the propensity to phase separate
can be much higher than that predicted by assuming a constant 
$\epsilon_{\rm r}$. In apparent agreement with this intuition,
earlier studies using modified RPA formulations with an 
$\epsilon_{\rm r}(\phi)$ that depends on IDP volume fraction $\phi$ 
show significant increases in LLPS propensities.\cite{linJML,njp2017}
Subsequently, however, a more nuanced analysis that takes into account
a ``self-energy'' term in the 
$\epsilon_{\rm r}\rightarrow\epsilon_{\rm r}(\phi)$ 
RPA formulation indicates that there is a trade-off 
between the enhanced effective electrostatic interactions among the IDPs
in the condensed phase (which tend to increase LLPS propensity) and
the enhanced interactions between the IDPs and the solvent in the
dilute phase (which tend to decrease LLPS propensity), resulting
in a relative minor overall effect on LLPS propensity.\cite{SumanPNAS} 

While the insights provided by these preliminary investigations are
valuable---not the least for formulating the questions needed to be 
tackled, our understanding of the effect of the biomolecule-dependent 
dielectric environment on the energetics of LLPS remains quite limited. This 
is because the $\epsilon_{\rm r}(\phi)$ in these formulations are obtained 
from effective medium theories\cite{bragg,jackson,markel2016}
based on an implicit treatment of water rather than having
dielectric properties emerge from an explicit model of water.
Indeed, a shortcoming of this approach is that predictions based on 
different effective medium theories can differ 
substantially.\cite{linJML,njp2017,SumanPNAS}
With this in mind, the present work aims to attain a more definitive 
physical picture by modeling sequence-dependent LLPS of polyampholytes with
an explicit polar solvent.
Recently, new insights have been gained into the interactions between 
salt ions and protein as well as the diffusive dynamics in condensed FUS
and LAF-1 IDR droplets by using the TIP4P/2005 model of 
water.\cite{JeetainAtom} 
The explicit water model, however, is absent in the initial thermodynamic
simulation to assemble the condensed phase.
Separately, explicit coarse-grained water has been applied successfully to 
simulate condensates of FUS IDR\cite{gerhard2020} using the MARTINI 
coarse-grained force field,\cite{martini2013} though polar versions
of the coarse-grained water model\cite{Smiatek2017} is not employed.
Since our goal is to better understand the general physical principles
rather than to achieve quantitative match with experiment, here we
take a complementary approach by using simple dipoles---each consisting
only of one positive and one negative charge---as a model for a 
water-like polar solvent.
In this regard, our construct shares similarities with a simple lattice 
dipole model of water restricted to an essentially one-dimensional 
configuration\cite{gerhard2009} as well as three-dimensional simple dipole
models for polar liquids.\cite{aluru2018}
Unlike our previous implicit-solvent 
formulations,\cite{linJML,njp2017,SumanPNAS}
here the different dielectric environments for the dilute 
and condensed phases arise physically from the polar nature of the 
dipoles rather than as a prescribed function of local 
solvent and IDP concentrations alone. Importantly, 
the model's simplicity allows equilibrated LLPS systems comprising of 
both explicit dipole solvent molecules and polyampholytes to be simulated.

Consistent with the trend observed in a recent implicit-solvent RPA 
analysis that include a self-energy term,\cite{SumanPNAS} 
polyampholyte phase properties simulated in explicit-chain models
with the present explicit dipole model of polar solvent are quantitatively 
similar to---albeit not identical with---those 
simulated without the explicit solvent but with a uniform relative 
permittivity corresponding to that of the pure polar solvent.
Whether the LLPS propensity of the explicit-solvent system is
slightly higher or slightly lower than that of the corresponding 
implicit-solvent 
system likely depends on the system's structural and energetic details.
Nonetheless, our results suggest, assuringly, that the error in employing an 
IDP-concentration-independent bulk-water $\epsilon_{\rm r}\approx 80$ 
in LLPS simulations\cite{jeetainACS,Schuster2020,singperry2017,dignon18,suman1,jeetainPNAS,koby2020,joanElife}
may be relatively small even when the dielectric environments inside
and outside of the condensed phase are significantly different.
To address the generality of this putative principle,
we have also developed corresponding field-theory formulations for
polyampholytes with a polar solvent modeled as permanent or polarizable 
dipoles that are treated explicitly in the theory.
Similar permanent\cite{orlandPRL2012,orlandJCP2013,ZGWangJCP2018}
and polarizable\cite{GHFJCP2016,FredricksonPRL2019} dipole models
were used to study dielectric properties of polar 
solutions\cite{orlandPRL2012,orlandJCP2013,ZGWangJCP2018,GHFJCP2016,FredricksonPRL2019} 
and the effect of electric field on phase stability of charged 
polymers;\cite{GHFJCP2020}
but have not been applied to tackle the question at hand.
Here we find that the trend of phase behaviors predicted using RPA of 
our field-theory formulation regarding explicit versus implicit solvent
are in agreement with that obtained by explicit-chain simulations, 
suggesting that a compensation between
more favorable polyampholyte-polar-solvent electrostatic interactions in 
a high-$\epsilon_{\rm r}$ dilute phase and more favorable effective 
inter-polyampholyte electrostatic interactions in a low-$\epsilon_{\rm r}$ 
condensed phase is a rather robust feature of polyampholyte LLPS systems.
Details of our methodology and findings are described below.
\\


\noindent
{\large\bf MODELS AND METHODS}\\

The formulation and computational methodology of our coarse-grained 
explicit-chain model of polyampholyte LLPS in the presence of explicit dipoles 
as a model polar solvent is adapted from---and follows largely---a
protocol developed recently for implicit-solvent
simulations of LLPS of chain molecules,\cite{dignon18,jeetainPNAS,panag2017}
an approach our group has applied previously to study LLPS of model 
polyampholytes\cite{suman2} and to address experimental data on IDR 
LLPS.\cite{SumanPNAS}
Following refs.~\citen{suman1,suman2}, but with a notation
more in line with that in ref.~\citen{kings2020},
let $n_{\rm p}$ be the total number of polyampholyte chains in the system 
which are labeled by $\mu,\nu = 1,2,\dots,n_{\rm p}$, let
$N$ be the number of monomers (beads) in each chain, labeled by
$i,j=1,2,\dots,N$, and 
$n_{\rm w}$ be the total number of dipoles (which may be viewed as
serving as model water molecules
and hence the subscript ``w'') which are labeled by
$\upsilon=1,2,\dots,n_{\rm w}$. 
Note that a dipole in the present formulation is equivalent to a 
polyampholytic dimer chain carrying a pair of opposite charges,
denoted here as $q_{\rm d}=\pm |q_{\rm d}|$, on its two monomer beads.

The total potential energy $U_{\rm T}$, expressed in the equation below
as a sum of contributions, is a function of the positions of the monomers 
in the polyampholyte chains and in the dipoles:
\begin{equation}
U_{\rm T} = U_{\rm bond} + U_{\rm LJ} + U_{\rm el} \; 
\label{eq:U-T}
\end{equation}
with $U_{\rm bond}$ being the bond-length term for chain connectivity
that applies to the polyampholytes as well as the dipoles:
\begin{equation}
U_{\rm bond} = \frac {K_{\rm bond}}{2} \Biggl[ \; 
\sum_{\mu=1}^{n_{\rm p}} \sum_{i=1}^{N-1}
(r_{\mu i,\mu i+1}-a)^2
+ \sum_{\upsilon=1}^{n_{\rm w}} (r_{\upsilon +,\upsilon -}-a)^2
\Biggr ]
\;
\label{eq:U-bond}
\end{equation}
where, in general, $r_{\mu i,\nu j}$
is the distance between
monomer $i$ in polyampholyte chain $\mu$
and monomer $j$ in polyampholyte chain $\nu$, 
$r_{\upsilon \pm,\upsilon^\prime \pm}$
is the distance between the position of the 
positive (+) or negative ($-$) charge on dipole $\upsilon$
and the position of the
positive (+) or negative ($-$) charge on dipole $\upsilon^\prime$,
and $a$ is the reference bead-bead bond length
that serves as the length scale of the model.
Here we set $K_{\rm bond}=75,000 \varepsilon/a^2$,
where $\varepsilon$ is an energy scale to be defined below.
This value of $K_{\rm bond}$, which is in line with the TraPPE
force field,\cite{TraPPE1,TraPPE2,TraPPE3,TraPPE4}  entails a very stiff
spring constant;\cite{Alan,suman2,panag2017}
thus the dipoles are essentially permanent (fixed) dipoles (with magnitude
of dipole moment $\mu \approx |q_{\rm d}|a)$ because they are
practically not polarizable by fluctuation of the bond length ($\approx a$) 
between a dipole's $\pm |q_{\rm d}|$ beads. 
$q_{\rm d}$ is given in units of the elementary protonic charge, $e$, below.
(The symbol $\mu$ for dipole moment magnitude is not to be 
confused with the polyampholyte label $\mu=1,2,\dots,n_{\rm p}$). 

$U_{\rm LJ}$ in the above Eq.~\ref{eq:U-T} accounts for
the non-bonded Lennard-Jones-like interactions, which in the present
model takes different forms depending on whether the monomer beads 
involved are both from polyampholyte chains (pp), both from dipoles (dd),
or one from a polyampholyte chain and the other from a dipole (pd):
\begin{equation}
U_{\rm LJ}= U_{\rm LJ,pp}+U_{\rm LJ,dd}+U_{\rm LJ,pd}
\; .
\label{eq:LJ-all}
\end{equation}
As before,\cite{suman2}
the polyampholyte-polyamopholyte term $U_{\rm LJ,pp}$ is the 
standard 12-6 LJ potential:
\begin{equation}
    U_{\rm LJ,pp} = 
4\varepsilon_{\rm LJ}
\sum_{\mu,\nu=1}^{n_{\rm p}}
\sum_{\substack{i,j=1\\ \null\hskip -0.8cm(\mu,i)\ne(\nu,j)}}^N
    \left[\left(\frac{a}{r_{\mu i,\nu j}}\right)^{12}-
        \left(\frac{a}{r_{\mu i,\nu j}}\right)^6\right]
\; ,
\label{eq:LJ}
\end{equation}
where $\varepsilon_{\rm LJ}$ is the well depth
and the bead diameter $a$ is equal to the reference bond length
in Eq.~\ref{eq:U-bond}. During simulation, $U_{\rm LJ,pp}$ is truncated
at $r_{\mu i,\nu j}=6a$ for computational efficiency
(individual terms in $U_{\rm LJ,pp}$ are set to zero 
for $r_{\mu i,\nu j}\ge 6a$).
To avoid freezing of the dipole solvent at relatively low temperatures and other
complications as well as to facilitate comparison with analytical theory, 
the dipole-dipole and polyampholyte-dipole contributions in Eq.~\ref{eq:LJ-all}
are chosen to be consisting of only excluded-volume (ex) repulsion terms, viz.,
\begin{equation}
U_{\rm LJ,dd}=
\sum_{\upsilon=1}^{n_{\rm w}-1}\sum_{\upsilon^\prime=\upsilon + 1}^{n_{\rm w}}
[ {\tilde{U}}_{\rm LJ}^{\rm ex}(r_{\upsilon +,\upsilon^\prime +})
+ {\tilde{U}}_{\rm LJ}^{\rm ex}(r_{\upsilon -,\upsilon^\prime -})
+ {\tilde{U}}_{\rm LJ}^{\rm ex}(r_{\upsilon +,\upsilon^\prime -})
+ {\tilde{U}}_{\rm LJ}^{\rm ex}(r_{\upsilon -,\upsilon^\prime +}) ]
\; 
\label{eq:LJ-dd}
\end{equation}
and
\begin{equation}
U_{\rm LJ,pd}=
\sum_{\mu=1}^{n_{\rm p}} \sum_{i=1}^N \sum_{\upsilon=1}^{n_{\rm w}}
[ {\tilde{U}}_{\rm LJ}^{\rm ex}(r_{\mu i,\upsilon +}) 
+ {\tilde{U}}_{\rm LJ}^{\rm ex}(r_{\mu i,\upsilon -}) ]
\; ,
\label{eq:LJ-pd}
\end{equation}
where the individual purely repulsive terms in the above summations take
the Weeks-Chandler-Andersen form:\cite{WCA}
\begin{equation}
{\tilde{U}}_{\rm LJ}^{\rm ex}(r)\equiv \left\{
\begin{array}{cc}
{\tilde{U}}_{\rm LJ}(r) + \varepsilon_{\rm LJ} \; , &
\quad\quad {\rm for\ }
        r \leq 2^{1/6}a \\
    0 \; , & \quad\quad {\rm for\ } r > 2^{1/6}a
\end{array}
\right.
        \label{eq:ULJex}
\end{equation}
which is obtained, as specified here, by performing a cutoff and a shift on
an individual standard LJ term
${\tilde{U}}_{\rm LJ}(r)\equiv 4\varepsilon_{\rm LJ}[(a/r)^{12}-(a/r)^6]$
in Eq.~\ref{eq:LJ}.

Similar to Eq.~\ref{eq:LJ-all}, the electrostatic 
interaction $U_{\rm el}$ in Eq.~\ref{eq:U-T} is written as a sum 
of pp, dd, and pd components:
\begin{equation}
U_{\rm el} =
U_{\rm el,pp}+U_{\rm el,dd}+U_{\rm el,pd} 
\; .
\label{eq:Uel-all}
\end{equation}
The polyampholyte-polyampholyte contribution is given by
\begin{equation}
U_{\rm el,pp} =
\sum_{\mu,\nu=1}^{n_{\rm p}}
\sum_{\substack{i,j=1\\ \null\hskip -0.8cm(\mu,i)\ne(\nu,j)}}^N
\frac {\sigma_{i}\sigma_{j}e^2}{4\pi\epsilon_0\epsilon_{\rm r}
r_{\mu i,\nu j}} 
\; ,
\label{eq:U-elpp}
\end{equation}
where $\sigma_{i}$ is the charge of the $i$th monomer bead in units of
elementary protonic charge $e$, ($\sigma_{i}$ depends on the
polyampholyte sequence but is independent of $\mu,\nu$),
and $\epsilon_0$ is vacuum permittivity.
For simulation with explicit dipoles, relative permittivity 
$\epsilon_{\rm r}$ is set to unity for the $U_{\rm el,pp}$ term 
in Eq.~\ref{eq:U-elpp}; but $\epsilon_{\rm r}> 1$ can be included in 
implicit-solvent simulations. 
The other contributions in Eq.~\ref{eq:Uel-all}
which involve the dipoles are given by:
\begin{equation}
U_{\rm el,dd} =
\sum_{\upsilon=1}^{n_{\rm w}-1}
\sum_{\upsilon^\prime=\upsilon + 1}^{n_{\rm w}}
\frac {q_{\rm d}^2 e^2}{4\pi\epsilon_0}
\Biggl (
\frac {1}{r_{\upsilon +,\upsilon^\prime +}} 
+
\frac {1}{r_{\upsilon -,\upsilon^\prime -}} 
-
\frac {1}{r_{\upsilon +,\upsilon^\prime -}} 
-
\frac {1}{r_{\upsilon -,\upsilon^\prime +}} 
\Biggr )
\; ,
\label{eq:U-eldd}
\end{equation}
where the charges $q_{\rm d}$ on the dipoles are in units of
elementary protonic charge $e$, and
\begin{equation}
U_{\rm el,pd} =
\sum_{\mu=1}^{n_{\rm p}} \sum_{i=1}^N
\sum_{\upsilon=1}^{n_{\rm w}}
\frac {\sigma_{i}q_{\rm d}e^2}{4\pi\epsilon_0}
\Biggl (
\frac {1}{r_{\mu i,\upsilon +}} - \frac {1}{r_{\mu i,\upsilon -}} 
\Biggr )
\; .
\label{eq:U-elpd}
\end{equation}
As in ref.~\citen{suman2}, the energy scale $\varepsilon$ introduced above
in conjunction with $K_{\rm bond}$ is chosen to be the strength of electrostatic
interactions at a monomer-monomer separation $a$ for the polyampholytes, i.e.,
$\varepsilon=e^2/(4\pi\epsilon_0\epsilon_{\rm r}a)$, and the well depth
$\varepsilon_{\rm LJ}$ of LJ-like interactions in Eq.~\ref{eq:LJ}
is set to be equal to $\varepsilon$, i.e.,
$\varepsilon_{\rm LJ}=\varepsilon$, as in Fig.~4a of ref.~\citen{suman2}.
These specifications complete the description of the total potential energy
function $U_{\rm T}$ in Eq.~\ref{eq:U-T} used in our simulations.

Fig.~1 depicts the $N=50$ model polyampholyte sequences and the 
model dipole solvent molecule used in this study.
The three overall neutral sequences (each 
containing 25 positive and 25 negative
beads), sv15, sv20, and sv30, were introduced by 
Das and Pappu.\cite{rohit2013}
These sequences have been used in recent explicit-chain simulations of
sequence-dependent IDP binding and LLPS.\cite{Alan,suman2}
Here, they serve to assess whether the 
sequence-dependent LLPS observed in explicit-chain, implicit-solvent 
simulations\cite{suman2,koby2020} holds for explicit-chain-simulated
LLPS in the presence of explicit dipole solvent molecules, bearing in mind that
these sequences cover a considerable variation of charge
patterns as quantitatively characterized by the intuitive blockiness parameter
$\kappa$ (ref.~\citen{rohit2013}; $\kappa=0.1354$, $0.2721$, and
$1.0$, respectively, for sv15, sv20, and sv30) and 
the theory-derived ``sequence charge decoration'' parameter\cite{kings2015}
(SCD; $-$SCD $=4.35$, $7.37$, and $27.8$, respectively, for sv15, sv20,
and sv30).

Regarding sv sequences, it should be noted that $U_{\rm LJ,pp}$ in 
Eq.~\ref{eq:LJ}, which contains an attractive interaction, is used here
to facilitate comparison with previous explicit-chain 
simulation results of sv sequences employing the same potential.\cite{suman2}
While the pure repulsive ${\tilde{U}}_{\rm LJ}^{\rm ex}$ in Eq.~\ref{eq:ULJex} 
without nonelectrostatic attraction for polyampholyte-polyampholyte 
interaction may appear more in line with the analytical theory to be 
developed below, we found that as a pure polymer system, sv15 fails to 
phase separate under ${\tilde{U}}_{\rm LJ}^{\rm ex}$ (ref.~\citen{suman2}).
Effects of other LJ-type potentials with a nonzero but weakened attraction
such as the ``with 1/3 LJ'' model
in ref.~\citen{suman2} are worthy of further exploration but are beyond 
the scope of our present effort.

\begin{center}
   \includegraphics[width=0.66\columnwidth]{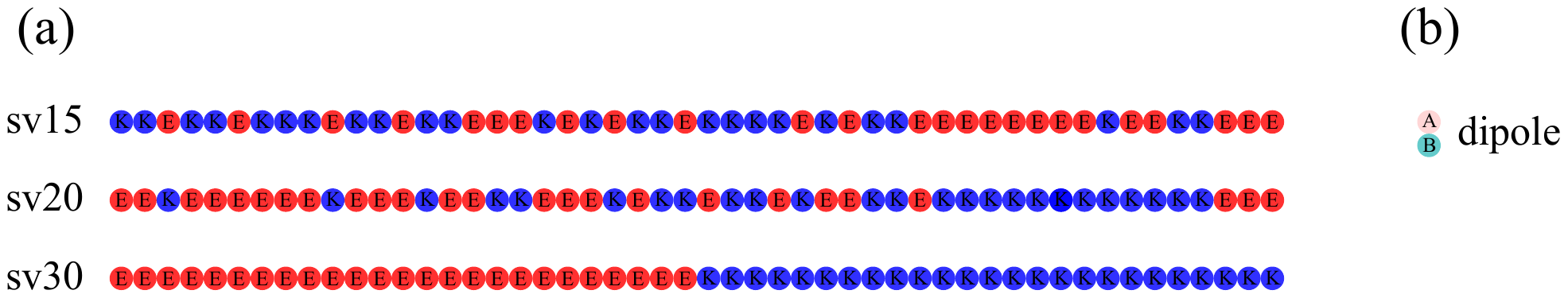}
\end{center}
\vskip 0 mm
{\footnotesize{\bf Fig.~1:} Model polyampholytes and dipole solvent molecules.
(a) The $N=50$ model sequences studied in the present work. As in
ref.~\citen{rohit2013}, positively and negatively charged monomer beads
are labeled as ``K'' (lysine) and ``E'' (glutamic acid), respectively, and
depicted in blue and red. (b) Dipole as model polar solvent molecule.
Underscoring that the charges on the solvent dipoles can
be different from those on the model polyampholytes in (a), i.e.,
$|q_{\rm d}|$ does not necessarily equal unity, the positively charged
(``B'', cyan) and negatively charged (``A'', pink) beads in (b) do not
share the K/E and blue/red labels in (a).
}
$\null$\\

As in our previous studies,\cite{suman2,SumanPNAS} 
molecular dynamics simulations of the present model are performed 
by the GPU-version of the HOOMD-blue simulation 
package\cite{Anderson,Glaser} using a recently developed methodology 
for simulating phase behaviors\cite{dignon18,SumanPNAS} as well as 
binding\cite{Alan} of IDPs. The computational protocol has been 
described in detail elsewhere.\cite{SumanPNAS,dignon18,suman2,panag2017}
Electrostatic interactions are treated using the PPPM method\cite{LeBard}
implemented in the HOOMD package.
For our simulation of polymers (polyampholytes) and dipoles, $n_{\rm p}=100$ 
polymer chains are first randomly placed in a relatively large
simulation box of dimensions $40a\times 40a\times 40a$. This system
is energy-minimized for a period of $500\tau$ with a timestep of
$0.001\tau$, where $\tau\equiv \sqrt{ma^2/\varepsilon}$ with $m$ being
the mass of each monomer bead. 
For simplicity, similar to the models in ref.~\citen{suman2},
all monomer beads making up the polyampholyte chains and the dipole dimers 
are assigned the same mass. The initial equilibration is then
performed for a period of $1,000\tau$ at $T^*=4.0$. The reduced
temperature $T^*\equiv k_{\rm B}T/\varepsilon$, where 
$k_{\rm B}$ is Boltzmann constant and $T$ is absolute temperature, 
contains a factor of $\epsilon_{\rm r}$ via the above definition 
for $\varepsilon$. For direct comparisons of temperatures in different 
dielectric environments on the same footing, however, it is often useful to
use the vacuum reduced temperature ($T^*$ for $\epsilon_{\rm r}=1$), 
denoted in the present work as
\begin{equation}
T^*_0\equiv \frac {T^*}{\epsilon_{\rm r}}=
\frac {4\pi\epsilon_0 k_{\rm B}T a}{e^2}
\; . 
\label{eq:T_red0}
\end{equation}
The velocity-Verlet algorithm is utilized to propagate the equations of 
motion. Periodic boundary conditions are applied to all three dimensions. 
The system is then compressed isotropically at $T^*_0=4.0$ 
for $5,000\tau$ using linear scaling to a sufficiently higher density 
of $\sim 0.73 m/a^3$ by confining the polymers to a box of size 
$19a\times 19a\times 19a$.
We next expand the box 15 times along one of the spatial dimensions
(labeled as $z$) at a much lower temperature of $T^*_0=0.5$, resulting in 
final box dimensions of $19a\times 19a\times 285a$. 
After this procedure, the system is equilibrated for 
$1,000\tau$ at the same $T^*_0=0.5$ using Langevin dynamics with a weak 
friction coefficient of $0.1 m/\tau$ (ref.~\citen{panag2017}). 
The Packmol package\cite{Martinez} 
is subsequently used to insert $n_{\rm w}=12,000$ dipoles 
into the simulation box
while keeping the positions of all the polymer beads fixed.\cite{JeetainAtom} 
During the insertion process, a suitable distance criterion is set 
to avoid steric repulsion. After inserting the dipoles, we equilibrate the 
system again for a longer period of $30,000\tau$ at several desired
temperatures $T^*_0\ge 3.0$ at which the dipoles are dynamic and well
dispersed throughout the simulation box.  At lower temperatures ($T^*_0<2.5$), 
we observe that the dipoles do not move much over extended periods of 
time---they appear to be essentially frozen---and thus are not a good 
model for liquid solvent. For this reason, $T^*_0< 3.0$ is not used 
for production runs. 
Finally, each subsequent production run for $T^*_0\ge 3.0$ is for a duration of
$100,000\tau$, during which snapshots are saved every 
$10\tau$ for the determination of various ensemble properties, including 
density profiles of polymers and dipoles in the simulation box. 
Coexistence curves (phase diagrams) of the polymers are determined 
from the polymer density distributions as described before.\cite{suman2}
Density of dipoles ($\rho_{\rm w}$) and density of polyampholytes ($\rho$) 
are reported by the number densities of monomer beads in the two types 
of molecules, respectively, in units of $1/a^3$.

For the study of pure polymers, i.e., implicit-solvent systems 
of polyampholytes with no dipoles, the procedure is
identical to that in ref.~\citen{suman2}, viz., $n_{\rm p}=500$
polymer chains are initially placed randomly in a cubic box of length $70a$.
After energy minimization and initial equilibration, this box is
compressed to a cubic box of length $33a$. The simulation box 
for the system is then expanded eight times along one of the axes 
(named $z$ as before)\cite{suman2} to reach a final dimensions of
$33a\times 33a\times 264a$. As for the polymer plus dipole simulations
described above, equilibration is performed under periodic boundary
conditions in the expanded box for $30,000\tau$, which is then followed
by a production run for $100,000\tau$ with snapshots saved every $10\tau$
for subsequent analysis. Further details regarding simulation temperatures 
and boundary conditions are provided in ref.~\citen{suman2}. 

For the study of pure dipole solvent, i.e., systems that contain dipoles 
but no polymers, the main purpose of the simulations is to determine the 
effective relative permittivity of bulk solvent as modeled by the dipoles 
under various conditions. We use $n_{\rm w}=10,000$ dipoles for each
simulation. The size of the cubic periodic simulation box is adjusted
to model different dipole densities ($\rho_{\rm w}$). We first focus on 
$\rho_{\rm w}=$ $0.205$ and $\rho_{\rm w}=0.252$, which we deem sufficiently 
dense to exhibit liquid-like characteristics but not too dense to 
render the systems computationally intractable. The length of the
cubic boxes utilized for these simulations are $46a$ and $43a$ respectively.
The simulations are conducted for a broad range of temperatures, namely 
$T^*_0=3.0$, $4.0$, $5.0$, $6.0$, and $7.0$, as well as for a broad range of 
dipole moments, with $q_{\rm d}=\pm 1$, $\pm 2$, $\pm 3$, $\pm 4$, and $\pm 5$.
To probe the dependence of the effective relative permittivity on both
density and temperature, we further fix $q_{\rm d}=\pm 3$, vary 
$\rho_{\rm w}$ from
$0.205$ to $\approx 0.55$ at intervals of $\approx 0.05$ ($\rho_{\rm w}=0.205$,
$0.252$, $0.301$, $0.350$, $0.401$, $0.451$, $0.500$, and $0.552$), and
simulate every system for five temperatures from $T^*_0=3.0$ to $7.0$ as 
above. Lastly, to arrive at a set of conditions to maintain a constant
$\rho_{\rm w}$ as well as a temperature-independent effective relative
permittivity, we fix $\rho_{\rm w}=0.252$
and use theoretically fitted values of $q_{\rm d}$ (see below)
to conduct simulations at $T^*_0=3.0$, $4.0$, $5.0$, $6.0$, $7.0$, and $8.0$
(the corresponding lengths of the cubic simulation boxes, in units of $a$, are:
$46.0$, $43.0$, $40.5$, $38.5$, $36.8$, $35.4$, $34.2$, and $33.1$).
For every simulation described above,
we first energy-minimize to rid the system of
significant steric clashes for a duration of 
$100\tau$, then equilibrate the system
at the desired temperature using Langevin dynamics with the same 
low friction coefficient of $0.1 m/\tau$ for $500\tau$.
The subsequent production run is performed
for $4,000\tau$ and snapshots are collected every
$1.0\tau$ for analysis.

As noted above, we use the same friction coefficient for explicit- and 
implicit-solvent systems.  Equilibrium properties such 
as phase coexistence of the present explicit-chain models are
independent of the choice of Langevin friction coefficient
because friction coefficients affect dynamics
but not thermodynamic distributions. 
\\

\noindent
{\large\bf RESULTS}\\

{\bf Effective Permittivities Of Pure Dipole Solvents.} We begin by 
determining baseline dielectric constants of the bulk solvents in our model
by conducting simulations of pure dipole systems under a variety 
of conditions as outlined above, utilizing the following expression for 
the effective relative permittivity:\cite{SumanPNAS,vanGunsteren2001}
\begin{equation}
\epsilon_{\rm r}= 1 + \frac {\langle{\bf M}_{\rm T}\cdot{\bf M}_{\rm T}\rangle 
- \langle{\bf M}_{\rm T}\rangle \cdot 
\langle{\bf M}_{\rm T}\rangle}{3\Omega\epsilon_0 k_{\rm B}T}
\label{eq:ep-M}
\end{equation}
where ${\bf M}_{\rm T}$ is the total dipole moment vector and 
$\Omega$ is the volume
of the system, and $\langle \dots \rangle$ represents averaging over
the simulated ensemble, effectuated here by averaging over snapshots.
The total dipole moment is given by ${\bf M}_{\rm T}=e\sum_i q_i{\bf R}_i$,
where the summation is over monomer beads of the system/subsystem of interest;
$q_i$ and ${\bf R}_i$ are, respectively, the charge and position vector 
of the $i$th bead. 
Our primary focus here is the dielectric contribution from the dipole
solvents, in which case the summation is over the beads of the dipole
dimers. In situations where the contribution from the polyampholytes to the 
dielectric environment is also taken into consideration, the summation 
would encompass all monomer beads in the system.
(Note that the magnitude $M_{\rm T}$ 
in Eq.~S12 of ref.~\citen{SumanPNAS} for $\epsilon_{\rm r}$ should be 
replaced by the vector ${\bf M}_{\rm T}$.)

\begin{center}
   \includegraphics[width=0.56\columnwidth]{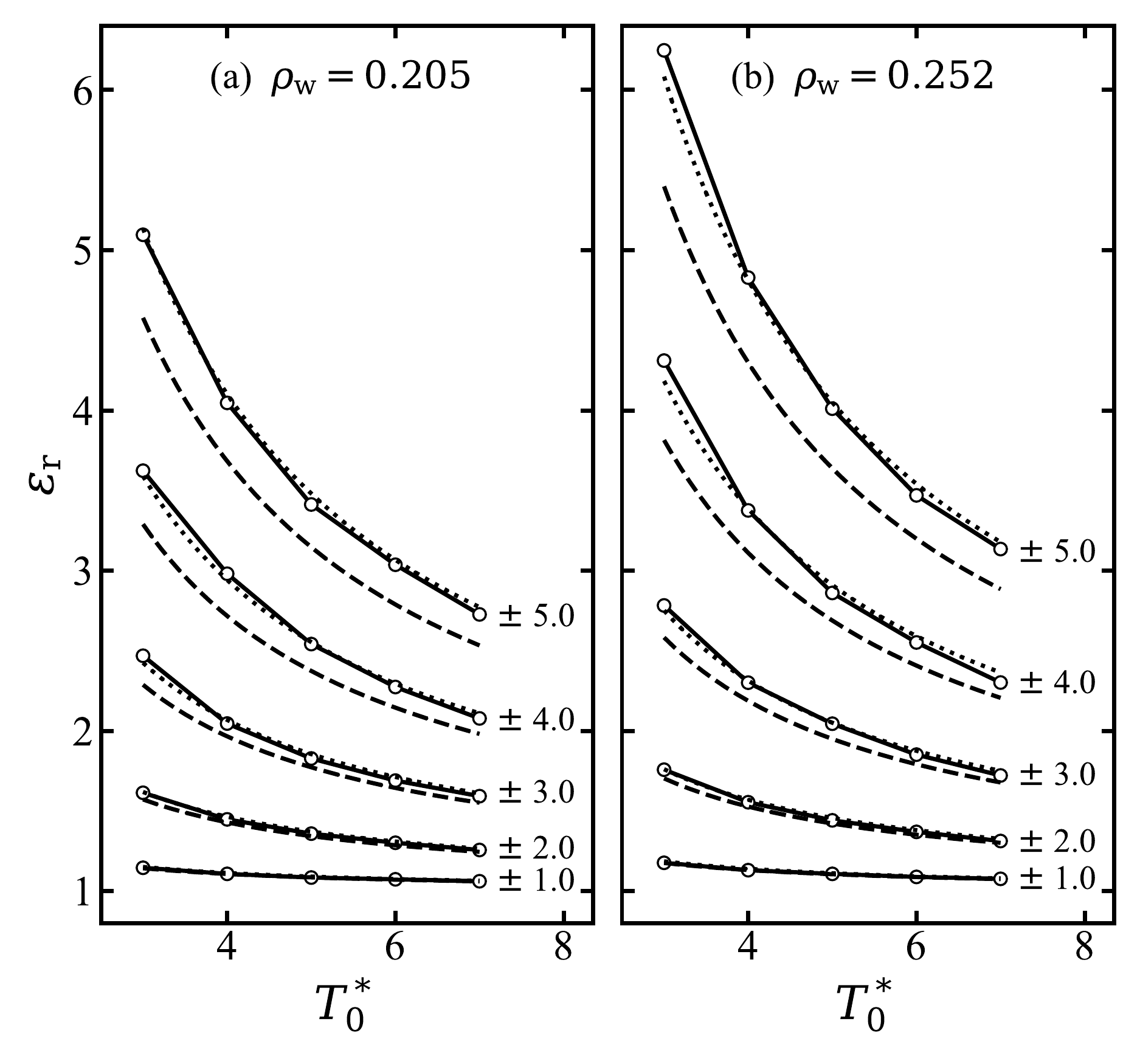}
\end{center}
\vskip -5 mm
{\footnotesize{\bf Fig.~2:} Effective relative permittivity of pure dipoles
as a model of polar solvent. Effective $\epsilon_{\rm r}$ is calculated
by Eq.~\ref{eq:ep-M} with $\rho_{\rm w}=0.205$ (a) and
$0.252$ (b) for dipoles with the indicated $q_{\rm d}$ values (shown
as $\pm |q_{\rm d}|$ next to the curves) as functions of reduced
temperature $T^*_0$.
Solid lines connecting the simulated data points (circles)
are merely a guide for the eye.
Dashed curves representing Eq.~\ref{eq:epr_dipole} are for idealized
uncorrelated dipoles. Dotted curves show the empirical fit
for the different $|q_{\rm d}|$ values by
Eq.~\ref{eq:epr_fit}.
}
$\null$\\

Fig.~2 shows the variation of effective relative permittivity of pure
dipole solvents with respect to dipole density ($\epsilon_{\rm r}$ for
$\rho_{\rm w}=0.205$ and $0.252$ are compared), temperature $T^*_0$, and 
the charges $q_{\rm d}$ on the dipoles that make up the model polar solvent.
As expected, since $\epsilon_{\rm r}$ is positively correlated with the 
ease at which the solvent can be polarized, the simulated data in Fig.~2 
indicate that $\epsilon_{\rm r}$ increases with increasing $|q_{\rm d}|$ and
$\rho_{\rm w}$. This is because that corresponds to having 
stronger charges on the dipoles (hence a larger dipole moment 
$\mu\approx |q_{\rm d}|a$) and the presence, in a given volume, of a 
larger number of dipoles whose orientations can be perturbed by 
introducing other charges into the system. 
As $T^*_0$ increases, however, $\epsilon_{\rm r}$ decreases because 
biased distributions of dipole orientations are disfavored by thermal 
agitation, as reflected partly by the $1/T$ dependence 
of the second term in Eq.~\ref{eq:ep-M}.

Inasmuch as the orientations of dipoles can be assumed to be
uncorrelated, the analytical formula
\begin{equation}
\epsilon_{\rm r} = 1 + \frac {4\pi l_{\rm B}\mu^2}{3}
\left(\frac {\rho_{\rm w}}{2}\right )
,
\;
\label{eq:epr_dipole}
\end{equation}
where $l_{\rm B}=e^2/(4\pi\epsilon_0 k_{\rm B}T)$
is the vacuum Bjerrum length and
$(\rho_{\rm w}/2)$ is the number density of dipoles (because $\rho_{\rm w}$ 
is defined to count two monomer beads per dipole as stated above), 
applies for a collection of fixed dipoles 
with $\mu$ being the magnitude of the individual dipole moment (see, e.g., 
refs.~\citen{orlandPRL2012,orlandPRL2007}). This formula does not provide 
a sufficiently accurate description of our simulation data (dashed curves
in Fig.~2), however, 
because although the magnitude of the dipole moment for a given 
$|q_{\rm d}|$ in our model may be considered practically fixed at 
$\mu=|q_{\rm d}|a$ because of the stiff spring constant $K_{\rm bond}$, 
the orientations of the dipoles are correlated to a degree
because of inter-dipole electrostatic interactions. As expected, 
the correlation among dipole orientations increases with increasing
$|q_{\rm d}|$, leading to larger mismatches seen in Fig.~2 between 
the simulated $\epsilon_{\rm r}$ data points and the dashed curves
representing Eq.~\ref{eq:epr_dipole}. Nonetheless, 
we find that the following empirical formula, 
which takes a form similar to Eq.~\ref{eq:epr_dipole} but with 
an extra multiplicative factor linear in $\mu$, viz.,
\begin{equation}
\epsilon_{\rm r} = 1 + \frac {4\pi l_{\rm B}\mu^2}{3}
\left (\frac {\rho_{\rm w}}{2}\right)
(C_1 + C_2\mu)
\;
\label{eq:epr_fit}
\end{equation}
is capable of providing a reasonably good global fit for all the 
simulated $\epsilon_{\rm r}$ values in Fig.~2 (dotted curves) when
$C_1 = 1.0321441$ and $C_2 = 0.02471877$, thus enabling estimation of
$\epsilon_{\rm r}$ values of dipole systems beyond those we have 
directly simulated.

\begin{center}
   \includegraphics[width=0.53\columnwidth]{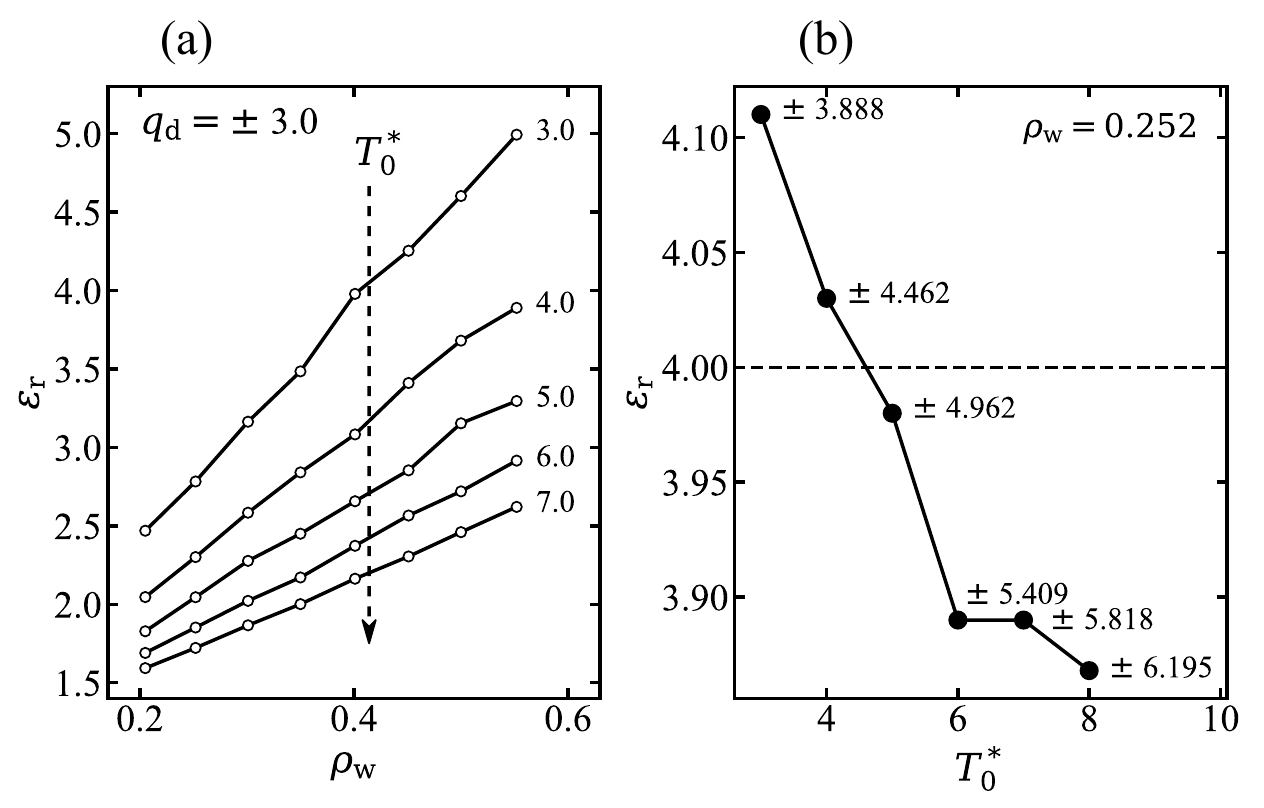}
\end{center}
\vskip -6 mm
{\footnotesize{\bf Fig.~3:} Dependence of effective relative permittivity
of model dipole solvent on density, dipole moment, and temperature.
(a) Variation of effective $\epsilon_{\rm r}$ (Eq.~\ref{eq:ep-M})
with respect to dipole density $\rho_{\rm w}$ and reduced temperature
$T^*_0$ (marked on the right of the curves) for dipoles sharing
the same $|q_{\rm d}|=3.0$.
The vertical dashed arrow across the curves indicates increasing temperature.
(b) At constant $\rho_{\rm w}=0.252$, computed effective $\epsilon_{\rm r}$
(Eq.~\ref{eq:ep-M}) for $q_{\rm d}$s
estimated by Eq.\ref{eq:epr_fit} ($q_{\rm d}$ values indicated beside
simulated data points) to result in an essentially constant
$\epsilon_{\rm r}\approx 4.0$ (marked by dashed horizontal line) at
different temperatures (horizontal scale for $T^*_0$).
Note that the vertical scale covers only a narrow range of
$\epsilon_{\rm r}$ from $\approx 3.85$ to $\approx 4.10$.
Solid lines connecting the simulated data
points (circles in (a) and (b)) are merely a guide for the eye.
}
$\null$\\

Further data in Fig.~3a on the dependence of simulated $\epsilon_{\rm r}$ value
on temperature and dipole density $\rho_{\rm w}$ suggest an approximate linear 
relationship between $\epsilon_{\rm r}$ and $\rho_{\rm w}$. Consistent 
with the observation above that the dipoles are well dispersed, 
and therefore fluid-like, for $T^*_0\ge 3.0$, the variation of 
$\epsilon_{\rm r}$ with $\rho_{\rm w}$ and $T^*_0$ is 
smooth in this set of simulations.  Making use of the trend
in Fig.~3a and the empirical Eq.~\ref{eq:epr_fit},
we obtain a set of $\mu$ values for dipole solvents that would maintain 
an essentially constant $\epsilon_{\rm r}$ over a range of different 
temperatures (Fig.~3b). Because $\epsilon_{\rm r}$ for dipoles with 
a constant $\mu$ decreases with increasing $T^*_0$ (Fig.~2), $\mu$ 
has to increase with increasing $T^*_0$ to maintain an essentially 
temperature-independent $\epsilon_{\rm r}$ (Fig.~3b). 
We will use these $\mu$ values to construct
temperature-independent solvent dielectric environments
for the simulation study of polyampholyte LLPS below.
\\

{\bf Phase Behaviors Of Polyampholytes In Dipole Solvent.} With the baseline
effective relative permittivities of dipole models of polar solvent 
established, we now turn to LLPS behaviors of the polyampholyte chains
immersed in these model solvent molecules. For most of these studies,
we set the effective bulk-solvent $\epsilon_{\rm r}\approx 4.0$. This
relatively small $\epsilon_{\rm r}$ is chosen for computational 
tractability. We refrained from assigning highly positive and negative charges
to the model solvent molecules or having very high densities 
of these molecules in the model solvent. Both of these features
would increase effective $\epsilon_{\rm r}$ but would decrease 
sampling efficiency. As shown below, insights are gained by
the present model into physical principles regarding 
electrostatics-driven LLPS in polar solvent,
$\epsilon_{\rm r}\approx 4.0$ being much smaller than 
$\epsilon_{\rm r}\approx 80$ for water notwithstanding.

Fig.~4 is an overview of our polyampholytes-and-dipoles systems,
showing the density profiles of the polyampholytes 
and the dipoles for the three sequences 
sv15, sv20, and sv30, each at four different simulation temperatures.
A condensed polyampholyte-rich droplet, corresponding to a local peak in
polymer bead density $\rho$ (red curves), is seen in every panel of this
figure. Expectedly, each of these droplets coincides with a local minimum 
in dipole density (blue curves). As temperature increases from top
to bottom in Fig.~4, effects of configurational entropy gain in
significance. Consequently, the packing of polyampholytes in the droplet 
loosens up (lower and broadened peak, red curves)
and a concomitant relative increase in dipole density ensues (shallower 
but wider dip, blue curves). In other words, more solvent molecules 
reside in the polyampholyte-rich phase as temperature increases.

Fig.~4 shows further that the effective $\epsilon_{\rm r}$ contributed
by the dipole solvent decreases inside the polyampholyte-rich droplet 
(dips along green curves at $z$-positions of the droplets), 
tracking the decrease in $\rho_{\rm w}$.
The $z$-dependent relative permittivity 
in Fig.~4 is computed using the formulation in ref.~\citen{HansenJCP2005} 
for ``slab geometry'' in which one spatial dimension ($z$) is distinct from
the other two spatial dimensions ($x,y$).
The $\epsilon_{\rm r}(z)$ plotted in Fig.~4 is the component of the 
diagonal permittivity tensor parallel to the latter two dimensions
i.e., the $x$--$x$, or equivalently the $y$--$y$ compoent, given by
the expression
\begin{equation}
\epsilon_{{\rm r}\parallel}(z)
= 1 + \frac {1}{2\epsilon_0 k_{\rm B}T}\Bigl [
\langle {\bf m}_\parallel(z)\cdot{\bf M}_{{\rm T}\parallel}\rangle 
-\langle {\bf m}_\parallel(z)\rangle\cdot
\langle{\bf M}_{{\rm T}\parallel}\rangle 
\Bigr ]
\;
\label{eq:epr_slab}
\end{equation}
where the two-dimensional vectors ${\bf m}_\parallel(z)$ and 
${\bf M}_{{\rm T}\parallel}$ are both confined to the $x$--$y$ plane.
${\bf m}_\parallel(z)$ is the $x$--$y$ component of the local dipole 
moment density (dipole moment within the local volume sampled divided
by the local volume) and ${\bf M}_{{\rm T}\parallel}$ 
is the $x$--$y$ component of the dipole moment of the entire system, 
and $\langle\dots\rangle$ denotes averaging over snapshots
(Eq.~13 of ref.~\citen{HansenJCP2005}).
Note that expressions for local $\epsilon_{\rm r}$
involves not only local but also global dipole 
moments,\cite{SternFeller2003,HansenJCP2005,Zhu_etal2020} and 
Eq.~\ref{eq:ep-M} can be recovered from Eq.~\ref{eq:epr_slab} by recognizing
that ${\bf m}_\parallel(z)\rightarrow {\bf M}_{{\rm T}\parallel}/\Omega$
when the sampling volume for ${\bf m}_\parallel(z)$ encompasses the 
total volume $\Omega$, and that when the system is isotropic 
after subtracting an overall average dipole moment
$\langle{\bf M}_{{\rm T}\parallel}\rangle$, 
$\langle |{\bf M}_{{\rm T}\parallel}-\langle{\bf M}_{{\rm T}\parallel}\rangle|^2
\rangle$ $=$
$(2/3)\langle |{\bf M}_{\rm T}-\langle{\bf M}_{\rm T}\rangle|^2\rangle$.
In our calculation of ${\bf m}_\parallel(z)$, we consider the dipoles
in slices of thickness $10a$ in the $z$-direction, and include only those
dipoles whose centers of mass are inside the bin.
The $z$--$z$ component of the permittivity tensor---the orthogonal
component---may also be obtained and is expected to yield values similar 
to that given by Eq.~\ref{eq:epr_slab}; but the more limited statistics 
available to this calculation due to the restriction to one spatial 
dimension may present computational 
challenges.\cite{HansenJCP2005} For this reason we do not pursue it
here, while recognizing that Fig.~4, as it stands, illustrates reasonably
well that the solvent dielectric environment experienced
by polyampholyte chains inside the condensed droplet can be quite
different from that outside in the dilute phase.

At temperatures significantly higher than those simulated in Fig.~4, 
condensed
droplets disassemble. An example is provided by the snaphots in Fig.~5 for
sequence sv20 at two temperatures. Consistent with
the $\rho_{\rm w}$ profiles in Fig.~4, at a moderate temperature
of $T^*_0=5.5$, Fig.~5a--c show that although a condensed polyampholyte
droplet is in place, a substantial number of dipole solvent molecules 
are found inside the droplet.

Using methods described above and in previous
studies,\cite{SumanPNAS,dignon18,suman2,panag2017}
we determine the phase diagrams for sequences
sv15, sv20, and sv30, each using five different solvent models (Fig.~6).
The upper critical solution temperature ($T^*_{0,{\rm cr}}$)---above which 
phase separation is impossible---is estimated for each system (Table~1).
When comparison is made across different sequences, LLPS propensity
is seen to increase from sv15 to sv20 to sv30 for all five solvent
models, as quantified by a monotonic increase in $T^*_{0,{\rm cr}}$ 
from left to right for every row in Table~1, consistent with and
generalizing previous findings that LLPS propensity tend to increase 
with increasingly negative values\cite{kings2015,Alan} of the SCD charge 
pattern parameter.\cite{lin2017,suman1,suman2,joanJPCL}

As far as solvent model is concerned, Fig.~6 compares three explicit-solvent
models [(i), (ii), and (iii)] and two implicit-solvent models [(iv) and (v)].
The implicit-solvent model results for sv20 are simulated here anew,
those for sv15 and sv30 are taken from 
ref.~\citen{suman2}. Between these two models, the lower LLPS propensity
of model (v) than model (iv)---for all three sequences---follows
directly from the definition that the LLPS-driving effective electrostatic 
interactions among the polyampholyte chains are weakened 
by a factor of $\epsilon_{\rm r}=4.0$ in model (v) than in model (iv).

$\null$\\
\begin{center}
   \includegraphics[width=0.50\columnwidth]{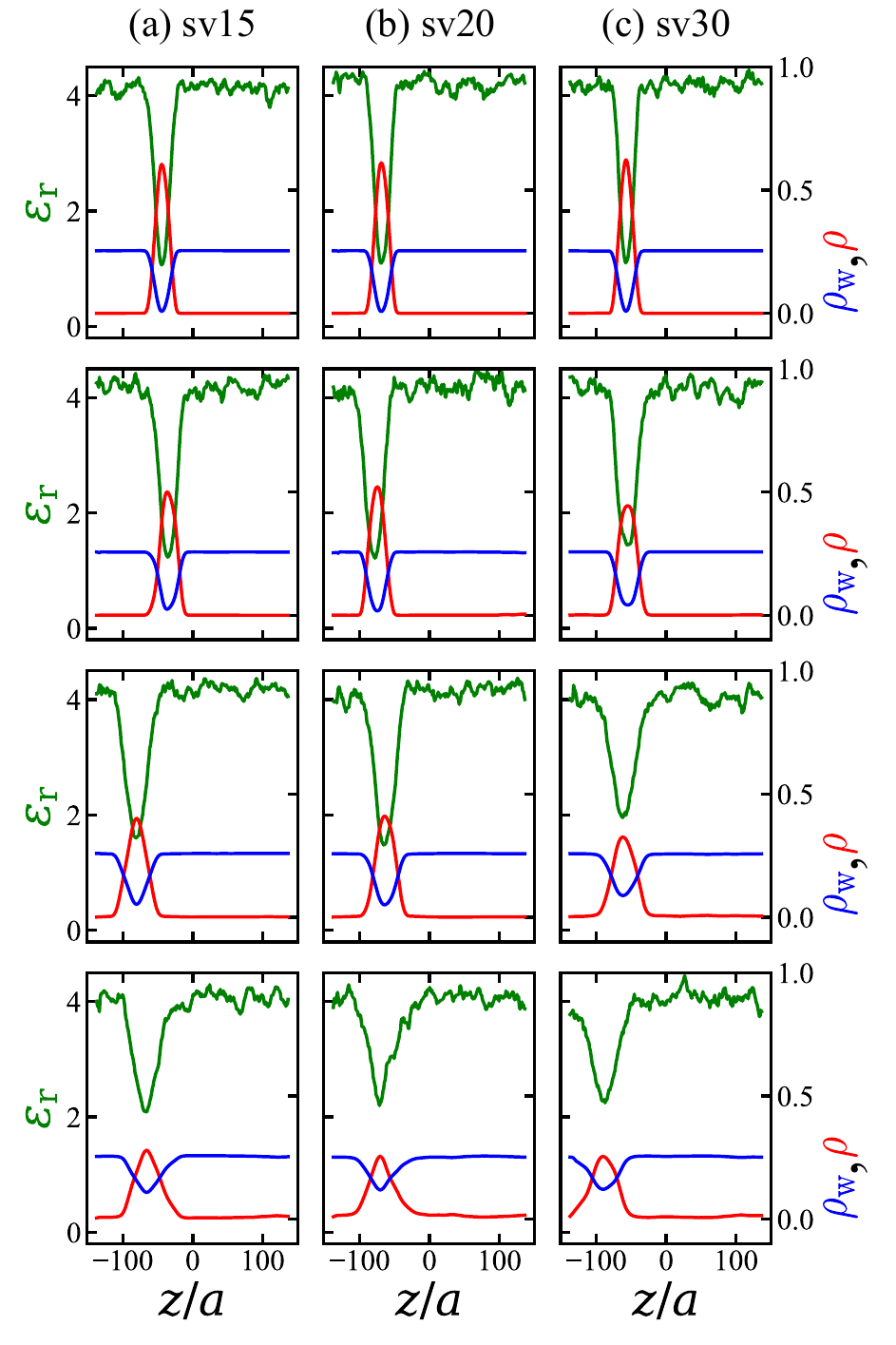}
\end{center}
\vskip -4 mm
{\footnotesize{\bf Fig.~4:}
Density and dielectric profiles of phase-separated polyampholytes
in dipole solvents. Simulation results for systems each containing
$n_{\rm p}=100$ polyampholyte chains and $n_{\rm w}=12,000$ dipoles
are shown for sequences
(a) sv15, (b) sv20, and (c) sv30 at temperatures (top to bottom)
$T^*_0=3.0$, $4.0$, $5.0$, and $5.5$ (a),
$T^*_0=3.0$, $4.0$, $5.0$, and $6.0$ (b), and
$T^*_0=3.0$, $5.0$, $6.0$, and $6.5$ (c), respectively.
Densities of dipoles ($\rho_{\rm w}$, blue curves) and
polyampholytes ($\rho$, red curves) as a function of the horizontal
coordinate $z$ of the simulation box are given by the vertical scale on
the right,
with $\rho_{\rm w}\approx 0.252$ outside of the polyampholyte-rich droplet
(peak region of the red curve) in every case.
Local effective relative permittivity $\epsilon_{\rm r}(z)$ arising
from the dipole molecules (green curves, vertical scale on the left)
is calculated by using Eq.~\ref{eq:epr_slab} for
$\epsilon_{{\rm r}\parallel}(z)$.
The $\rho_{\rm w}(z)$, $\rho(z)$, and $\epsilon_{\rm r}(z)$ profiles are
computed by averaging over successive bins of width $10a$ and plotting
the averages at the midpoint of each bin.
The effective $\epsilon_{\rm r}$ for bulk dipole solvent is
maintained at an essentially constant value of
$\approx 4.0$ at different $T^*_0$s for all the systems considered
in this figure by using the $q_{\rm d}$ values
in Fig.~3b and $q_{\rm d}=\pm 5.191$ for $T^*_0=5.5$ and
$q_{\rm d}=\pm 5.618$ for $T^*_0=6.5$.
}
$\null$\\

\begin{center}
   \includegraphics[width=1.00\columnwidth]{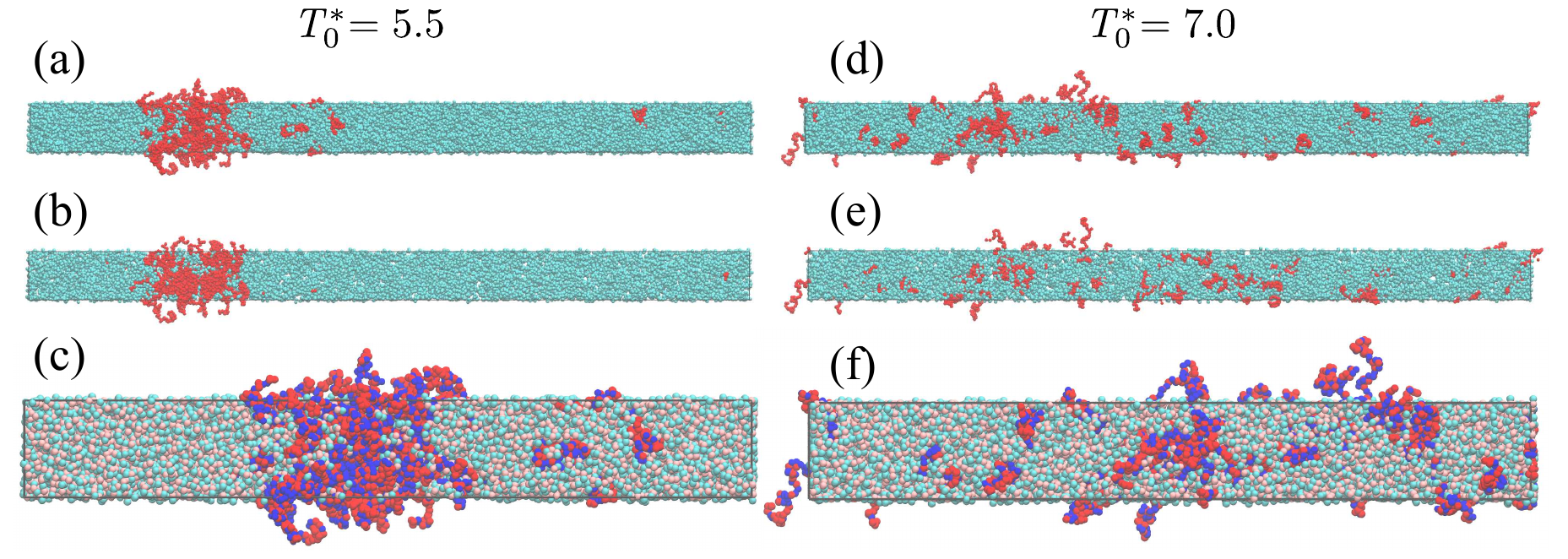}
\end{center}
\vskip -4 mm
{\footnotesize{\bf Fig.~5:} Polyampholyte phase separation in explicit
dipole solvent. Snapshots of $n_{\rm p}=100$ sv20 polyampholyte chains
and $n_{\rm w}=12,000$ dipoles depict a phase-separated condensed
polyampholyte-rich droplet at low temperature 
($T^*_0=5.5$, left column, (a)--(c))
and a lack of phase separation
at a higher temperature ($T^*_0=7.0$, right column, (d)--(f)).
Chains and dipoles at the boundaries of the periodic simulation boxes are
unwrapped. (a) and (d) show an overview of the entire simulation box,
whereas (b) and
(e) show a cross-sectional view of a thin slice parallel to the page.
(c) and (f) are enlarged views of the left halves of (a) and (d),
respectively.
In (a), (b), (d), and (e), polyampholytes are depicted in red, and
dipole solvent molecules are depicted in cyan. In (c) and (f), the positively
and negatively charged beads in the polyampholytes are in blue and red,
whereas the positively and negatively charged beads in the dipoles are in
cyan and pink, respectively (same color code as that in Fig.~1).
}
$\null$\\

The LLPS propensities of the present explicit-solvent models are all 
substantially higher than that of the corresponding implicit-solvent models
for the same sequence because, by construction, the non-electrostatic 
LJ-like part of dipole-polyampholyte interactions is purely 
repulsive (Eqs.~\ref{eq:LJ-pd} and \ref{eq:ULJex}). These repulsive
interactions, probably in conjunction with related configurational
entropy effects, amount to
a hydrophobicity-like,\cite{ChanDill97rev} general solvophobic
driving force\cite{ray1971} against solvation of the polyampholytes
in the dilute phase and thus favoring their condensation. 
While this feature may be viewed as artifactual, as the LJ-like 
interactions of realistic water models are not purely repulsive and
thus engender more subtle solvation effects,
the large offset between the coexistence curves of implicit-solvent 
and explicit-solvent models in Fig.~6 is largely inconsequential to the 
questions of physical principle at hand. As argued below, it is 
appropriate for the present purpose to focus our comparison on LLPS 
properties of the explicit-solvent models.

Among the three explicit-solvent models in Fig.~6, the LLPS propensity 
of model (i) is always higher than
that of model (iii) because they share the same
nonpolar solvent (``dipoles'' with their charges switched off) but
LLPS-driving electrostatic interactions among the polyampholytes are
weakened by a factor of $\epsilon_{\rm r}=4.0$ for model (iii) relative to 
model (i). In this respect, comparison between models (i) and (iii) is 
similar to the comparison between models (iv) and (v).


{{\bf Table 1.} Upper Critical Solution Temperature
$T^*_{0,{\rm cr}}$ for Polyampholyte LLPS Simulated Using
Different Solvent Models in Fig.~6. Values
of the SCD Sequence Charge Pattern Parameter\cite{kings2015} for the
Sequences\cite{rohit2013} sv15, sv20, and sv30 are Provided in Parentheses.}
\begin{center}
\begin{tabular}{cccc} 
\hline 
$\;$ solvent $\;$ & $\quad$ sv15 $\quad$ & $\quad$ sv20 $\quad$ 
& $\quad$ sv30 $\quad$\\
$\;$ model $\;$ & $\quad$ (SCD $=-4.35$) $\quad$ & 
$\quad$ (SCD $=-7.37$) $\quad$ & $\quad$ (SCD $=-27.8$) $\quad$ \\
 \hline 
(i) & $6.2$ & $6.6$ & $7.1$\\
(ii) & $5.7$ & $6.2$ & $7.0$\\
(iii) & $5.8$ & $5.9$ & $6.3$\\
(iv) & $3.9$ & $4.0$ & $5.0$\\
(v) & $3.6$ & $3.6$ & $3.8$\\
\hline
\end{tabular}  
\end{center}

$\null$\\
\begin{center}
   \includegraphics[width=0.71\columnwidth]{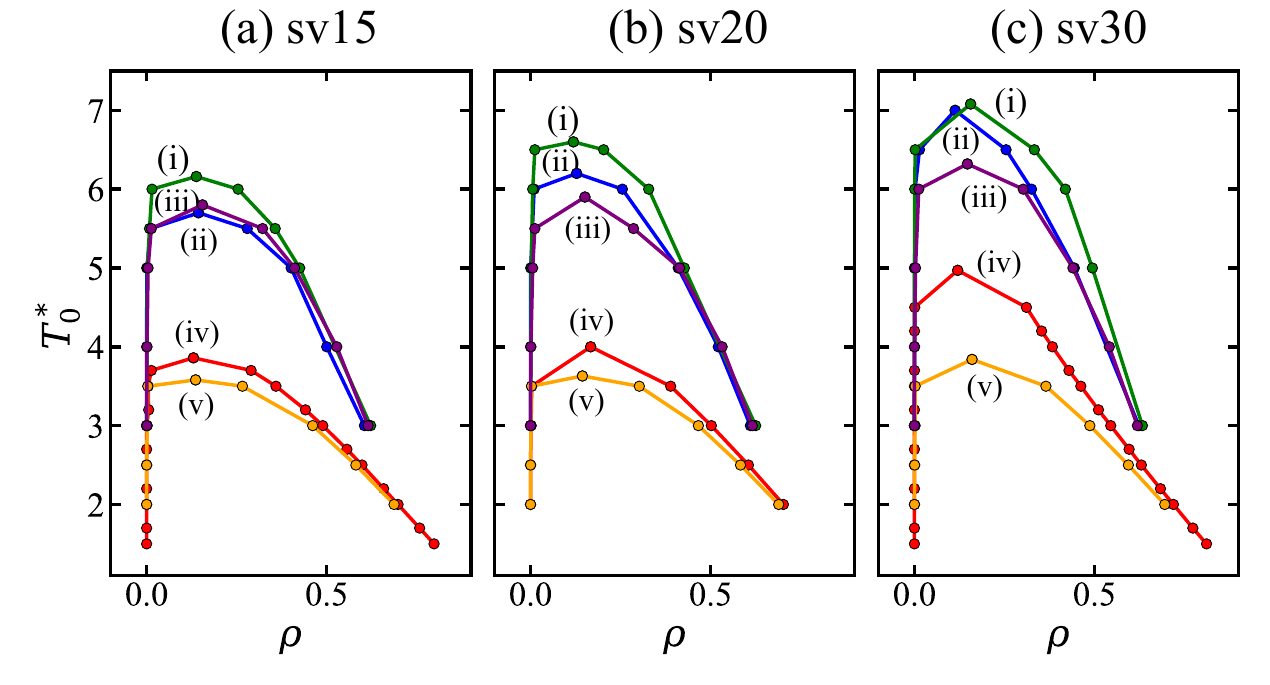}
\end{center}
\vskip -4 mm
{\footnotesize{\bf Fig.~6:} Phase diagrams of polyampholytes in
implicit solvents and explicit dipole and nonpolar solvents.
Coexistence curves (i), (ii), and (iii) for sequences sv15 (a), sv20 (b),
and sv30 (c) are determined by simulations of systems each consisting
of $n_{\rm p}=100$ polyampholyte chains and $n_{\rm w}=12,000$ dipoles
($|q_{\rm d}|>0$) or uncharged dimers ($q_{\rm d}=0$).
The charges $q_{\rm d}$ on the dipoles at different temperatures are
adjusted by Eq.~\ref{eq:epr_fit} such that the effective
relative permittivity arising from the dipoles are essentially constant,
at $\epsilon_{\rm r}\approx 4.0$, for all temperatures.
Corresponding coexistence curves (iv) and (v), determined by
implicit-solvent simulations (no explicit solvent molecules) using
$n_{\rm p}=500$ polyampholyte chains,\cite{suman2} are included for comparison.
$T^*_0$ is vacuum reduced temperature, $\rho$ is density of
polyampholyte monomer beads in units of $1/a^3$. Lines joining computed
data points (circles) are merely a guide for the eye.
The five different solvent models [(i)--(v)] are:
(i) Explicit nonpolar solvent, i.e., ``dipoles'' with
$q_{\rm d}=0$, $\epsilon_{\rm r}=1$ for polyampholyte electrostatics
(green curves).
(ii) Explicit dipole solvent, with different $|q_{\rm d}|>0$ for different
$T^*_0$ to maintain an essentially $T^*_0$-independent effective bulk-solvent
$\epsilon_{\rm r}\approx 4.0$, $\epsilon_{\rm r}=1$ for all electrostatic
interactions (blue curves).
(iii) Explicit nonpolar solvent, same as (i) but $\epsilon_{\rm r}=4.0$ is
used for polyampholyte electrostatics (magenta curves).
(iv) Implicit-solvent, $\epsilon_{\rm r}=1$ (red curves).
(v) Implicit-solvent, $\epsilon_{\rm r}=4.0$ (orange curves).
The highest $T^*_0$ value of a coexistence curve for a given system
is its upper critical solution temperature $T^*_{0,{\rm cr}}$. The
$T^*_{0,{\rm cr}}$ values for all the systems considered in this
figure are provided in Table~1.
}

For our purpose, the most physically
informative comparison in Fig.~6 is between model (ii) and
model (iii). The two models share the same repulsive LJ-like 
polyampholyte-solvent interactions, which serve as a baseline for
comparing, on one hand, the LLPS behavior of polyampholytes immersed in 
explicit dipole solvent molecules with dipole charges tuned to provide an 
essentially temperature-independent effective bulk-solvent 
$\epsilon_{\rm r}\approx 4.0$ with all charges in the system 
interacting via the full Coulomb potential, and on the other hand,
the LLPS behavior of polyampholytes immersed in nonpolar solvent
molecules interacting via a Coulomb potential with strength
reduced by $\epsilon_{\rm r}=4.0$.
At a conceptual level, this comparison serves to assess the accuracy,
or potential pitfall, of the common practice of using the bulk-water
$\epsilon_{\rm r}$ in implicit-solvent simulations of 
LLPS,\cite{jeetainACS,Schuster2020,singperry2017,dignon18,suman1,jeetainPNAS,koby2020,joanElife}
although the temperature dependence of 
implicit-solvent effective interactions\cite{jeetainACS,Roland2019}
and bulk-water $\epsilon_{\rm r}$
(which varies from $\approx 88.7$ to $55.7$
from $0^\circ$ to $100^\circ$C)\cite{waterdielectric}  
should also be taken into account in real-world applications.

Comparisons between the blue (ii) and magenta (iii) curves in Fig.~6
indicate that there is, in general, an appreciable difference in 
predicted LLPS properties between the two approaches. The observed differences  
are not large. It can be very minor for the sequence with a low 
$-$SCD value (sv15) but are larger for the sequence with a high $-$SCD
value (sv30). The critical temperature $T^*_{0,{\rm cr}}$ for the
model with explicit dipole solvent molecules and full electrostatic
interactions (blue curve) is higher 
than the corresponding $T^*_{0,{\rm cr}}$ for the model with nonpolar
solvent molecules but reduced electrostatic interactions (magenta curve) 
for sv20 and sv30. However, the reverse is true for sv15,
albeit the difference in this case is very small. Taken together,
these observations suggest that although there is a tendency 
for implicit-solvent simulations of LLPS using bulk-water 
$\epsilon_{\rm r}$ to underestimate LLPS propensity,
there is no absolute rule against it giving a small
overestimation of LLPS propensity, as that may well 
depend on the structural and energetic details of the system. 
\\


{\bf Analytical Theory of Pure Dipole Solvent Systems.}
To assess the generality of the above observations from
molecular dynamics simulations, we have also developed corresponding 
analytical theories that treat both polyampholyte and dipole degrees 
of freedom, paying particular attention to excluded volume effects of the 
polyampholyte chains and the dipoles as model solvent molecules.
Analytical formulations are useful for conceptual understanding 
and hypothesis evaluation. As a theoretical tool, they are complementary
to molecular dynamics simulations. Although analytical theories 
lack the structural and energetic details of molecular dynamics models,
computation needed for numerical solutions to analytical 
formulations such as RPA\cite{SumanPNAS,kings2020} and 
field-theoretic simulations (FTS)\cite{joanJPCL,joanPNAS,joanJCP,Pal2021}
are often significantly more efficient than molecular dynamics 
simulations, thus allowing more physically informative
variations in modeling setup to be explored.

As for the explicit-chain, explicit-solvent simulation study above, our
analytical development begins
with the baseline case of pure dipoles as a model of polar solvents. 
Consider a system of volume $\Omega$ containing $n_{\mathrm{w}}$ 
dipoles subject to a local incompressibility condition to account
for dipole excluded volume,
\begin{equation} \label{eq:incomp_cond}
\rho_{\rm w}({\bf{r}}) = \frac{1}{v} \, ,
\end{equation}
where $\rho_{\mathrm{w}}({\bf{r}})$
is the number density at the spatial position ${\bf{r}}$
and $v$ is the volume assigned to each dipole. (Note that each dipole is
counted once by the $\rho_{\rm w}$ here whereas the bead density for the
dipoles in the molecular dynamics simulations, denoted by the same 
symbol $\rho_{\rm w}$, counts each dipole twice because it contains two beads;
hence the $\rho_{\rm w}$ value in the present analytical theory formulation
is equivalent to $(\rho_{\rm w}/2)$ in the molecular dynamics results 
reported above).
With only dipoles in the system,
$v^{-1}=n_{\mathrm{w}} / \Omega$. We let ${\bf{r}}_i$ and $\bm{\mu}_i$ denote the
position and dipole moment of individual 
dipole $i$, and model each dipole as a soft
sphere following a Gaussian distribution $\Gamma({\bf{r}})$ of width $\bar{a}$.
The local number density $\rho_{\rm w}({\bf{r}})$ and dipole moment density 
$\dipole({\bf{r}})$ (i.e. the polarization) can
then be re-expressed in their respective smeared forms:
\begin{subequations}
\begin{align}
\rho_{\rm w }({\bf{r}}) &= \sum_{i=1}^{n_{\mathrm{w}}} \Gamma({\bf{r}}-{\bf{r}}_i)            \\
\dipole({\bf{r}}) &= \sum_{i=1}^{n_{\mathrm{w}}} \bm{\mu}_i \Gamma({\bf{r}}-{\bf{r}}_i)
\end{align}
\label{eq:smearing}
\end{subequations}
with the Gaussian smearing function
$\Gamma({\bf{r}}) = \exp(- {\bf{r}}^2/2 \bar{a}^2 )/(2\pi\bar{a}^2)^{3/2}$.
This smearing procedure was introduced in
ref.~\citen{Wang2010} 
and is a convenient way of handling ultraviolet (short-distance)
divergences arising from point
interactions. A spatially varying $\dipole({\bf{r}})$ gives rise to a bound
charge density $\rho_{\rm c}({\bf{r}}) = - \bm{\nabla} \cdot \dipole({\bf{r}})$. We
consider two distinct cases where the dipoles are either given freely
orientable dipole moments of fixed, i.e., permanent, magnitude, 
$|\bm{\mu}_i |= \muwr $ (``fixed
dipoles''), or where every dipole is associated with a harmonic energy
contribution $\kappa_{\rm d} \bm{\mu}_i^2/2$ (``polarizable dipoles'')
where $\kappa_{\rm d}$ is the corresponding spring constant.

The canonical partition function for this system is
\begin{equation} \label{eq:part_func_pure_dipole}
\Z = \frac{1}{n_{ \rm w }! } \prod_{i=1}^{ n_{\rm w} } \int \dint {\bf{r}}_i \int \dint \bm{\mu}_i \, {\rm e}^{-\UU} \delta \left[ \rho_{\rm w }({\bf{r}}) - \frac{1}{v} \right] \, ,
\end{equation}
where the functional $\delta$-function implements the incompressibility 
condition in Eq.~\ref{eq:incomp_cond} and $\UU$ is the Hamiltonian. In the case
of fixed dipoles, the integrations over $\bm{\mu}_i$ are reduced to solid angle
integrals, and $\UU$ is the total electrostatic potential energy,
\begin{equation}
\UU = \frac{1}{2 \Omega} \sum_{{\bf{k}} \neq 0} \hat{\rho}_{\rm c}({\bf{k}}) 
\frac{4 \pi l_{\rm B } }{{\bf{k}}^2} \hat{\rho}_{\rm c }(-{\bf{k}}) = 
\frac{\lb }{2} \int \dint {\bf{r}} \int \dint {\bf{r}}' 
\frac{\rho_{\rm c}({\bf{r}}) \rho_{\rm c}({\bf{r}}')}{|{\bf{r}}-{\bf{r}}'|}
\label{eq:UU-basic-def}
\end{equation}
where $\hat{\rho}_{\rm c}({\bf{k}}) = \ii {\bf{k}} \cdot \hat{\dipole}({\bf{k}}) $ is
the Fourier transform of $\rho_{\mathrm{c}}({\bf{r}})$, $\lb$ is the vacuum
Bjerrum length as defined above, and $\ii^2=-1$. 
Note that the ${\bf{k}} = \bm{0}$ mode can be dropped in the above summation 
because overall charge neutrality of the system entails
$\hat{\rho}_{\rm c}({\bf{k}}=\bm{0})=0$.
In the case of polarizable dipoles, $\UU$ contains the additional term
$\sum_{i=1}^{n_{\rm w}} \beta \kappa_{\rm d} \bm{\mu}_i^2/2$ where  
$\beta \equiv 1/\kB T$.

A formula for the relative permittivity $\epsilon_{\rm r}$ arising from
the dipoles can be
obtained by imagining adding a small test charge to the system and looking at
the induced electrostatic potential far away from the test charge.
As derived in ref.~\citen{Chandler1977}, this consideration yields a
relation between $\epsilon_{\rm r}$ and the charge-charge correlation function:
\begin{equation}
\frac{1}{\epsilon_{\rm r} } = \lim_{{\bf{k}}\rightarrow \bm{0}} 
\left( 1 - \frac{4 \pi \lb }{{\bf{k}}^2 \Omega} 
\langle \hat{\rho}_{\rm c}({\bf{k}}) \hat{\rho}_{\rm c}(-{\bf{k}}) \rangle \right) 
\, ,
\label{eq:epr-basic}
\end{equation}
where $\langle \dots \rangle$ denotes the thermal average over $\lbrace {\bf{r}}_i, \bm{\mu}_i \rbrace$.

Following a standard procedure using Hubbard-Stratonovich 
transformations,\cite{Edwards1965,Fredrickson2006} 
the partition function in Eq.~\ref{eq:part_func_pure_dipole} can be turned into
a statistical field theory where the $\lbrace {\bf{r}}_i, \bm{\mu}_i \rbrace$
variables are traded in favor of charge- and number density conjugate fields
$\psi({\bf{r}})$ and $w({\bf{r}})$, leading to
\begin{equation}
\Z = \frac{\Omega^{n_{\rm w}} }{n_{\rm{w}}! } \int \DD w \DD \psi \, {\rm e}^{-
\HH } \, 
\label{eq:Z-integral}
\end{equation}
up to an inconsequential overall multiplicative constant, where
\begin{equation}
\HH[ w, \psi ] = - \nw\ln\Qw + 
\frac{1}{2\Omega}\sum_{\kk\neq\kzero}
\frac{ \kk ^2}{4\pi\lb}\hat{\psi}(\kk)\hat{\psi}(-\kk)  
- \frac{ \ii }{v} \hat{w}(\bf{0}) \, ,
\label{eq:H_0}
\end{equation}
and
\begin{equation}
\int \DD w\DD\psi =  \prod_{\kk\neq\kzero} \sqrt{\frac{\kk^2 }{4\pi\lb v}}\int \dint\hat{w}(\kk) \int \dint\hat{\psi}(\kk)
        \label{eq:field_int_factor}
\end{equation}
in $\kk$-space representation, with $\hat{\psi}(\kk)$ and
$\hat{w}(\kk)$ being the Fourier transforms, respectively, of
$\psi({\bf{r}})$ and $w({\bf{r}})$.
The field Hamiltonian $\HH$ is thus expressed in terms of the 
Fourier transformed fields $\hat{w}$ and $\hat{\psi}$. 
In Eq.~\ref{eq:H_0}, $\Qw$ 
is a single dipole partition function,
whose form follows from the Hubbard-Stratonovich 
transformation.\cite{GHFJCP2016} It is
\begin{equation}
\label{Qwfixed}
\Qw =  \frac {1}{\Vol}\int \dint {\bf{r}}   \exp\left\{
- \frac{ \ii }{\Vol}\sum_{\kk} \hat{w}(\kk) \hat{\Gamma}(-{\bf{k}}) {\rm e}^{ - \ii \kk\cdot {\bf{r}} }
\right\}   
\frac{\sinh\left[ (1/\Vol)\sum_{\kk\neq\kzero} |\kk| \muwr 
\hat{ \psi }( \kk) \hat{\Gamma}(-{\bf{k}}) {\rm e}^{ - \ii \kk\cdot  {\bf{r}} } 
\right]}{(1/\Vol)\sum_{\kk\neq\kzero} |\kk| \muwr  
\hat{\psi}(\kk) \hat{\Gamma}(-{\bf{k}}) {\rm e}^{-  \ii \kk\cdot {\bf{r}} } }
\end{equation}
for fixed dipoles, and
\begin{equation}
\label{Qwpolarizable}
\Qw =  \frac{1}{\Vol} \int \dint {\bf{r}} \exp\left\{
- \frac{  \ii }{\Vol}\sum_{\kk} \hat{w}(\kk) \hat{\Gamma}(-{\bf{k}}){\rm e}^{-  \ii \kk\cdot {\bf{r}} }
\right\}  \exp\left\{
\frac{\left[ \sum_{\kk\neq\kzero}\kk \hat{\psi}(\kk) \hat{\Gamma}(-{\bf{k}}) {\rm
e}^{- \ii \kk\cdot {\bf{r}} } \right]^2}{2\beta\kappa_{\rm d}\Vol^2} \right\}
\end{equation}
for polarizable dipoles, where the Fourier-transformed smearing
function $\hat{\Gamma}({\bf{k}}) = \exp(- \bar{a}^2 {\bf{k}}^2 / 2) $. 
The second factor in the integrands of Eqs.~\ref{Qwfixed} and
\ref{Qwpolarizable} (involving the $\sinh$ function in Eq.~\ref{Qwfixed}
and the $\exp$ function in Eq.~\ref{Qwpolarizable}) 
results from the dipole moment integrals $\int \dint
\hat{\bm{n}} \, \exp( - \ii \mu \hat{\bm{n}} \cdot \bm{\nabla} \breve{\psi})$ 
(where $\hat{\bm{n}}$ is the unit vector along the dipole moment) and $\int
\dint \bm{\mu} \, \exp(- \beta \kappa \bm{\mu}^2 / 2 - \ii \bm{\mu} \cdot
\bm{\nabla}\breve{\psi} )$, respectively, with $\breve{\psi}(\bm{r}) \equiv
\int \dint \bm{r}' \Gamma(\bm{r}-\bm{r}') \psi(\bm{r}')$. For the case of fixed
dipoles, the integral is most easily computed in the frame of reference where
the $z$-component of $\hat{\bm{n}}$ points along $\bm{\nabla}\breve{\psi}$. In
calculating these integrals, overall multiplicative constants have been dropped
as they are irrelevant for the computations that will follow.  
Recognizing that the charge-charge correlation function in 
Eq.~\ref{eq:epr-basic} 
may be rewritten in a field-theory representation
involving the correlation function of the $\psi$ field,\cite{Fredrickson2006}
\begin{equation}
\frac{1}{\Vol} \langle \hat{\rho}_{\rm c}(\kk) \hat{\rho}_{\rm c}(-\kk) \rangle 
= \frac{ \kk^2 }{ 4 \pi \lb } \left[ 1 - \frac{ \kk^2 }{ 4 \pi \lb } 
\frac{ \langle \hat{\psi}(\kk) \hat{\psi}(-\kk) \rangle }{\Vol} \right] 
\; ,
\end{equation}
and substituting this expression into 
Eq.~\ref{eq:epr-basic} leads to the field-theory expression for 
the relative permittivity,
\begin{equation} 
\label{eq:eps_r_field_def}
\frac{1}{\epsilon_{\rm r} } = \lim_{{\bf{k}}\rightarrow \bm{0}} \frac{{\bf{k}}^2}{4
\pi \lb \Vol} \langle \hat{\psi}(-{\bf{k}}) \hat{\psi}( {\bf{k}}) \rangle \, ,
\end{equation}
where now $\langle \dots \rangle$ denotes an average over field 
configurations weighted by $\exp(-\HH)$.

We next proceed to the calculation of this $\psi$ correlation function in order 
to compute $\epsilon_{\rm r}$. As a first approximation---which
corresponds to RPA---we compute it analytically accounting only for 
Gaussian fluctuations in the fields. 
Introducing
$\ket{\Psi(\kk)} = [\hat{w}(\kk), \hat{\psi}(\kk) ]^{\rm T}$,
where the superscript ``T'' denotes (vector) transposition,
the field Hamiltonian in this approximation is
\begin{equation} 
\label{eq:rpa_field_hamiltonian}
\HH \approx \frac{1}{2 \Vol } \sum_{\kk\neq \kzero} 
{\bra{\Psi(-\kk)} \RPAM(\kk) 
\ket{\Psi(\kk)}} \, ,
\end{equation}
which contains only terms quadratic in the fields, and where
\begin{equation}
\RPAM(\kk) \to \RPAM_k
=
\begin{pmatrix}
\hat{\Gamma}_k^2/v & 0 \\
0 & ({k^2}/{4 \pi \lb})\left(1 + \chi_{\rm D} \hat{\Gamma}_k^2 \right)
\end{pmatrix} \, 
\label{eq:Matrix}
\end{equation}
with $k = |\kk|$, $\hat{\Gamma}_k$ standing for $\exp(-\bar{a}^2k^2/2)$, 
and
\begin{equation}
\chi_{\rm D} = \left\{
\begin{aligned}
\frac{4\pi\lb \muwr^2}{3 v } & \quad \text{(fixed dipoles)} \\
\frac{ 4\pi K}{ v } \quad & \quad \text{(polarizable dipoles)}
\end{aligned}
\right.
\label{eq:RPAchiD}
\end{equation}
where $K \equiv \lb / \beta \kappa_{\rm d}$ is a temperature-independent factor
corresponding to the molecular polarizability in our reduced units. 
When Eq.~\ref{eq:eps_r_field_def} is computed using
the RPA Hamiltonian Eq.~\ref{eq:rpa_field_hamiltonian}, the $\psi$ correlation 
is simply the $\psi\psi$ element of $\RPAM^{-1}$, which is equal
to the reciprocal of the $\psi\psi$ component of $\RPAM$ (bottom-right
element) in Eq.~\ref{eq:Matrix}. It follows that 
\begin{equation} 
\label{eq:eps_r_RPA_pure_dipoles}
\epsilon_{\rm r} = 1 + \chi_{\rm D} \, 
\end{equation}
in RPA, yielding the standard formula in classical electrodynamics for 
uncorrelated dipoles. Eq.~\ref{eq:eps_r_RPA_pure_dipoles} with the 
$\chi_{\rm D}$ expression for fixed dipoles in Eq.~\ref{eq:RPAchiD} is 
equivalent to Eq.~\ref{eq:epr_dipole} for analyzing
molecular dynamics simulation data on pure dipole systems in Fig.~2.

Going beyond RPA, we can calculate the $\psi$ correlation function 
more accurately by using FTS on a lattice.\cite{FredricksonGanesanDrolet2002}
In FTS, the continuum fields $w({\bf{r}})$ and $\psi({\bf{r}})$ are approximated by
discrete field variables defined on the sites of a cubic lattice with periodic
boundary conditions. Due to the complex nature of $\HH$, field averages (such
as the $\psi$ correlation function) are typically computed in FTS using the
Complex-Langevin (CL) method,\cite{Parisi1983,Klauder1983}---a technique
inspired by stocahstic quantization of field 
theories\cite{ParisiWu1981,HSCMartyGhost,HSCMartyGravity}---where a fictitious 
time coordinate is introduced such that $w({\bf{r}})\rightarrow w({\bf{r}},t)$ and
$\psi({\bf{r}}) \rightarrow \psi({\bf{r}},t)$ and the fields are
now complex.
The fields evolve in CL time according to the Langevin equation
\begin{equation} \label{eq:CL_eqs}
\frac{\partial \varphi({\bf{r}},t) }{\partial t} = - \frac{\delta \HH}{\delta \varphi({\bf{r}},t) } + \eta_{\varphi}({\bf{r}},t) \, , \quad \varphi = w, \psi \, ,
\end{equation}
where $\eta_{\varphi}$ is a real-valued Gaussian noise with zero mean and
$\langle \eta_{\varphi}({\bf{r}},t) \eta_{\varphi'}({\bf{r}}',t') \rangle = 2
\delta_{\varphi,\varphi'} \delta({\bf{r}}-{\bf{r}}') \delta(t-t')$. In the CL
method, field averages can be shown to correspond to asymptotic CL time
averages. In this work, we use the first order semi-implicit method of
ref.~\citen{Lennon_etal_2008} to solve Eq.~\ref{eq:CL_eqs} numerically.

Fig.~7 shows $\epsilon_{\rm r}$ computed by FTS for several representative 
parameter values. The FTS $\psi$ correlation functions were computed on 
a $24^3$ lattice with a
lattice spacing $\Delta x = \bar{a} = 1/\sqrt{6}$ and a time step 
$\Delta t = 10^{-3}$. After an initial equilibration period of $2 \times 10^6$
time steps, the fields were sampled every 1,000 steps for $\sim 8 \times
10^6$ steps. 
Fig.~7
shows the mean and standard deviation among three independent simulations that
were performed for each value of $\lb$.  All simulations used $v=0.1$. The
results for
fixed dipoles were obtained for $\muwr=0.2$ and 
$\muwr=0.35$. The polarizable dipoles were simulated using $K=0.151197$
which gives $\epsilon_{\rm r}=20$ according to
Eqs.~\ref{eq:RPAchiD} and \ref{eq:eps_r_RPA_pure_dipoles}. 
With the relative permittivity of the polymer material set to unity in 
the following analytical theory development, the temperature-independent 
$\epsilon_{\rm r}$ value of 20 corresponds approximately to the
ratio of the relative permittivities of $\approx 80$ for water and $\sim 4$ 
for proteins (see, e.g., refs.~\citen{linJML,njp2017,SumanPNAS,honig} and
Discussion below).
For the parameters considered,
Fig.~7
shows that the FTS results follow the RPA formulas in
Eqs.~\ref{eq:RPAchiD} and \ref{eq:eps_r_RPA_pure_dipoles} reasonably well.


It is instructive to compare our predicted $l_{\rm B}$-independent
$\epsilon_{\rm r}$ for polarizable dipoles in Fig.~7 to the results of a recent 
field-theoretic perturbation theory for polymers of polarizable 
beads.\cite{GHFJCPchi2018} Interestingly, while Figs.~8b and 9b of 
ref.~\citen{GHFJCPchi2018}
show that field fluctuations beyond Gaussian order affect the value 
of $\epsilon_{\rm r}$ for the compressible case 
($B=1$ in the notation of their system) leading to an $\epsilon_{\rm r}$
that varies with $l_{\rm B}$, their one-loop contribution to $\epsilon_{\rm r}$
vanishes for the incompressible case ($B=\infty$) when there is 
only one kind of beads in the system (their $\alpha_A=\alpha_B$) leading
to an $\epsilon_{\rm r}$ essentially independent of $l_{\rm B}$
(Eq.~29 of ref.~\citen{GHFJCPchi2018}). Because our pure dipole systems
is incompressible and the $\alpha_A=\alpha_B$ case in 
ref.~\citen{GHFJCPchi2018} is akin to our polarizable dipole system,
the finding in ref.~\citen{GHFJCPchi2018}
that the one-loop contribution vanishes
for $\alpha_A=\alpha_B$, $B=\infty$ provides additional rationalization
for the observation in Fig.~7 of good agreement between RPA and FTS as well 
as an $l_{\rm B}$-independent $\epsilon_{\rm r}$ for polarizable dipoles. 

\begin{center}
   \includegraphics[width=0.46\columnwidth]{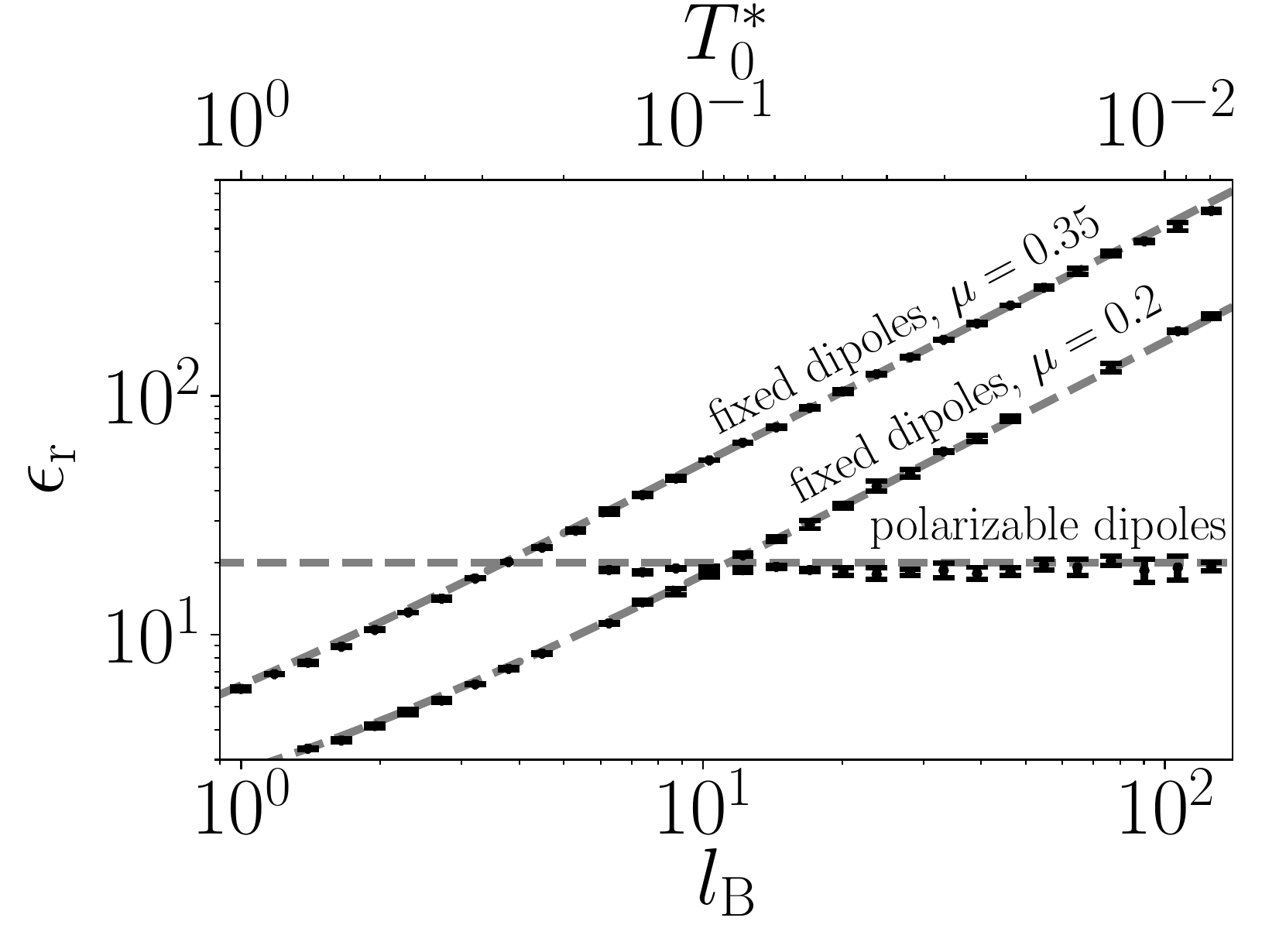}
\end{center}
\vskip -4 mm
{\footnotesize{\bf Fig.~7:}
Relative permittivity computed by FTS for incompressible pure dipoles.
Each error bar indicates the standard deviation from three independent
FTS runs. Dashed lines correspond to RPA Eq.~\ref{eq:eps_r_RPA_pure_dipoles}.
The reduced temperature, $T^*_0$ (top horizontal scale), used in the
analytical theories presented in this work is defined by
$T^*_0=l/l_{\rm B}$, where $l=\sqrt{6}\bar{a}$ is the reference polymer
bond length of the polyampholyte chains\cite{kings2020} to be discussed below.
The magnitude of the dipole moment, $\mu$, is in units of $el$.
$l$ is equivalent to the reference bond length $a$
used in Eq.~\ref{eq:T_red0} for $T^*_0$ in the molecular dynamics study.
}
$\null$\\


{\bf Analytical Theory for Mixture of Dipole and Nonpolar Solvents.}
Before introducing polymers to the pure dipole systems described above,
it is instructive to first consider the simpler case of an incompressible 
mixture of fixed dipoles and nonpolar (and nonpolarizable) particles
as a model of nonpolar solvent molecules. For this purpose,
we add $n_{\rm n}$ Gaussian-smeared particles with positions 
${\bf{r}}_{{\rm n}, i}$ and modify the incompressibility condition in
Eq.~\ref{eq:incomp_cond} to
\begin{equation}
\rho_{\rm w}({\bf{r}}) + \rho_{\rm n}({\bf{r}}) = \frac{1}{v}  \, ,
\end{equation}
where $ \rho_{\rm n}({\bf{r}}) = \sum_{i=1}^{n_{\rm{n}}}
\Gamma({\bf{r}}-{\bf{r}}_{{\rm n},i})$. The canonical partition function for this
system is
\begin{equation} 
\label{eq:part_func_dipole_neutral}
\Z = \frac{1}{n_{ \rm w }! n_{ \rm n }! }\prod_{i=1}^{ n_{\rm n} }  \int \dint
{\bf{r}}_{{\rm n},i} \prod_{i=1}^{ n_{\rm w} } \int \dint {\bf{r}}_i \int \dint
\bm{\mu}_i \, {\rm e}^{-\UU} \delta \left[ \rho_{\rm w }({\bf{r}}) + \rho_{\rm
n}({\bf{r}}) - \frac{1}{v} \right] \, , 
\end{equation}
where $\UU$ is unchanged. Accordingly, the field theory version of 
the partition function is given by
\begin{equation}
\Z = \frac{ \Vol^{ n_{\rm w} + n_{\rm n} } }{ n_{\rm w}! n_{\rm n}! }  
\int \DD w \int \DD \psi \, {\rm e}^{- \HH } \, ,
\end{equation}
where now 
\begin{equation} \label{eq:sol_mix_field_hamiltonian}
\HH[ w, \psi ] = - \nw\ln\Qw - n_{\rm n} \ln\Qn + \frac{1}{2\Omega}\sum_{\kk\neq\kzero}\frac{ \kk ^2}{4\pi\lb}\hat{\psi}(\kk)\hat{\psi}(-\kk)  - \frac{ \ii }{v} \hat{w}(\bf{0})
\end{equation}
includes a term with the partition function $\Qn$ for a single 
nonpolar particle, viz.,
\begin{equation}
\Qn =  \frac{1}{\Vol} \int \dint {\bf{r}}  \exp\left\{
- \frac{  \ii }{\Vol}\sum_{\kk} \hat{w}(\kk) \hat{\Gamma}(-\kk){\rm e}^{-  \ii \kk\cdot {\bf{r}} }
\right\} \, .
\end{equation}

With this setup, once again 
we can apply Eq.~\ref{eq:eps_r_field_def} to compute the bulk 
relative permittivity of mixtures of the two solvents, both in RPA and FTS. 
In RPA, 
\begin{equation}
\epsilon_{\rm r} = 1 + \chi_{\rm D} \phi_{\rm w} \, ,
\end{equation}
where $\phi_{\rm w}  = v n_{\rm w} / \Vol$ is the dipole volume fraction. The
corresponding FTS calculation follows the same recipe as under the previous
subheading, but now with the field Hamiltonian in
Eq.~\ref{eq:sol_mix_field_hamiltonian}. Comparisons between $\epsilon_{\rm
r}$ computed in RPA and FTS are shown in 
Figs.~8 and 9
for $\muwr=0.2$ and $\muwr=0.35$, respectively,
for a wide range of $\lb$ values. Common to all runs are the parameter values
$v=0.1$ and $\bar{a}=1/\sqrt{6}$. FTS follows RPA quite well except at high
$\lb$ and intermediate $\phi_{\rm w}$ where the FTS $\epsilon_{\rm r}$ can be
substantially lower than the RPA $\epsilon_{\rm r}$ (see, e.g.,
Fig.~9c).

This drop in $\epsilon_{\rm r}$ may be understood by envisioning
the system entering into an inhomogeneous state where a dipole-rich
and a dipole-depleted phase coexist. Obviously, by construction,
this phase separation is entirely driven by the electrostatic dipole-dipole 
interactions; and the system may be viewed as a special case of a
binary mixture of two types of dipoles distinguished only by their dipole
moments, which has been considered in ref.~\citen{GHFJCP2016}. In
this reference, the virial expansion of such a system
was studied using RPA, and it was shown that there is an effective Flory $\chi$
parameter, proportional to the square of the difference in dipole moment, that
can induce phase separation. By directly calculating $\Z$ in RPA, we can
subsequently compute the binodal curve using the free energy $f =- \ln \Z \, /
\Vol$. In RPA, the field Hamiltonian still follows
Eq.~\ref{eq:rpa_field_hamiltonian}, but now with
\begin{equation}
\RPAM_k = \begin{pmatrix}
\hat{\Gamma}_k^2/v & 0 \\
0 & ({k^2}/{4 \pi \lb})\left(1 + \chi_{\rm D} \hat{\Gamma}_k^2 \phi_{\rm w} 
\right)
\end{pmatrix} \, ,
\end{equation}
which leads to the free energy
\begin{equation}
f = \phi_{\rm w} \ln  \phi_{\rm w} + (1- \phi_{\rm w}) \ln( 1 -  \phi_{\rm w} ) + \frac{v}{4 \pi^2} \int_0^{\infty} \dint k \, k^2 \ln \left[ 1 + \chi_{\rm D}  \phi_{\rm w} \hat{\Gamma}^2 \right]  \, 
\end{equation}
up to an additive constant. 
Fig.~10
shows the phase
diagram computed for the two values of the dipole moment in Figs.~8 and 9 
using this free
energy. Dotted horizontal lines indicate the $\lb$ values considered in FTS.
Indeed, we see that the parameter values for which FTS and RPA do not
quite agree with regard to $\epsilon_{\rm r}$ ($\lb=300$ for 
$\muwr=0.2$ in Fig.~8 and $\lb=150$ for $\muwr=0.35$ in Fig.~9)
coincide well with parameter values in Fig.~10 for which RPA predicts
phase separation with significant differences in dilute- and condensed-phase
$\phi_{\rm w}$s.

$\null$\\
\begin{center}
   \includegraphics[width=0.68\columnwidth]{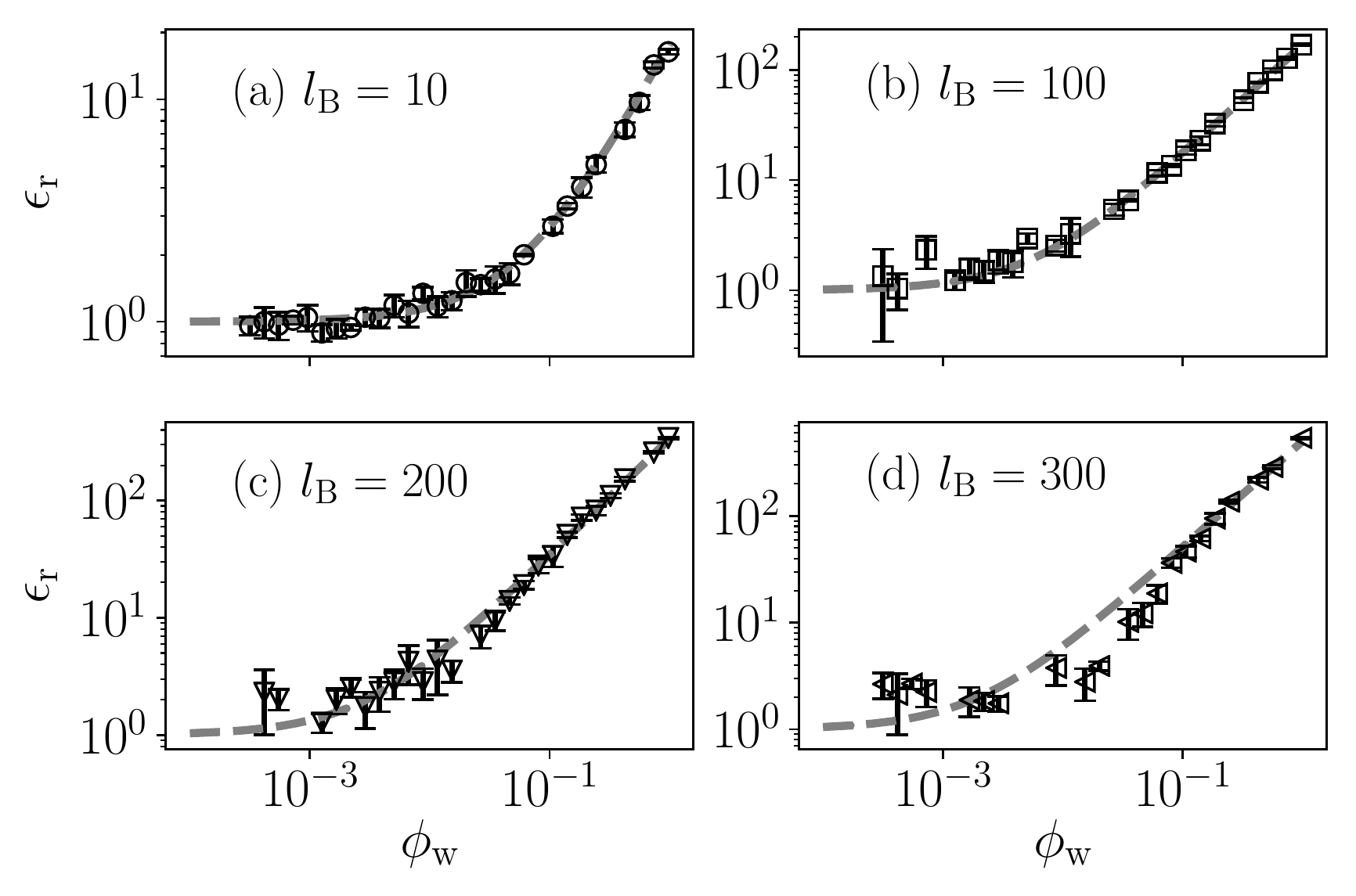}
\end{center}
\vskip -4 mm
{\footnotesize{\bf Fig.~8:}
Relative permittivity $\epsilon_{\rm r}$ computed in FTS and RPA
for an incompressible mixture of fixed dipoles and nonpolar particles
with $\muwr=0.2$ at several values of $\lb$ (in units of $l$). As in Fig.~7,
error bars are standard deviations of independent FTS runs, and dashed
curves are RPA predictions.
}
$\null$\\
\begin{center}
   \includegraphics[width=0.76\columnwidth]{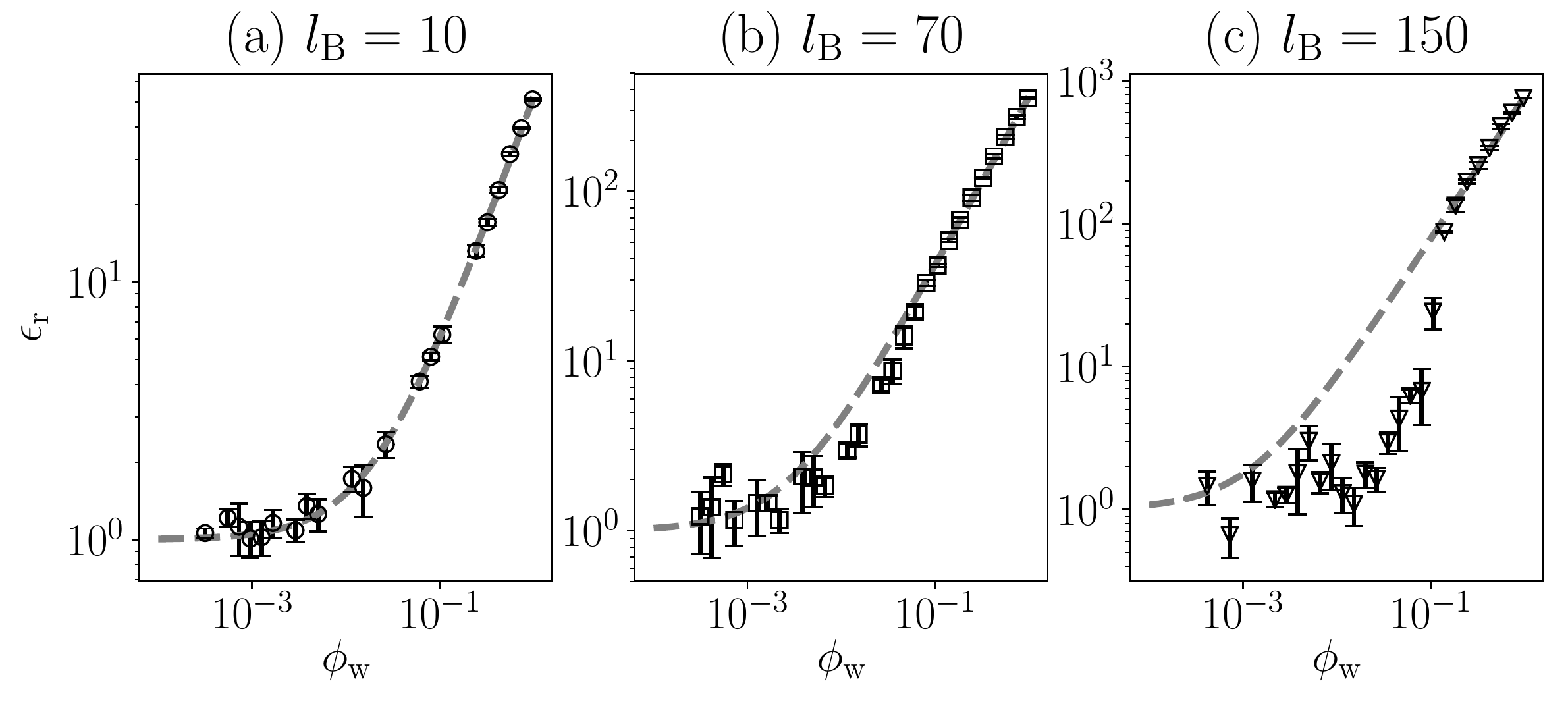}
\end{center}
\vskip -4 mm
{\footnotesize{\bf Fig.~9:}
Same as Fig.~8 but for $\muwr=0.35$.
}

$\null$\\
\begin{center}
   \includegraphics[width=0.50\columnwidth]{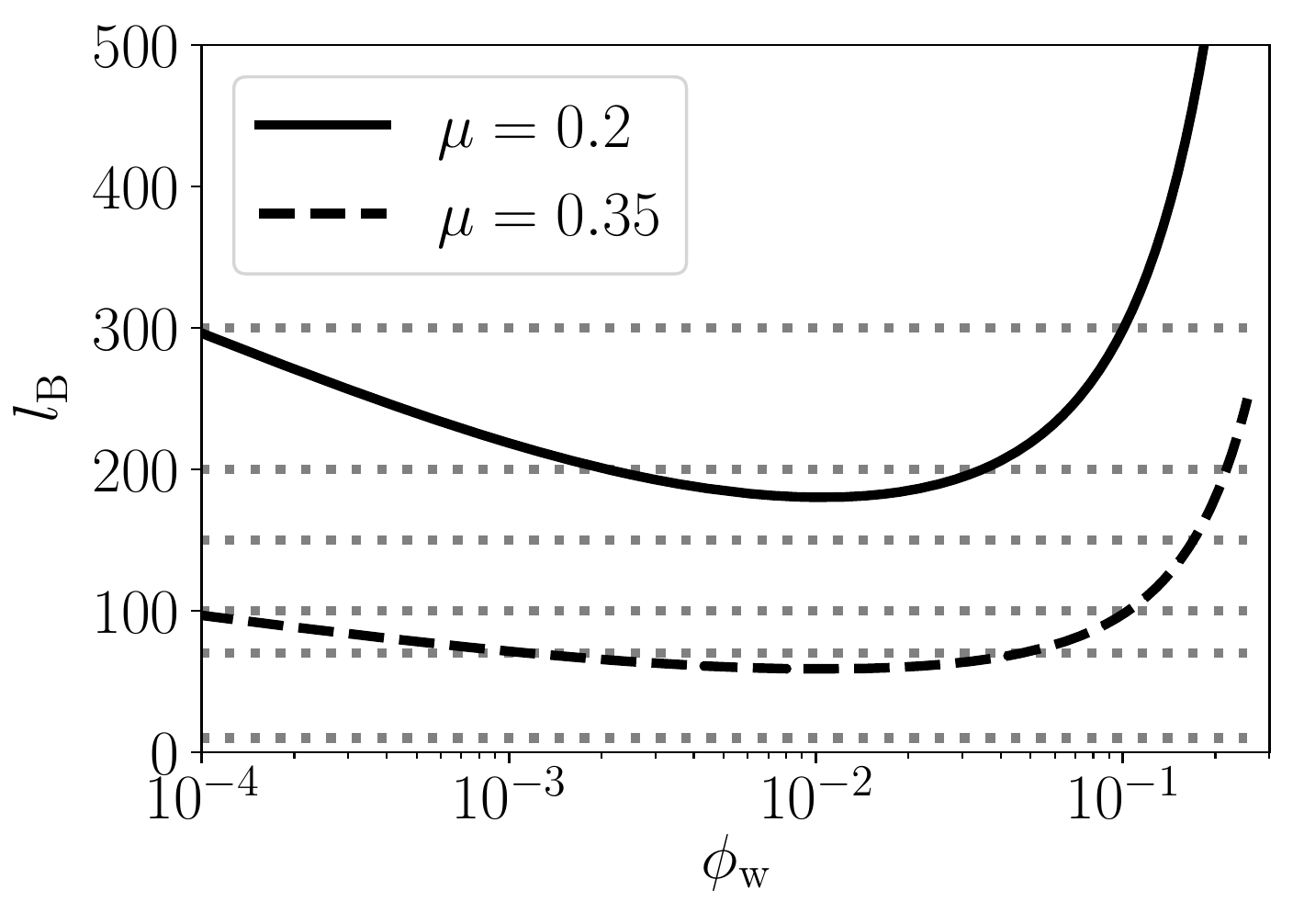}
\end{center}
\vskip -4 mm
{\footnotesize{\bf Fig.~10:}
Phase diagrams computed in RPA for incompressible mixtures of
fixed dipoles and nonpolar particles.  The binodal (coexistence) curves
are shown for two values of the dipole moment, $\muwr=0.2$ and $\muwr=0.35$,
$v=0.1$, and $\bar{a}=1 / \sqrt{6}$ (length unit $=l$, as noted for Fig.~7).
The dashed horizontal lines indicate $\lb$ values considered in FTS.
}
$\null$\\


{\bf Analytical Theory for Polyampholytes in Dipole Solvent.}
We now consider systems consisting of $\nw$ dipoles and, instead of 
nonpolar particles, $\np$ polyampholytes.
Following the notation in ref.~\citen{kings2020},
each polyampholyte (labeled by $\alpha$) is made up of
$N$ monomer beads (as in the above molecular dynamics simulations) 
at positions $\R_{\alpha,\tau}$, $\tau=1,2,\dots,N$, and carries 
charge sequence 
$\ket{\sigma} = [\sigma_1, \sigma_2, ... \sigma_N]^{\rm T}$. 
The incompressibility constraint for this system is given by
\begin{equation}
\rho_{\rm p}({\bf{r}}) + \rho_{\rm w}({\bf{r}}) = \frac{1}{v}  \, 
\label{eq:fullincomp}
\end{equation}
wherein the polymer matter (mass) density 
$ \rho_{\rm p}({\bf{r}}) = \sum_{\alpha=1}^{n_{\rm{p}}} \sum_{\tau=1}^{N} 
\Gamma({\bf{r}}-\R_{\alpha,\tau})$. The Fourier-transformed
density of polymer charges,
\begin{equation}
\hat{\charge}(\kk) =  \hat{\Gamma}(\kk) \sum_{\alpha=1}^\np
\sum_{\tau=1}^{N}\sigma_\tau {\rm e}^{\ii \kk \cdot \R_{\alpha,\tau}}
\; ,
\end{equation}
now contribute to the Fourier-transformed total charge density such that
\begin{equation}
\hat{\rho}_{\rm c}(\kk) = \hat{\charge}(\kk)  + \ii \kk\cdot \hat{\dipole}(\kk).
\end{equation}
The partition function of the polyampholytes-plus-dipoles system 
is given by
\begin{equation}
\Z = \frac{1}{\np!\nw!} \prod_{\alpha=1}^\np\prod_{\tau=1}^N \int \dint
\R_{\alpha,\tau}
\prod_{i=1}^\nw\int \dint \rr_i 
\int \dint \bm{\mu}_i
\;
{\rm e}^{-\TT[\R] - \UU[\R, \rr]} 
\;
\delta \left[ \rho_{\rm w}({\bf{r}}) + \rho_{\rm p}({\bf{r}}) -
\frac{1}{v} \right]  ,
\label{eq:Z-poly-dipole_0}
\end{equation}
where
\begin{equation}
\TT[\R] = \frac{3}{2l^2}\sum_{\alpha=1}^\np\sum_{\tau=1}^{N-1}\left( \R_{\alpha,\tau+1}- \R_{\alpha,\tau} \right)^2
\end{equation}
is the Hamiltonian term for Gaussian chain connectivity, 
$l$ is the reference bond length that we have used as a length
scale in Fig.~7 to define $T^*_0$ for the present analytical theories, 
and $\UU$ for electrostatic interactions 
takes the same general form as that given by Eq.~\ref{eq:UU-basic-def}.
The field-theory expression for the partition 
in Eq.~\ref{eq:Z-poly-dipole_0} is
\begin{equation} \label{eq:Z_field}
\Z = \frac{\Vol^{\np + \nw} }{ \np ! \nw !}  \int \DD w \DD \psi \, {\rm e}^{- \HH }
\end{equation}
where
\begin{equation} \label{eq:HH_field}
\HH[\psi, w] =  -  \np\ln\Qp - \nw\ln\Qw +
\frac{1}{2\Omega}\sum_{\kk\neq\kzero}\frac{ \kk^2}{4\pi\lb}\hat{\psi}(\kk)
\hat{\psi}(-\kk)  - \frac\ii{v} \hat{w}({\bf{0}}) \,   
\end{equation}
and, via Hubbard-Stratonovich transformation, the single-polymer 
partition function $\Qp$ is given by
\begin{equation}
\label{Qpsv}
\Qp =  \frac{1}{\Vol} \prod_{\tau=1}^N  \int  \dint \R_\tau \exp
\left\{
-\frac{3}{2l^2} \sum_{\tau=1}^{N-1}\left( \R_{\tau+1}- \R_{\tau} \right)^2
- \frac{ \ii }{\Vol}\sum_{\kk}\sum_{\tau=1}^N\left[ \hat{w}(\kk) +  \sigma_\tau \hat{\psi}(\kk) \right] \hat{\Gamma}(-\kk) {\rm e}^{-\ii \kk\cdot\R_\tau}
\right\} \, .
\end{equation}
As in the above analysis of pure dipoles and dipoles-plus-nonpolar-particles
systems, we may now expand the field Hamiltonian to second order in 
the fields to arrive at an RPA theory. 
Again, this leads to the general RPA form in 
Eq.~\ref{eq:rpa_field_hamiltonian}, 
now with a different expression for the kernel $\RPAM_k$ for the system
described by Eq.~\ref{eq:Z-poly-dipole_0}:
\begin{equation}
\RPAM_k=
\begin{pmatrix}
( \phi \gmm{k}+ \phi_{\rm w} ) \hat{\Gamma}_k^2 / v &\phi \gmc{k} \hat{ \Gamma}_k^2 / v  \\
 \phi \gmc{k} \hat{ \Gamma}_k^2 / v &  (\phi \gcc{k} + {\cal D} \phi_{\rm w}
k^2 ) \hat{ \Gamma}_k^2 / v  + (k^2/4\pi\lb) 
\end{pmatrix} \, ,
        \label{eq:RPAM_poly}
\end{equation}
where ${\cal D}=\muwr^2/3$ for intrinsic dipoles and 
${\cal D}=1/(\beta\kappa_{\rm d})$ for induced (polarizable) dipoles. 
Here $\phi = v N \np / \Vol $ is the polymer volume fraction 
(thus $\phi_{\rm w} = 1-\phi$ is the dipole volume fraction). 
The factors $\gmm{k}$, $\gmc{k}$ and $\gcc{k}$ stem from the single-polymer 
partition function in Eq.~\ref{Qpsv},
\begin{subequations}
\begin{align}
\gmm{k} &= \frac{1}{N} \sum_{\nu,\tau=1}^N \exp(-l^2 k^2|\nu\!-\!\tau|/6) \, ,  \\
\gmc{k} &= \frac{1}{N} \sum_{\nu,\tau=1}^N \sigma_\nu \exp(-l^2  k^2|\nu\!-\!\tau|/6) \, , \\
\gcc{k} &= \frac{1}{N} \sum_{\nu,\tau=1}^N \sigma_\nu \sigma_\tau\exp(- l^2 k^2 |\nu\!-\!\tau|/6) \, .
\end{align}
\label{eq:gcc_etc}
\end{subequations}

Applying the RPA form of the Hamiltonian $\HH$
in Eq.~\ref{eq:rpa_field_hamiltonian} to the expression
for $\Z$ in Eq.~\ref{eq:Z-integral} 
and performing the $\int \DD w\DD\psi$
integrals by including the prefactors of
$\int \dint\hat{w}(\kk) \int \dint\hat{\psi}(\kk)$
in Eq.~\ref{eq:field_int_factor} lead to the following expression
for the free energy $f=-\ln \Z/\Omega$ in RPA:
\begin{equation}
f  = \frac{\phi}{N}\ln\phi + (1\!-\!\phi)\ln(1\!-\!\phi) +
\frac{v}{4 \pi^2} \int_0^\infty \dint k \, k^2 
\ln{\left[  \frac{4\pi\lb v}{k^2}\det \RPAM_k \right]} \; ,
        \label{eq:f}
\end{equation}
where $\RPAM_k$ is here given by Eq.~\ref{eq:RPAM_poly}.
We rewrite the argument of the logarithm in the above integrand in the form
of
\begin{equation}
 \frac{(4\pi\lb v/k^2)\det \RPAM_k}{\left. (4\pi\lb v/k^2)\det \RPAM_k\right|_{\phi=0} }  = 1 + {\cal A}_k \phi + {\cal B}_k \phi^2
\end{equation}
to arrive at a form of $f$ in which irrelevant constant terms are subtracted,
with
\begin{subequations}
\begin{align}
{\cal A}_k &= \gmm{k} - 1 + \frac{4 \pi \lb }{v k^2 (1 + \chi_{\rm D} \hat{\Gamma}_k^2) } \hat{\Gamma}_k^2 (\gcc{k} - k^2 {\cal D}) \, ,  \\
{\cal B}_k &= \frac{4 \pi \lb }{v k^ 2 (1 + \chi_{\rm D} \hat{\Gamma}_k^2 ) } \hat{\Gamma}_k^2 \left[ (\gmm{k} - 1) (\gcc{k} - k^2 {\cal D}) - \left(\gmc{k}\right)^2 \right]  \, .
\end{align}
\label{eq:AkBk}
\end{subequations}

\vskip -4mm
\begin{center}
   \includegraphics[width=0.63\columnwidth]{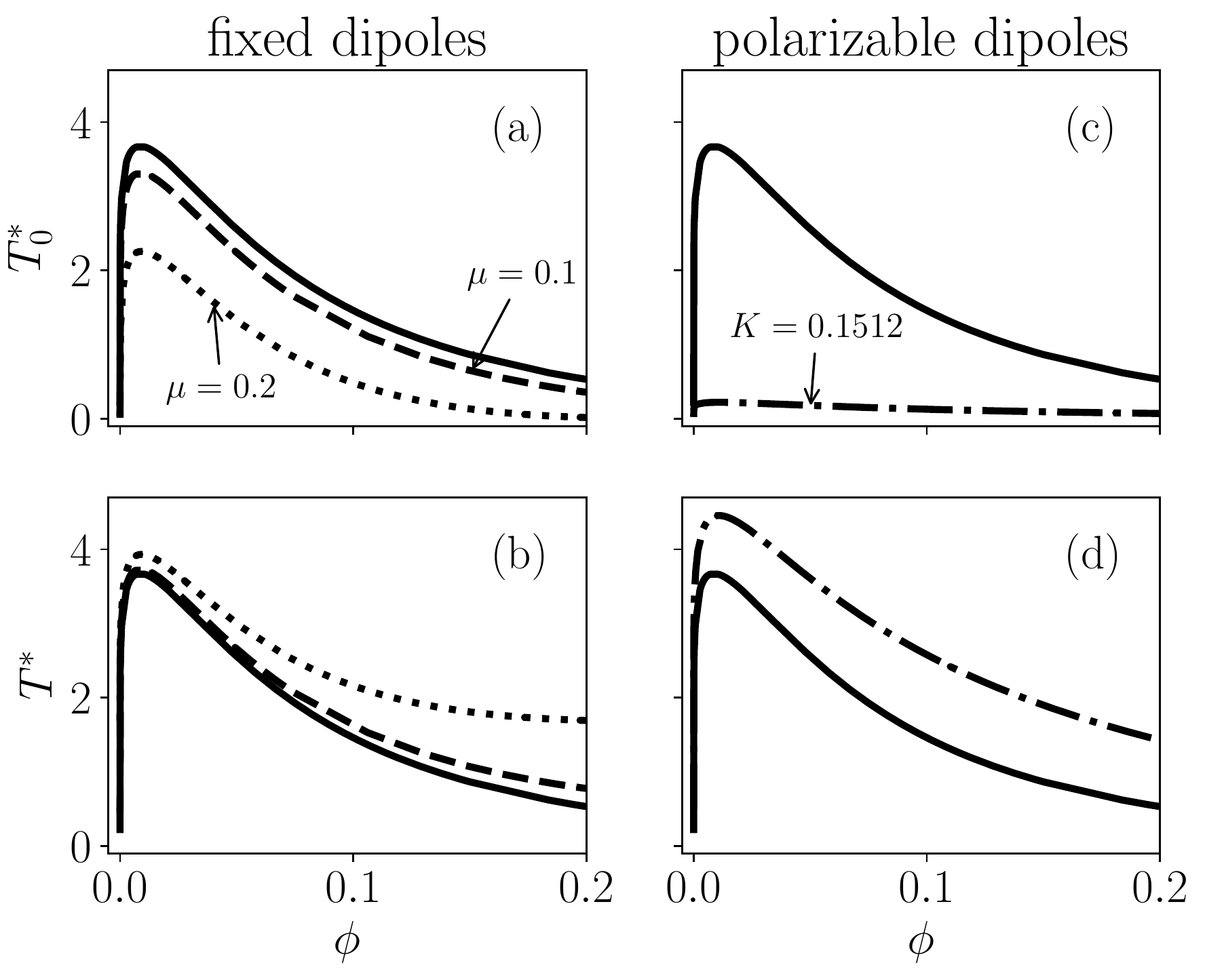}
\end{center}
\vskip -4 mm
{\footnotesize{\bf Fig.~11:}
RPA phase diagrams of incompressible mixtures of
polyampholytes and dipoles.
Results are for sequence sv15 with $v=0.1$ and $\bar{a}=1/\sqrt{6}$
(length unit $=l$). $\phi$ is polymer volume fraction,
$T^*_0=l/l_{\rm B}$ is vacuum reduced temperature
as in Fig.~7, and $T^* = T^*_0 \epsilon_{\rm r}(\phi=0)$ is reduced
temperature that takes into account bulk-solvent relative permittivity
(which is temperature dependent for fixed dipoles).
Several solvent models are compared. Solid curves (all panels) are
for nonpolar solvent ($\muwr=0$);
dashed and dotted curves [(a) and (b)] are for fixed-dipole solvents with
$\muwr=0.2$ and $\muwr=0.1$, respectively; and dashed-dotted
curves [(c) and (d)] are for polarizable-dipole solvent with
$K=0.1512$ (which, according to Eq.~\ref{eq:eps_r_RPA_pure_dipoles},
entails a temperature-independent bulk-solvent
$\epsilon_{\rm r}(\phi=0)=20$).
Note that the $\lb$ values considered in this figure are much
lower than that required for the dipoles to phase separate on their own
(Figs.~8--10).
}
$\null$\\

At this juncture, we are in a position to use Eq.~\ref{eq:f} to 
compute phase diagrams in RPA. 
Fig.~11 shows such phase diagrams in the
$(\phi,T^*_0)$ plane for polyampholyte chains with charge sequence
sv15 immersed in fixed dipoles (Fig.~11a) and polarizable 
dipoles (Fig.~11c). Not surprisingly, compared to the case with
nonpolar solvent (solid curves), LLPS propensity is lowered
in the presence of polar solvents (dashed, dotted, and dashed-dotted 
curves in Figs.~11a and c) because the LLPS-driving effective electrostatic 
interactions among the polyampholytes are weakened due to screening, 
and this LLPS-suppressing
effect increases with increasing polarity of the solvent ($T^*_{0,{\rm cr}}$
lower for $\muwr=0.2$ than for $\muwr=0.1$ in Fig.~11a).
This comparison corresponds to that between models (i) and (ii) 
in Fig.~6a for the corresponding molecular dynamics results.
Figs.~11b and d show the corresponding phase 
diagrams in the $(\phi,T^*)$ plane, where $T^*=T^*_0\epsilon_{\rm r}(\phi=0)$
with $\epsilon_{\rm r}(\phi=0)$ being the relative permittivity of
the bulk-solvent.
These plots offer a comparison between the phase behavior of polyampholytes 
interacting via a homogeneous background dielectric constant (solid curves)
and the phase behavior of polyampholytes interacting with an incompressible
dipole solvent (thus allowing for dipole-concentration-dependent 
relative permittivity) with the same bulk-solvent relative permittivity.
Thus this comparison corresponds to that between molecular dynamics
models (iii) and (ii) in Fig.~6a.
Interestingly, whereas models (iii) and (ii) for sv15 in Fig.~6a exhibit 
essentially the same LLPS propensity---a feature that may be caused by the
relative weak electrostatic interactions in sv15 being largely overwhelmed 
by the attractive LJ interactions,
RPA predictions in
Figs.~11b and d indicate that interacting with an incompressible dipole
solvent enhances polyampholyte LLPS propensity relative to that predicted
for the corresponding situation with a homogeneous background dielectric
constant.
This enhancing effect is quite small for the fixed dipoles in 
Fig.~11b---the effect nonetheless increases with $\muwr$---but the effect 
is more appreciable for the case of polarizable dipoles in Fig.~11d.
As discussed previously,\cite{suman2}
because there are relatively strong non-electrostatic 
background attractive LJ interactions 
in the molecular dynamics model---interactions
that are absent in the present theoretical formulation,
the upward concavity on the condensed 
sides of the RPA coexistence curves in Fig.~11 is not observed for the 
molecular dynamics coexistence curves in Fig.~6.
\\

{\bf Comparison of Theory-Predicted Polyampholyte Phase Behaviors Under
Different Relative Permittivity Models.}
To clarify the difference between various permittivity models that
have been employed by our group to study polyampholyte LLPS using 
analytical theories,\cite{linJML,njp2017,SumanPNAS}
we take the mean-field limit for mass-density fluctuation,
i.e.~$w\to 0$, of the above formulation to focus on features of
the RPA electrostatic free energy. In such a limit, 
only the bottom-right component of the $\RPAM$ array 
in Eq.~\ref{eq:RPAM_poly} remains, viz.
\begin{equation}
\RPAM_k 
\underset{w\to 0}{\longrightarrow}
\frac{1}{v} \left[\phi \gcc{k} + (1-\phi) {\cal D} k^2 \right] 
\hat{ \Gamma}_k^2   + \frac{ k^2}{4\pi\lb}
\; .
\end{equation}
The RPA electrostatic free energy is then given by
\begin{equation}
 f_{\rm el} = \frac{v}{4\pi^2}\int_0^\infty\dint k\, k^2 \ln\left[ 1+ \chi_{\rm D}\hat{\Gamma}_k^2(1-\phi)
        + \frac{4\pi\lb}{v k^2}\gcc{k}\hat{\Gamma}_k^2\phi  \right]
\; ,
\label{eq:RPAfelw0}
\end{equation}
which may be rewritten as
\begin{equation}
\begin{aligned}
f_{\rm el} = &
 \frac{v}{4\pi^2}\int_0^\infty\dint k\, k^2
        \ln\left[ 1+ \chi_{\rm D}\hat{\Gamma}_k^2(1-\phi)  \right] +
 \frac{v}{4\pi^2}\int_0^\infty\dint k\, k^2
        \ln\left[ 1 + \frac{4\pi\lb\gcc{k}\hat{\Gamma}_k^2\phi}{v k^2[ 1+  \chi_{\rm D}\hat{\Gamma}_k^2(1-\phi) ]}\right]
        \label{eq:fel_sep1} \\
\equiv & f_{\rm el}^{\rm dd} + f_{\rm el}'
\end{aligned}
\end{equation}
in which the first term $f_{\rm el}^{\rm dd}$ accounts for dipole-dipole 
interaction between solvent molecules, and the second term includes 
charge-charge (polymer-polymer) and charge-dipole (polymer-solvent) 
interactions.
It should be noted that in the present system, the pure material 
susceptibilities are $\chi_{\rm D}^{\rm p} = 0$ for polymers 
(polyampholytes) and $\chi_{\rm D}^{\rm w} = \chi_{\rm D}$ for solvent 
molecules (dipoles). A concentration-dependent permittivity which for
analytical tractability is assumed to be dependent linearly on 
volume fractions\cite{SumanPNAS} can then be constructed as
\begin{equation}
\begin{aligned}
\epsilon_{\rm r}(\phi) = & \epsilon_{\rm p}\phi +  \epsilon_{\rm w}(1-\phi) \\
        = & (1+\chi_{\rm D}^{\rm p} )\phi + (1+\chi_{\rm D}^{\rm w} )(1-\phi) \\
        = & \phi + (1+\chi_{\rm D} )(1-\phi) \\
        = & 1 + \chi_{\rm D}(1-\phi).
\label{eq:linear_epr}
\end{aligned}
\end{equation}
Thus, in the case of vanishing Gaussian smearing, 
i.e.~$\bar{a}\to0$ and $\Gamma_k\to 1$, the second term 
in Eq.~\ref{eq:fel_sep1} can be cast in the following form:
\begin{equation}
\lim_{\hat{\Gamma}_k\to1} f_{\rm el}' =  f_{\rm el}^{\rm emp}
= \frac{v}{4\pi^2}\int_0^\infty\dint k\, k^2
        \ln\left[ 1 + \frac{4\pi\lb}{v\epsilon_{\rm r}(\phi)k^2}\gcc{k}\phi \right],
\end{equation}
where the superscript ``emp'' stands for ``empirical permittivity''. It
is noteworthy that $f_{\rm el}^{\rm emp}$ derived above is 
equivalent to the RPA electrostatic energy of the recent 
``with {`self-energy'}'' model used in Fig.~7B of Das et al.\cite{SumanPNAS} 
More specifically, with $v=l^3$, $f_{\rm el}^{\rm emp}$ is seen to be
identical to the expression in Eq.~S60 in the SI Appendix of 
ref.~\citen{SumanPNAS}, which in turn is based on the partition 
function $\Z_{\rm el}$ in Eq.~S51 of the same reference.

We now proceed to clarify how dipole-dipole interactions promote LLPS 
whereas dipole-polymer interactions mostly suppress LLPS in the present 
analytical models. First,
a small-$\phi$ Taylor expansion of the logarithmic integrand of 
$f_{\rm el}^{\rm dd}$ in Eq.~\ref{eq:fel_sep1} yields
\begin{equation}
\ln\left[ 1+ \chi_{\rm D}\hat{\Gamma}_k^2(1-\phi)  \right]
= \ln\left[ 1+ \chi_{\rm D}\hat{\Gamma}_k^2  \right]
-   \frac{\chi_{\rm D}\hat{\Gamma}_k^2}{1+ \chi_{\rm D}\hat{\Gamma}_k^2}\phi
        -\frac{1}{2} \left(  \frac{\chi_{\rm D}\hat{\Gamma}_k^2}{1+ \chi_{\rm D}\hat{\Gamma}_k^2} \right)^2\phi^2 + \mathcal{O}(\phi^3),
\label{eq:ph2_1}
\end{equation}
wherein
$\phi^2$ is seen to associate with a negative factor, 
indicating that dipole-dipole interactions contribute an effectively 
attractive polymer-polymer interaction. In other words, dipole-dipole
interactions contribute a positive Flory-Huggins $\chi$ parameter 
or a negative second virial coefficient and therefore promote 
polymer phase separation.
It should be noted as well that an equivalent
conclusion can be reached by considering a Taylor expansion 
with respect to the mean-field concentration of the solvent molecules, 
$(1-\phi)$, leading to the conclusion that the effective interactions 
among the dipoles are also attractive.

Second, we note that
the $f_{\rm el}'$ term in Eq.~\ref{eq:fel_sep1} for the 
``empirical permittivity" model may be dissected further by rewriting it 
in the following form:
\begin{equation}
f_{\rm el}'= \frac{v}{4\pi^2}\int_0^\infty\dint k\, k^2
        \ln\left[ 1 + \frac{4\pi\lb}{v k^2}\gcc{k}\hat{\Gamma}_k^2\phi\right]
+  \frac{v}{4\pi^2}\int_0^\infty\dint k\, k^2
\ln\left[1 -
\frac{\frac{4\pi\lb}{v k^2}\left( \frac{\overline{\chi}_{\rm D}}{1+\overline{\chi}_{\rm D}} \right)\gcc{k}\hat{\Gamma}_k^2\phi}{1 + \frac{4\pi\lb}{v k^2}\gcc{k}\hat{\Gamma}_k^2\phi}
\right].
\label{eq:decoupled0}
\end{equation}
Here the first term is the RPA theory with constant vacuum permittivity, 
i.e.~purely charge-charge (polymer-polymer) interactions, and the second term
accounts for charge-dipole (polymer-solvent) interactions, and we
have defined $\overline{\chi}_{\rm D}\equiv\chi_{\rm D}\Gamma_k^2(1-\phi)$.
A Taylor expansion of the second term with respect to $\phi$ gives
\begin{equation}
\ln\left[1 -
\frac{\frac{4\pi\lb}{v k^2}\left( \frac{\overline{\chi}_{\rm D}}{1+\overline{\chi}_{\rm D}} \right)\gcc{k}\hat{\Gamma}_k^2\phi}{1 + \frac{4\pi\lb}{v k^2}\gcc{k}\hat{\Gamma}_k^2\phi}
\right]
=
- \frac{A_k B_k}{B_k+1}\phi + \frac{A_k B_k(A_k B_k + 2 A_k + 2)}{2(B_k + 1)^2}\phi^2 + \mathcal{O}(\phi^3),
        \label{eq:charge-dipole}
\end{equation}
where
\begin{equation}
A_k \equiv \frac{4\pi\lb}{v k^2}\gcc{k}\hat{\Gamma}_k^2 \; , \; B_k =\chi_{\rm D}\hat{\Gamma}_k^2.
\end{equation}
Given $A_k, B_k\!\! > \!\!0 \; \forall k$, the polymer-dipole 
interactions entail a positive second virial coefficient associated
with the $\phi^2$ term in Eq.~\ref{eq:charge-dipole} that equals to
\begin{equation}
\frac{v}{4\pi^2}\int_0^\infty\dint k\, k^2 \frac{A_k B_k(A_k B_k + 2 A_k + 2)}{2(B_k + 1)^2} > 0 ,
\end{equation}
which acts as an effective excluded-volume-like repulsion between polymers 
and thus suppress phase separation.

The comparison above is between two models with the same polymer
relative permittivity ($\epsilon_{\rm p} \!=\! 1$) but different solvent 
relative permittivities 
($\epsilon_{\rm w}\!=\!1$ versus $\epsilon_{\rm w} \!=\! 1+\chi_{\rm D}$,
see Eq.~\ref{eq:linear_epr}).
Another comparison, of more direct interest to the question at hand, is between
an implicit-solvent model in which the polymers interact in a dielectric
environment with a constant $\epsilon_{\rm w}=1+\chi_{\rm D}$ (the 
relative permittivity of the polymer material itself remains the same at 
$\epsilon_{\rm p} \!=\! 1$)
and the full theory with the polymers immersed in dipole solvent wherein
the bulk-dipole relative permittivity is $\epsilon_{\rm w}=1+\chi_{\rm D}$.
This comparison may be analyzed by using a modified Bjerrum length,
$\lb^{\rm r} = \lb/\epsilon_{\rm w}$, as characteristic temperature factor,
to regroup Eq.~\ref{eq:decoupled0} for $f_{\rm el}'$
into a sum of two contributions,
\begin{equation}
f'_{\rm el} =
\frac{v}{4\pi^2}\int_0^\infty\dint k\, k^2
        \ln\left[ 1 + \frac{4\pi\lb^{\rm r}}{v k^2}\gcc{k}\hat{\Gamma}_k^2\phi\right]
+  \frac{v}{4\pi^2}\int_0^\infty\dint k\, k^2
        \ln\left[ 1 + \frac{\frac{4\pi\lb^{\rm r}}{v k^2}\gcc{k}\hat{\Gamma}_k^2\left( \frac{1+\chi_{\rm D}}{1+\overline{\chi}_{\rm D}} -1\right)\phi}{1 + \frac{4\pi\lb^{\rm r}}{v k^2}\gcc{k}\hat{\Gamma}_k^2\phi} \right]
\; ,
\label{eq:decouple2}
\end{equation}
such that the first term is the electrostatic free energy for 
polymers interacting with a constant implicit-solvent 
$\epsilon_{\rm w}=1+\chi_{\rm D}$.
The integrand of the second term in Eq.~\ref{eq:decouple2}, 
which is a polymer-solvent 
interaction term, can now be expanded with respect to $\phi$:
\begin{equation}
\begin{aligned}
\!\!\!\!
\ln\left[ 1 + \frac{\frac{4\pi\lb^{\rm r}}{v k^2}\gcc{k}\hat{\Gamma}_k^2\left( \frac{1+\chi_{\rm D}}{1+\overline{\chi}_{\rm D}} -1\right)\phi}{1 + \frac{4\pi\lb^{\rm r}}{v k^2}\gcc{k}\hat{\Gamma}_k^2\phi} \right]
= & -\frac{A_k^{\rm r} \chi_{\rm D}(\hat{\Gamma}_k^2-1)}{\chi_{\rm D}\hat{\Gamma}_k^2+1}\phi \\
& + \frac{A_k^{\rm r}\chi_{\rm D}\left\{ A_k^{\rm r} (\hat{\Gamma}_k^2-1)\left[\chi_{\rm D}(\hat{\Gamma}_k^2+1)+2\right]+2(1+\chi_{\rm D})\hat{\Gamma}_k^2\right\}}{2\left( \chi_{\rm D}\hat{\Gamma}_k^2+1\right)^2}\phi^2 \\
& +  \mathcal{O}(\phi^3)
\; ,
\end{aligned}
\end{equation}
where $A_k^{\rm r} = 4\pi\lb^{\rm r}/(vk^2)\gcc{k}\hat{\Gamma}_k^2$.
Now, the form of the the second virial coefficient, denoted 
$v_2^{{\rm pw, r}}$, that associates with the $\phi^2$ term,
\begin{equation}
v_2^{{\rm pw, r}} =
 \frac{v}{4\pi^2}\int_0^\infty\dint k\, k^2
\frac{A_k^{\rm r}\chi_{\rm D}\left\{ A_k^{\rm r} (\hat{\Gamma}_k^2-1)\left[\chi_{\rm D}(\hat{\Gamma}_k^2+1)+2\right]+2(1+\chi_{\rm D})\hat{\Gamma}_k^2\right\}}{2\left( \chi_{\rm D}\hat{\Gamma}_k^2+1\right)^2} \, ,
\label{eq:2nd_intricate}
\end{equation}
becomes rather intricate.
To analyze the sign of $v_2^{\rm pw, r}$, we substitute the 
definition of $A_k^r$ into the numerator of the integrand in
the above expression and rewrite it as
\begin{equation}
\frac{4\pi\lb^{\rm r}\chi_{\rm D}}{v}  \frac{\gcc{k}}{k^2}\hat{\Gamma}_k^4
\left[
        \frac{4\pi\lb^{\rm r}}{vk^2}  \gcc{k} 
\left( \hat{\Gamma}_k^2 -1\right)\left( \chi_{\rm D}\hat{\Gamma}_k^2 + 
\chi_{\rm D} + 2  \right) + 2(1+\chi_{\rm D})
\right]
\; .
\label{eq:numerator}
\end{equation}
For overall neutral polyampholytes, 
$\gcc{k} \ll 1$ when $k\to0$, $\gcc{k}\to1$ when $k\to\infty$,
and $\gcc{k}$ does not increase monotonically with $k$ but has a peak at some
intermediate $k$ for most cases (except, e.g., for the strictly alternating
sequence sv1, which monotonically increases from 0 to 1 when $k$ increases).
For $k \ll 1/\bar{a}$, $\hat{\Gamma}_k\to1$, the general expression in
Eq.~\ref{eq:numerator} becomes
\begin{equation}
\frac{8\pi\lb^{\rm r}\chi_{\rm D}(1+\chi_{\rm D})}{v}  \frac{\gcc{k}}{k^2}\left[
        -\frac{4\pi\lb^{\rm r}\overline{a}^2}{v}  \gcc{k} + 1
\right] > 0
\; ,
\end{equation}
and for $k \gg 1/\bar{a}$, $\hat{\Gamma}_k\to0$, the same general
expression becomes
\begin{equation}
\frac{8\pi\lb^{\rm r}\chi_{\rm D}(1+\chi_{\rm D})}{v}  
\frac{\gcc{k}}{k^2}\hat{\Gamma}_k^4 > 0
\; .
\label{eq:decouple2_3}
\end{equation}
Thus, Eq.~\ref{eq:numerator} contributes a repulsive effect
to the integral for $v_2^{\rm pw, r}$ in Eq.~\ref{eq:2nd_intricate}
in both the small-$k$ and large-$k$ regimes.
Moreover, the $\hat{\Gamma}_k^4$ factor in the same expression tends 
to reduce large-$k$ contributions to the $v_2^{\rm pw, r}$ integral.
Therefore, it is expected that as long as the peak of $\gcc{k}$ is at
$k\lesssim 1/\overline{a}$, effective polymer-solvent interactions 
should be repulsive.
However, in view of the intricate form of $v_2^{\rm pw, r}$
in Eq.~\ref{eq:2nd_intricate}, 
the possibility of effective polymer-solvent interactions 
being attractive in certain extreme scenarios cannot be precluded.

\begin{center}
   \includegraphics[width=0.63\columnwidth]{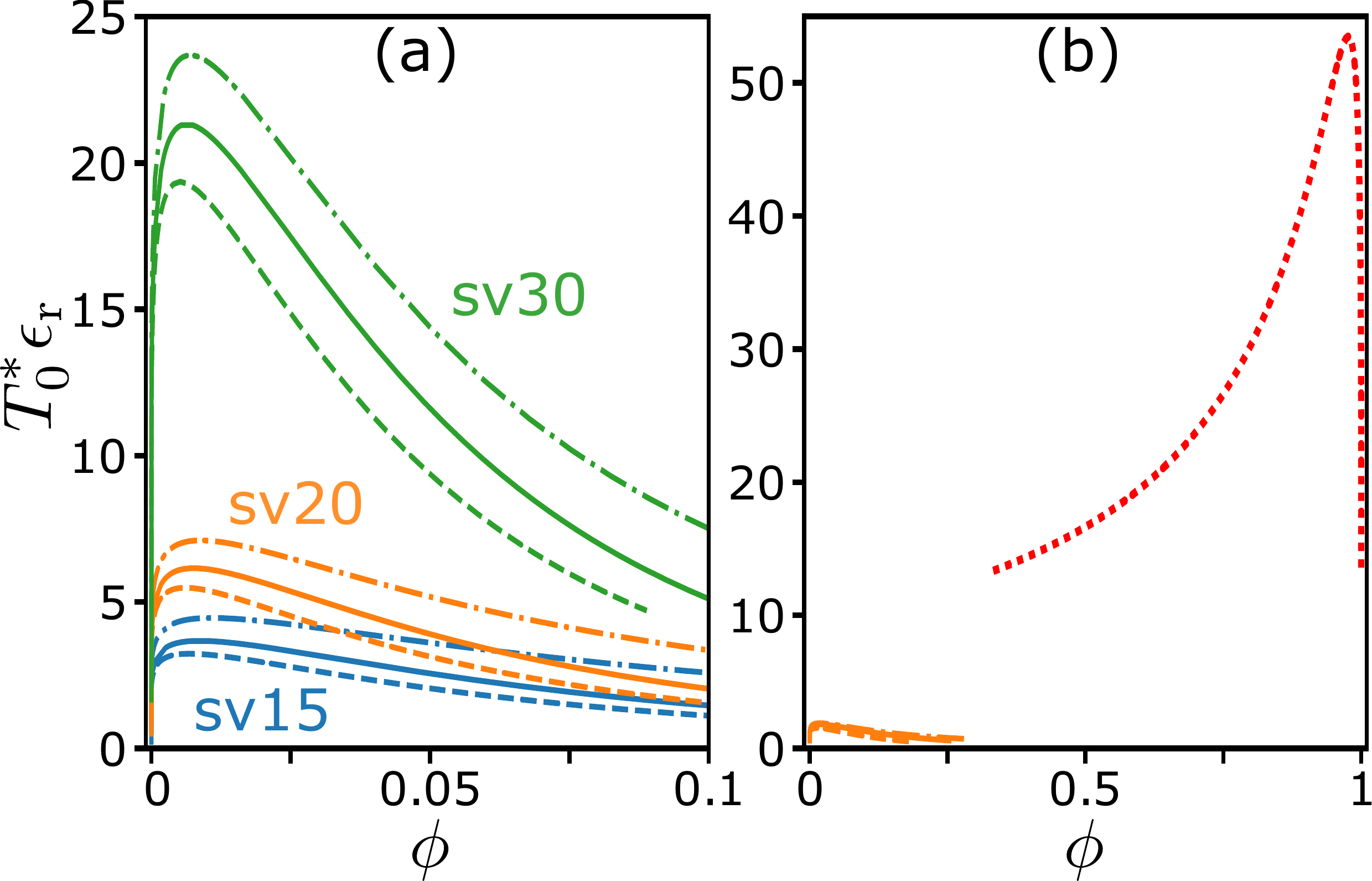}
\end{center}
\vskip -4 mm
{\footnotesize{\bf Fig.~12:}
Comparing RPA LLPS theories with different solvent permittivity models.
The vertical variable $T^*_0\epsilon_{\rm r}=T^*$, where $\epsilon_{\rm r}$
is bulk-solvent relative permittivity.
(a) Coexistence curves of sequences sv15, sv20, and sv30 are shown in
different colors.
Dashed-dotted curves are for the full explicit dipole solvent model,
solid curves are for the implicit-solvent constant permittivity model
(with nonpolar solvent).
Both set of curves are computed with the mass density conjugate field $w$
in accordance with the formulation given by Eqs.~\ref{eq:f}--\ref{eq:AkBk}
(dashed-dotted and solid curves for sv15 are the same as those in Fig.~11d).
Dashed curves are for the empirical permittivity model, computed in
the $w\to 0$ formulation wherein excluded volume is implemented by
incompressibility at the mean-field level (electrostatic free energy
given by the second term in Eq.~\ref{eq:fel_sep1}).
All coexistence curves in (a) are computed using $v=0.1$ (in units of $l^3$).
(b) Coexistence curves for sv20 with the same models in (a) depicted in
the same line styles in orange but all computed using the $w\to 0$ formulation
with $v=1.0$ (Eq.~\ref{eq:RPAfelw0} with $\lb\to
\lb^{\rm r}=\lb/\epsilon_{\rm w}$
for dashed-dotted curve, first term in
Eq.~\ref{eq:decouple2} for solid curve, and
second term in Eq.~\ref{eq:fel_sep1}, as in (a), for dashed curve).
The red dotted curve for a ``no self energy'' model is the coexistence
curve computed using an electrostatic free energy equals to that
of the empirical permittivity model (second term in Eq.~\ref{eq:fel_sep1})
minus the ``self-energy'' term in Eq.~\ref{eq:fel-self}.
}
$\null$\\
$\null$\\

Fig.~12a
compares the theory-predicted phase behaviors of sequences sv15, sv20, and sv30 
under three different relative permittivity models for the solvent.
All three sequences studied exhibit
the same ranking ordering of LLPS propensities for the three models
(as quanitified by their critical temperatures):
\begin{equation*}
\text{explicit dipole solvent $>$ 
constant permittivity $>$ empirical permittivity}
\; .
\end{equation*}
While recognizing that two of the solvent models investigated in Fig.~12a 
use the theoretical formulation with the matter conjugate field $w$ 
(Eqs.~\ref{eq:f}--\ref{eq:AkBk}), it is noteworthy that
the rank ordering of LLPS propensity in Fig.~12a is consistent with
the above $w\to 0$ analysis. First, because dipole-dipole interactions 
in the explicit dipole solvent model enhance phase separation relative 
to the empricial model (Eqs.~\ref{eq:fel_sep1} and \ref{eq:ph2_1}), 
explicit dipole solvent $>$ empirical permittivity. Second,
after subtracting the free energy for the constant permittivity model
from the free energy $f_{\rm el}'$ for the empirical permittivity model,
a LLPS-disfavoring repulsive term probably remains
(Eqs.~\ref{eq:decouple2}, \ref{eq:2nd_intricate}--\ref{eq:decouple2_3} and 
discussion that follows), i.e., it is likely that
constant permittivity $>$ empirical permittivity.
These analytical trends thus provide a partial rationalization of the rank
ordering of LLPS propensity in Fig.~12a, although the above $w\to 0$
consideration does not address the relative LLPS propensities
of explicit dipole solvent and constant permittivity models.

In previous RPA studies of LLPS by our group, a ``self-energy'' term was 
subtracted from the RPA electrostatic free energy for calculational 
convenience.\cite{linJML,njp2017} While the procedure is valid for
polyampholytes interacting in a medium of uniform relative permittivity 
because in that case the self-energy term does not affect phase 
behaviors,\cite{linJML} the procedure is problematic for systems
with polymer-concentration-dependent relative permittivity because
in those situations the self-energy term may capture part of the 
polymer-solvent interactions and therefore can impact phase equilibrium, 
as we have pointed out recently.\cite{SumanPNAS}
The present analysis allows for further evaluation of such a self-energy 
term.  Consider, in the theoretical framework developed above, 
the expression
\begin{equation}
f_{\rm self} \equiv
\int_0^\infty \frac{\dint k k^2}{4\pi^2} 
\frac{4\pi\lb}{\epsilon_{\rm r}(\phi)k^2 N}
\hat{\Gamma}_k^2\phi\sum_{\tau=1}^N\sigma_\tau^2
\; ,
\label{eq:fel-self}
\end{equation}
which can be seen to be equivalent to the self-energy 
${\cal G}_2({\tilde k})$ quantity in Eq.~69b of ref.~\citen{linJML}
when there is no salt or counterion ($\phi_s=\phi_c=0$),
no short-range cutoff of Coulomb interactions 
(${\tilde k}^2[1+{\tilde k}^2]\to k^2$), and $\eta=1$,
by noting that $k={\tilde k}/l$ ($l$ now corresponds to $b=a$ 
in ref.~\citen{linJML}), $\lb=l/T^*_0$,
$\hat{\Gamma}_k\to 1$ for $\bar{a}\to 0$, and that
$\sigma_i^2=|\sigma_i|$ for polyampholytes with $\sigma_i=\pm 1$.
An expansion of the factor $[\epsilon_{\rm r}(\phi)]^{-1}$
$=[1+\chi_{\rm D}(1-\phi)]^{-1}$ (Eq.~\ref{eq:linear_epr})
in the integrand for $f_{\rm self}$ in Eq.~\ref{eq:fel-self}
yields 
\begin{equation}
f_{\rm self} =
\int_0^\infty \frac{\dint k k^2}{4\pi^2} \frac{4\pi\lb}{k^2N}\hat{\Gamma}_k^2
\phi\sum_{\tau=1}^N\sigma_\tau^2
\left[ 1 - \chi_{\rm D}(1-\phi) 
+ \mathcal{O}\left( \chi_{\rm D}^2 \right) \right]
\; .
\label{eq:fselfexpansion}
\end{equation}
This form conveys two messages. First, while a part 
of $f_{\rm self}$ may be identified
with a part of $f_{\rm el}'$, e.g.,
the term linear in $\phi$ in the expansion of $f_{\rm self}$ is equal to
a part of the first term (linear in $\phi$) in the expansion of
the logarithmic integrand of the first integral
in Eq.~\ref{eq:decoupled0} (the
$\mu=\nu$ part of the summation for $g^{\rm cc}_k$ in Eq.~\ref{eq:gcc_etc}c),
and therefore may be viewed as capturing part of polymer-solvent
interactions, $f_{\rm self}$ by itself
does not coincide with the free energy arising from polymer-solvent
interactions. Second,
the $\phi^2$ term in the $f_{\rm self}$ expanion 
in Eq.~\ref{eq:fselfexpansion} is positive, indicating that $f_{\rm self}$
entails an effective polymer-polymer repulsion. It follows that subtracting
$f_{\rm self}$ should increase LLPS propensity.
This effect is illustrated by the coexistence curve 
in 
Fig.~12b
computed using a free energy with $f_{\rm self}$ subtracted (dotted red curve).
The dramatic increase in LLPS propensity of this model compared with
corresponding models without the self-energy subtraction (orange curves)
is in line with some of our previous results on models with self-energy
subtraction (Fig.~12 of ref.~\citen{linJML}). However, for
other heteropolymer models with less charge densities and augmented
Flory-Huggins interactions, subtracting an electrostatic self-energy 
term can lead to more modest---though still substantial---increases 
in LLPS propensity (Fig.~7C of ref.~\citen{SumanPNAS}).
Most importantly, the present analysis shows that
a self-energy term similar to the one in Eq.~\ref{eq:fel-self}
does not account neatly for polymer-solvent interactions, and in any event,
subtraction of such a term from the RPA electrostatic free energy is
physically unwarranted.
\\

\noindent
{\large\bf DISCUSSION}\\

{\bf FTS of Polyampholytes with Dipole Solvent.}
In addition to RPA, we have extended our FTS effort
beyond the study of pure dipoles and dipole and nonpolar solvent
mixtures (Figs.~7--9) to the investigation of
polyampholyte LLPS in the presence of dipoles.
In this endeavor, we find that the strict incompressibility constraint 
considered above for RPA (Eq.~\ref{eq:fullincomp}) is challenging
to realize numerically in the fully fluctuating FTS simulations of 
systems comprised of polyampholytes and explicit polar solvents.  
Therefore, we opt instead for a compressible model in which the Dirac 
$\delta[\rho_{\rm p}({\bf{r}}) + \rho_{\rm w}({\bf{r}}) - 1/v]$ for
incompressibility is replaced 
by an energy penalty term in the interaction Hamiltonian in
the form of
$ \int \dint {\bf{r}} (\rho_{\rm p}({\bf{r}}) + \rho_{\rm w}({\bf{r}})  - 1/v)^2 / 2
\gamma$. The resulting system is
incompressible in the $\gamma\rightarrow 0$ limit but
compressible for finite $\gamma$.
The corresponding field-picture Hamiltonian is given by
\begin{equation}
\HH[\psi, w] = \frac{1}{8\pi \lb}\int \dint {\bf{r}} (\bm{\nabla} 
\psi ({\bf{r}}) )^2 + \frac{ \gamma }{ 2 } \int \dint {\bf{r}} w( {\bf{r}} )^2 - 
\frac{\ii}{v} \int \dint {\bf{r}} w({\bf{r}}) - \nw \ln \Qw - \np \ln \Qp 
\; ,
\end{equation}
where the partition functions $\Qw$ and $\Qp$
for a single dipole and a single polymer 
chain, respectively, are given by Eq.~\ref{Qwfixed} and
Eq.~\ref{Qpsv}. We focus here on a solution comprised of a
single species of polyampholytes and a single species of dipoles as a model
of a polar solvent.
The equilibrium averages of thermodynamic variables of the system are obtained
by averaging over sufficiently large number of statistically independent
configurations of the fields $\psi$ and $w$.  Following the CL prescription 
in Eq.~\ref{eq:CL_eqs}, the field configurations are generated by numerically
solving the following two coupled differential equations:
\begin{equation}
\label{CLComp}
\begin{aligned}
\frac{\partial w({\bf{r}})}{\partial t} &= - \left[ \ii \left(\tilde{\rho}_{\rm{p}}({\bf{r}}) + \tilde{\rho}_{\rm{w}}({\bf{r}}) - \frac{1}{v} \right) +  \gamma w({\bf{r}}) \right] + \eta_{w}
\; ,\\
\frac{\partial\psi({\bf{r}})}{\partial t} &= - \left[ \ii \left(\tilde{c}({\bf{r}}) + \tilde{c}_{\rm{w}}({\bf{r}})\right) - \frac{1}{4\pi\lb}\bm{\nabla}^2\psi({\bf{r}}) \right]\ + \eta_{\psi}
\; .
\end{aligned}
\end{equation}
In Eq.~\ref{CLComp}, $t$ is a fictitious time,  $\eta_{w}$ and $\eta_{\psi}$
are real-valued random numbers drawn from a Normal distribution with zero mean
and variance $2\delta(t-t')\delta({\bf{r}} - {\bf{r}} ')$, and the field operators
for polymer bead density ($\tilde{\rho}_{\rm{p}}$),  polymer charge density
($\tilde{c}$), solvent bead density ($\tilde{\rho}_{\rm{w}}$) and solvent
charge density ($\tilde{c}_{\rm{w}}$) are given by
\begin{equation}
\begin{aligned}
\tilde{\rho}_{\rm{p}}({\bf{r}}) & = \frac{\ii \np}{\Qp \Omega} 
\frac{\delta \Qp}{\delta w}
\; , \quad \tilde{c}({\bf{r}}) = \frac{\ii \np}{\Qp \Omega} 
\frac{\delta \Qp}{\delta \psi}
\; ,  \\
\tilde{\rho}_{\rm{w}}({\bf{r}}) &= \frac{\ii \nw}{\Qw \Omega} 
\frac{\delta \Qw}{\delta w}
\; , \quad \tilde{c}_{\rm{w}}({\bf{r}}) = \frac{\ii \nw}{\Qw \Omega} 
\frac{\delta \Qw}{\delta \psi}
\; .
\end{aligned}
\end{equation}
The individual solvent beads are overall charge neutral, but polar in nature
with each having a fixed magnitude of dipole moment equals to $\muwr$.  
$\tilde{c}_{\rm{w}}$ can be
re-expressed by defining a solvent polarization density operator such that
$\tilde{c}_{\rm{w}} = - \bm{\nabla} \cdot \widetilde{\dipole}({\bf{r}})$ where
\begin{equation}
\label{Psnap}
\widetilde{\dipole}({\bf{r}}) = \frac{\ii \nw}{\Qw \Omega} {\rm e}^{-\ii \breve{w}({\bf{r}})} \frac{ \bm{\nabla} \breve{\psi} ({\bf{r}}) } { \left[\bm{\nabla} \breve{\psi} ({\bf{r}}) \right] ^2 } \left[ \cos \left( \muwr \mathopen| \bm{\nabla} \breve{\psi} ({\bf{r}}) \mathclose| \right) - \frac{\sin \left( \muwr \mathopen| \bm{\nabla} \breve{\psi} ({\bf{r}}) \mathclose| \right)}{\muwr \mathopen| \bm{\nabla} \breve{\psi} ({\bf{r}}) \mathclose|} \right]
\; ,
\end{equation}
and the smeared fields in this expression, each marked by a diacritic breve,
are defined by spatial convolution as
$\breve{\varphi} = \Gamma \star \varphi$, $\varphi= w, \psi$. We integrate
Eq.~\ref{CLComp} numerically by using the first-order semi-implicit time
integration method of ref.~\citen{Lennon_etal_2008} over a cubic box of volume
$\Omega = (32 \bar{a})^3$. To achieve good numerical precision, we discretize
the box into a $32\times32\times32$ grid.  The volume of a single bead is taken
to be $v = 0.1$, the bulk polymer bead density is set at $0.47$, $\muwr = 0.1$
and $\gamma = 10$.  Although not rigorously identical to 
matter or charge densities, the CL time snapshots of density 
operators are an intuitive
and instructive way to visualize phase separation in FTS
systems.\cite{joanElife,joanJPCL,Pal2021} 
Fig.~13 shows representative snapshots of the
field operators of different observables for two different temperatures to
illustrate that the present polyampholytes plus dipoles system 
can indeed undergo phase separation.
As temperature is lowered from a relatively high temperature 
$(\textit{T}^*_0 = 50)$ to a relatively low temperature 
$(\textit{T}^*_0 = 0.2)$, polymer material (red) distributed rather uniformly
throughout the simulation box at high temperature  (Figs.~13d--f)
is seen to coalesce into a droplet-like structure at 
low temperature (Figs.~13a--c). This FTS behavior is similar to that observed
for molecular dynamics simulations in Fig.~5 for the same polyampholyte
sequence sv20.


Technically, the cost of introducing compressibility is that we now 
have two independent chemical potentials corresponding to the 
polyampholytes and the solvent.
This makes computation of thermodynamically consistent FTS phase
diagrams difficult because the approach applied in 
refs.~\citen{joanElife}~and~\citen{Delaney2017} 
is not workable for the present system and one needs to employ Gibbs
ensemble FTS similar to methodologies used in 
refs.~\citen{joanJPCL}~and~\citen{Riggleman2010}. 
Even so, the expected extreme dilute nature of the low-density phase in
our FTS system of interest, as suggested by the RPA results in Figs.~11 and 12, 
would likely make it challenging to arrive at a
set of parameters for which the Gibbs ensemble FTS is numerically stable. 
Work is now underway to overcome these challenges.

\begin{center}
   \includegraphics[width=0.68\columnwidth]{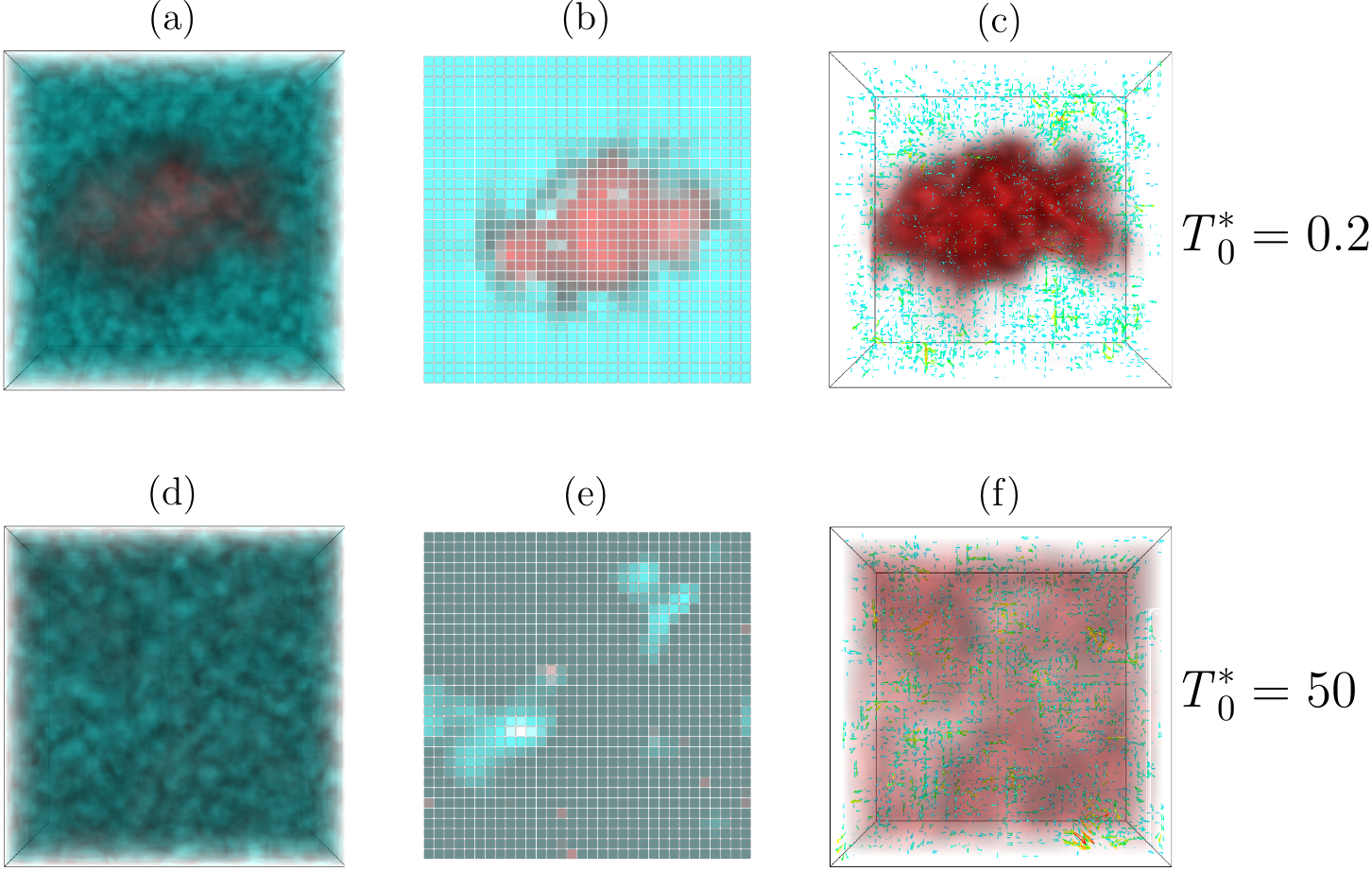}
\end{center}
{\footnotesize{\bf Fig.~13:}
Illustrative FTS snapshots of the density operators at low
and high temperatures.
All results shown are for polyampholyte sequence sv20,
dipole $\muwr = 0.1$, excluded-volume parameter
$v=0.1$, and $\gamma = 10$.
(a) Temperature
$\textit{T}^*_0 = 0.2$. Real non-zero parts of $ \tilde{\rho}_{\rm{p}}(\bf{r})$
and $\tilde{\rho}_{\rm{w}}(\bf{r})$. A droplet-like structure 
of sv20 molecules
(shown in red) is buried inside the polar solvent (shown in cyan). (b) A
cut-plane view of (a)
of a thin slice of the simulation box parallel to the page.
(c) sv20 droplet (in red) is depicted together
with the real part of solvent (dipole) polarization operator
$\widetilde{\dipole}(\bf{r})$ as arrows.
(d)--(f): Same as (a)--(c), but at $\textit{T}^*_0 =
50$. Here, the polyampholytes and solvent are seen to distribute
in an essentially uniform manner throughout the simulation box.
}
$\null$\\

{\bf The Distinctive Solvent Dielectric Environments in the Dilute 
and Condensed Phases Likely Contribute to a Minor Enhancement of 
Polyampholyte LLPS Propensity.} 
As reported above, our molecular dynamics and analytical RPA theories
indicate, at least for the models considered, that allowing
the solvent-contributed effective $\epsilon_{\rm r}$ inside a condensed 
polyampholyte-rich droplet to decrease by using a physically
plausible explicit polar solvent model (Fig.~4) leads in most cases to an 
enhancement of polyampholyte LLPS propensity relative to that predicted by
commonly utilized implicit-solvent formulations in which 
the polyampholytes interact electrostatically via $\epsilon_{\rm r}$ 
of the bulk solvent (Figs.~6, 11, and 12). However,
contrary to earlier suggestions that this effect can be 
dramatic,\cite{linJML,njp2017} the enhancement effect observed here 
is only minor or at most moderate.
We may quantify the enhancement effect by the ratio of the critical 
temperature of the explicit-solvent model to that of 
the implicit-bulk-solvent, constant $\epsilon_{\rm r}$ model, or
the percentage increase of the former over the latter.
For the molecular dynamics results in Fig.~6,
this percentage increase in critical temperature,
$[T^*_{0,{\rm cr}}({\rm ii})-T^*_{0,{\rm cr}}({\rm iii})]/T^*_{0,{\rm cr}}({\rm iii})$, 
obtained from the $T^*_{0,{\rm cr}}$ values in Table~1 
is approximately $-2\%$ for sequence sv15,
$5\%$ for sv20, and $11\%$ for sv30.
These changes are minor, but nonetheless exhibit an increasing trend
with increasing blockiness of the sequence charge pattern as characterized
by $-$SCD, from essentially no change in $T^*_{0,{\rm cr}}$
for sv15 ($-$SCD $=4.35$) to a minor increase of $11\%$ for sv30 
($-$SCD $=27.8$). In this context, it is interesting to note that for sv30, 
$T^*_{0,{\rm cr}}$ of the explicit-solvent model (ii) is almost identical 
to that of the explicit nonpolar solvent model with $\epsilon_{\rm r}=1$ 
(i), suggesting in this case a relatively strong enhancement of 
polyampholyte LLPS by the heterogeneous dielectric environment entailed
by the explicit-solvent model.

The corresponding RPA results in Fig.~12a also show minor to at most
moderate increases.
The $T^*_{0,{\rm cr}}\epsilon_{\rm r}$ values for the
explicit dipole, constant permittivity, and empirical permittivity
models in Fig.~12a are, respectively,
$4.46$, $3.37$, and $3.24$ for sv15,
$7.10$, $6.16$, and $5.49$ for sv20, and
$23.7$, $21.3$, and $19.4$ for sv30.
For these models, while the increase in $T^*_{0,{\rm cr}}\epsilon_{\rm r}$
of the explicit dipole model over that of the constant permittivity model
(difference between the two $T^*_{0,{\rm cr}}\epsilon_{\rm r}$ values)
tends to increase with $-$SCD ($1.09$ for sv15, $0.94$ for sv20, and
$2.40$ for sv30), the perecentage increase in $T^*_{0,{\rm cr}}$ decreases
with $-$SCD: approximately $32\%$ for sv15, $15\%$ for sv20, and $11\%$
for sv30.
To what extent are these differences between the molecular dynamics and
RPA models attributable to their different bulk-solvent $\epsilon_{\rm r}$
values (4.0 for molecular dynamics, 20.0 for RPA) and
their substantially different condensed-phase polymer volume fractions
(much lower for RPA than for molecular dynamics)  
remains to be investigated.
These uncertainties notwithstanding, the results in Figs.~6, 11, and 12
suggest that when all interactions among polyampholytes and polar solvents
are appropriately taken into account, the enhancement of polyampholyte
LLPS by the concentration-dependent dielectric heterogeneity of the polar
solvent is likely minor. At the very least, one can be certain that the 
physical effect is much more modest than when part of the solvent-polymer
interactions are neglected as in the case of the unphysical
``no self energy'' model depicted by the dotted red curve in Fig.~12b.
\\

{\bf The Dielectric Environment Experienced by Client Molecules in a
Condensed Polyampholyte Droplet Can be Complex.}
We have focused so far on the $\epsilon_{\rm r}$ contributed by the 
polar solvent and its concentration dependence (Figs.~4, 8, and 9) 
in order to gauge the solvent's role in the effective LLPS-driving 
electrostatic interactions among the polyampholytes.
In these formulations, the charges on the polyampholytes are treated 
explicitly, not as a part of the solvent-contributed dielectric 
background.\cite{linJML,njp2017,SumanPNAS} Once the polyampholyte-rich 
droplet---as a model biomolecular condensate---is formed, however, the 
dielectric environment entailed by the droplet as a whole---including 
contributions from both the polar solvent and the polyampholyte chains---is 
of potential biophysical and biochemical interest because, among its 
many effects, it would affect how other molecules, sometimes referred 
to as clients, partition into and interact within the droplet.

\begin{center}
\vskip - 16mm
   \includegraphics[width=0.56\columnwidth]{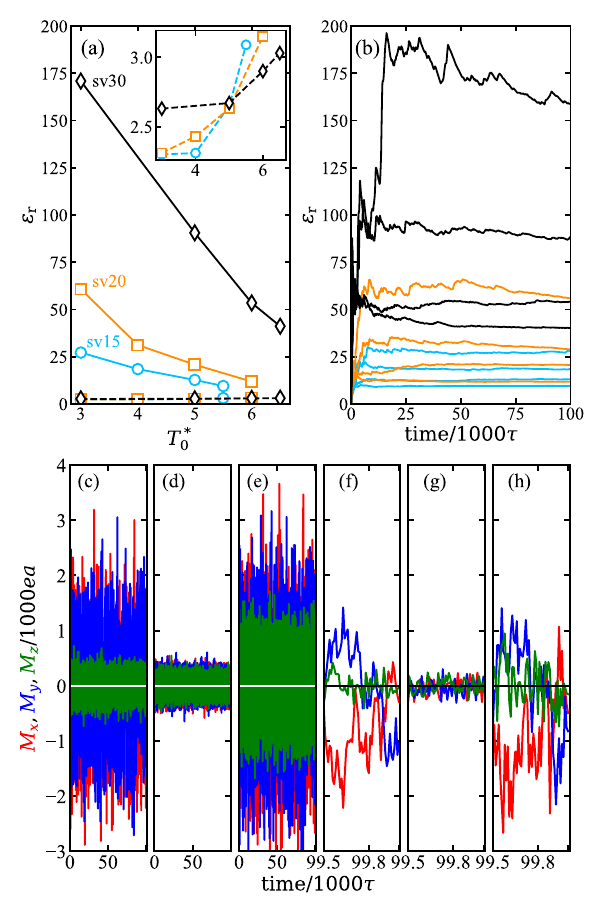}
\end{center}
\vskip -6 mm
{\footnotesize{\bf Fig.~14:}
Effective relative permittivity of condensed polyampholyte-rich droplets
in dipole solvent computed by molecular dynamics simulations for the 
systems in Fig.~4.
(a) For polyampholyte sequences sv15 (circles), sv20 (squares) and 
sv30 (diamonds), effective $\epsilon_{\rm r}$ inside each droplet 
(data points connected by solid lines) is obtained by applying 
Eq.~\ref{eq:epr_slab} to local and global dipole moments contributed from 
both the polyampholytes and the dipole solvent molecules, wherein local
dipole contributions are averaged over long simulation trajectories [shown
in (b)] and averaged over the volume of the droplet delimited
by an interval of width $\Delta z$ in the $z$-direction of the simulation 
box.  Included for comparison are effective $\epsilon_{\rm r}$
for each droplet contributed by the dipole solvent alone
(data points connected by dashed lines at the bottom, shown also in 
the inset using a zoomed-in vertical scale for $\epsilon_{\rm r}$). The 
widths $\Delta z/a$ of the droplets used for this calculation, listed by 
increasing temperature (same $T^*_0$s as those in Fig.~4), are: 40, 45, 
60, 80 for sv15; 40, 45, 60, 85 for sv20; and 45, 60, 70, 80 for sv30.
(b) Computed $\epsilon_{\rm r}$ as a function of sampling time for the
systems in (a) using the same color code for the sequences, with
$T^*_0$ increasing from the bottom to top curve for each sequence.
The time variable (horizontal axis) of Langevin molecular dynamics 
is in units of $\tau$ (Models and Methods).
(c)--(h) The variations of the $x$, $y$, and $z$ components of the dipole 
moment ${\bf M}$, $M_x$, $M_y$, and $M_z$, over a $10^5\tau$ simulation 
trajectory are traced in red, blue, and green, respectively,
in units of $ea$ ($e$ is elementary protonic charge; $a$ is reference
bond length) for the sv30, $T^*_0=6.0$ system in (a) and (b) as an example.
As a guide for the eye,
${\bf M}={\bf 0}$ is marked by a white (c, d, e) or black (f, g, h)
horizontal line.
Shown in (c)--(e) are the total dipole moments of (c) polyampholytes
and of (d) dipole solvent molecules inside the droplet (c and d), and
(e) the total dipole moment of the simulation box as a whole, all for 
the entire duration of the simulated trajectory. 
(f)--(h) The last $500\tau$ of the ${\bf M}$ variations in (c)--(e),
respectively, are depicted using in a zoomed-in horizontal 
scale for time $=99,500\tau$ to $100,000\tau$.  
}
$\null$\\

Taking a first step to address this interesting question, we compute,
using our molecular dynamics model, the overall effective relative 
permittivities of droplets that account for contributions from the polar
solvent as well as the polyampholytes themselves (Fig.~14). In the context
of biomolecular modeling, this overall effective $\epsilon_{\rm r}$ 
would approximate the relative permittivity experienced by client molecules
(test charges) inside a biomolecular condensate. Fig.~14a shows that this
overall droplet $\epsilon_{\rm r}$ increases dramatically with $-$SCD
of the sequences. For the three sequences considered, the overall droplet
$\epsilon_{\rm r}$s estimated from largely equilibrated trajectories
barring one exception (Fig.~14b) far exceed the model solvent's 
effective $\epsilon_{\rm r}$
of $4.0$. For sv30, the overall droplet $\epsilon_{\rm r}$s at low
temperatures likely exceed even the $\approx 80$ value for bulk water,
although the simulation at $T^*_0=3.0$ for this sequence has not fully 
equilibrated (top black curve in Fig.~14b).
These high values of overall droplet $\epsilon_{\rm r}$ is likely caused
by the highly charged nature of the model polyampholytes we studied. 
Intuitively, one may expect less dramatic dielectric effects in biomolecular
condensates scaffolded by natural IDP polyampholytes that are less charged.
By comparison, $\epsilon{\rm_r}$ of folded proteins has been estimated by
experiment on dry protein powder\cite{harvey1972} and an
early theoretical treatment\cite{honig} to be $\approx 2$--$4$,
whereas subsequent experiments on dry synthetic 
polypeptides\cite{hole1976} suggest higher values of up to 20 and 
experimental pK$_{\rm a}$ measurements of buried charged 
residues indicate $\epsilon_{\rm r}\approx 10$--$12$
in the largely nonpolar core of a folded protein.\cite{bertrand0,bertrand} 
Simulations using explicit atomic models of 
water\cite{vanGunsteren2001,kings2017} affording $\epsilon_{\rm r}\approx
15$--$40$ for natural folded proteins are roughly in line with the more 
recent experiments though the simulated $\epsilon_{\rm r}$ value
for some folded protein can be as high 
as $\approx 58$ (ref.~\citen{kings2017}). 
Whether condensates of dynamic, highly charged natural IDPs can
result in significantly higher $\epsilon_{\rm r}$ values
and related questions regarding the spatial heterogeneity 
of the dielectric environment within condensed droplets remain to be
further explored.

As an illustration of the relationship between dielectric properties and
the dynamics of a condensed polyampholyte-rich droplet, fluctuations 
of various components of dipole moment, ${\bf M}$, in the sv30, 
$T^*_0=6.0$ system are provided in Figs.~14c--h. We note that the 
${\bf M}$ fluctuation of the dipole solvent is essentially isotropic as 
expected (nearly identical amplitudes in the $x$, $y$, and $z$ 
directions in Fig.~14d), whereas the ${\bf M}$ fluctuation of the 
polyampholytes is significantly smaller in amplitude in the $z$
direction than in the $x$ and $y$ directions (green curve has narrower
range than the red and blue curves in Fig.~14c). This is because in our
simulation system, the droplet has two polyampholyte-solvent
interfaces along the $z$ directions but no such interfaces along
the $x,y$ directions because of periodic boundary conditions (Fig.~5), 
hence for the finite-size (small) droplet in the present model, fluctuation 
in the $z$ direction is relatively much more constrained.

The zoomed-in curves in Figs.~14f--h focusing on the large-time
equilibrated regime indicate that the ${\bf M}$ 
fluctuations of dipole solvent molecules (Fig.~14g) and 
polyampholytes (Fig.~14f) involve very different time scales, exhibiting
much slower ${\bf M}$ variation with respect to the Langevin 
molecular dynamics time (horizontal variable in Fig.~14c-h, denoted 
as $t_{\rm MD}$ below) for the polyampholytes than for the 
dipole solvent. We quantify this difference by calculating the
temporal correlation 
$\langle {\bf M}(t_{\rm MD}+t_{{\rm MD},0})\cdot{\bf M}(t_{{\rm MD},0})\rangle_{t_{{\rm MD},0}}/\langle {\bf M}(t_{{\rm MD},0})\cdot{\bf M}(t_{{\rm MD},0})\rangle_{t_{{\rm MD},0}}\approx \exp(-t_{\rm MD}/\tau_{\bf M})$ from
the simulated trajectory,
finding that the relaxation time $\tau_{\bf M}\approx 141.0\tau$ 
for the polyampholytes is $\sim 350$ times that of the
$\tau_{\bf M}\approx 0.406\tau$ for the dipole solvent.
This result suggests, perhaps not too surprisingly, 
that the polyampholyte chain molecules and the small dipole solvent 
molecules have very different dielectric response times.
Whereas investigations of dielectric dispersion of aqueous protein 
systems have primarily concerned with the dispersive effects of
protein-water interactions,\cite{bone1982} the present results suggest that
dielectric disperson caused by chain dynamics is an aspect of the
material properties of biomolecular condensates\cite{banerjee2021}
that deserves attention as well because of its biophysical, 
biochemical, and potentially novel functional implications.
\\


\noindent
{\large\bf CONCLUSION}\\

In summary, our explicit-chain molecular dynamics as well as analytical
theory suggest that the heterogeneous solvent dielectric environment
entailed by polyampholyte LLPS---with less solvent inside the condensed
phase than outside---entails in most cases a minor but nonetheless 
appreciable enhancement of LLPS propensity relative to that predicted 
by polyampholytes interacting in a uniform dielectric environment carrying
the bulk-solvent relative permittivity. While these observations
are based on simplified models, they suggest that the common approach
of modeling polyampholyte LLPS with a uniform $\epsilon\approx 80$
likely leads to a minor underestimation of LLPS propensity. The extent
of the expected small error, however, has to be ascertained by futher
investigation, using simulations with realistic water models if possible.
Further advances building on the analytical formulation developed
here should also provide a complementary approach to address this basic
question. Although concentration-dependent solvent permittivity likely
incurs only a minor effect on the stability of polyampholyte condensates, 
we find that the overall dielectric environment contributed by both 
the polar solvent and the polyampholyte scaffold inside a condensate
can be highly sensitive to the sequence charge pattern of the polyampholyte,
likely resulting in a complex combination of dielectric responses 
characterized by vastly different time scales. All told,
much about the complexity of the internal dielectric environment
of biomolecular condensates and its biophysical and biochemical 
ramifications remain to be investigated.
\\

{\bf Acknowledgements.}
We thank Julie Forman-Kay for helpful discussions. Financial support
for this work was provided by Canadian Institutes of Health
Research grant NJT-155930 and Natural Sciences and Engineering
Research Council of Canada Discovery grant RGPIN-2018-04351. 
We are grateful for the computational resources provided by 
Compute/Calcul Canada.
\\

\noindent The authors declare no conflict of interest.

\vfill\eject \noindent
{\Large\bf References}\\

\vfill\eject 

\begin{center}
   \includegraphics[height=44.5mm]{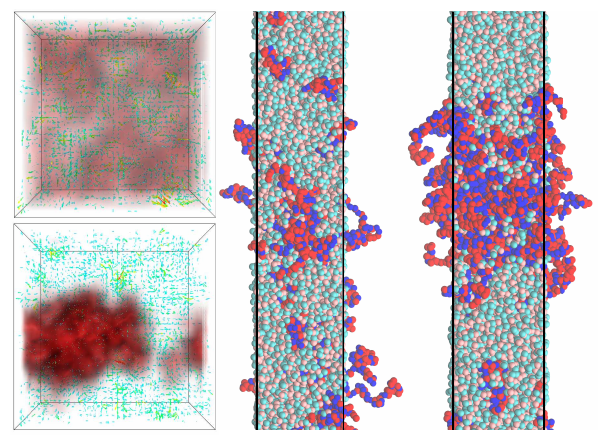}
\end{center}
\centerline{\bf TOC graphics}

\vfill\eject


\begin{thebibliography}{999}

\bibitem{rosen2017}
Banani,~S.~F.; Lee,~H.~O.; Hyman,~A.~A.; Rosen,~M.~K. 
Biomolecular condensates: organizers of cellular biochemistry.
{\it Nat. Rev.  Mol. Cell Biol.} {\bf 2017}, {\it 18}, 285--298.

\bibitem{CellBiol}
Hyman, A. A.; Weber, C. A.; J\"ulicher, F.
Liquid-liquid phase separation in biology.
{\it Annu. Rev. Cell Dev. Biol.} {\bf 2014}, {\it 30}, 39--58.

\bibitem{NatPhys}
Brangwynne, C. P.; Tompa, P.; Pappu, R. V.
Polymer physics of intracellular phase transitions.
{\it Nat. Phys.} {\bf 2015}, {\it 11}, 899--904.

\bibitem{linPRL}
Lin, Y.-H.; Forman-Kay, J. D.; Chan, H. S.
Sequence-specific polyampholyte phase separation in membraneless organelles.
{\it Phys. Rev. Lett.}, 2016, {\bf 117}, 178101.

\bibitem{tanja2019}
Alberti, S.; Gladfelter, A.; Mittag, T.
Considerations and challenges in studying liquid-liquid phase separation
and biomolecular condensates.
{\it Cell} {\bf 2019}, {\it 176}, 419--434.

\bibitem{brangwynne2009}
Brangwynne,~C.~P.; Eckmann,~C.~R.; Courson,~D.~S.; Rybarska,~A.; Hoege,~C.;
  Gharakhani,~J.; J\"ulicher,~F.; Hyman,~A.~A. Germline P granules are liquid
  droplets that localize by controlled dissolution/condensation. 
{\it Science} {\bf 2009}, {\it 324}, 1729--1732.

\bibitem{Rosen12}
Li,~P.; Banjade,~S.; Cheng,~H.~C.; Kim,~S.; Chen,~B.; Guo,~L.; Llaguno,~M.;
  Hollingsworth,~J.~V.; King,~D.~S.; Banani,~S.~F.; Russ, P. S.;
Jiang, Q.-X.; Nixon, B. T.; Rosen, M. K.
Phase transitions in the assembly of multivalent signalling proteins.
{\it Nature} {\bf 2012}, {\it 483}, 336--340.

\bibitem{McKnight12}
Kato,~M.; Han,~T.~W.; Xie,~S.; Shi,~K.; Du,~X.; Wu,~L.~C.; Mirzaei,~H.;
Goldsmith,~E.~J.; Longgood,~J.; Pei,~J.;
Grishin, N. V.; Frantz, D. E.; Schneider, J. W.; Chen, S.; Li, L.;
Sawaya, M. R.; Eisenberg, D.; Tycko, R.; McKnight, S. l. 
Cell-free formation of RNA granules: low complexity sequence domains 
form dynamic fibers within hydrogels.
{\it Cell} {\bf 2012}, {\it 149}, 753--767.

\bibitem{Nott15}
Nott,~T.~J.; Petsalaki,~E.; Farber,~P.; Jervis,~D.; Fussner,~E.;
  Plochowietz,~A.; Craggs,~T.~D.; Bazett-Jones,~D.~P.; Pawson,~T.;
  Forman-Kay,~J.~D.; Baldwin, A. J.
Phase transition of a disordered nuage
 protein generates environmentally responsive membraneless organelles.
{\it Mol. Cell} {\bf 2015}, {\it 57}, 936--947.

\bibitem{tanja2015}
Molliex,~A.; Temirov,~J.; Lee,~J.; Coughlin,~M.; Kanagaraj,~A.~P.; Kim,~H.~J.;
Mittag,~T.; Taylor,~J.~P. Phase separation by low complexity domains promotes
stress granule assembly and drives pathological fibrillization.
{\it Cell} {\bf 2015}, {\it 163}, 123--133.

\bibitem{parker2015}
Lin, Y.; Protter, D. S. W.; Rosen, M. K.; Parker, R. 
Formation and maturation of phase-separated liquid droplets by RNA-binding 
proteins. {\it Mol. Cell} {\bf 2015}, {\it 60}, 208--219.

\bibitem{Michnick2016}
Bergeron-Sandoval, L.P.; Safaee, N.; Michnick, S. W.
Mechanisms and consequences of macromolecular phase separation.
{\it Cell} {\bf 2016}, {\it 165}, 1067--1079.

\bibitem{cliff2017}
Shin, Y.; Brangwynne, C.P. Liquid phase condensation in cell physiology and
disease. {\it Science} {\bf 2017}, {\it 357}, eaaf4382.

\bibitem{Monika2018Rev}
Boeynaems,~S.; Alberti,~S.; Fawzi,~N.~L.; Mittag,~T.; Polymenidou,~M.;
  Rousseau,~F.; Schymkowitz,~J.; Shorter,~J.; Wolozin,~B.; Van Den~Bosch,~L.;
Tompa, P.; Fuxreiter, M.
Protein phase separation: a new phase in cell biology.
{\it Trends Cell Biol.} {\bf 2018}, {\it 28}, 420--435.

\bibitem{weber2020}
Ladouceur, A.-M.; Parmar, B. S.; Biedzinski, S.; Wall, J.;
Tope, S. G.; Cohn, D.; Kim, A.; Soubry, N.; Reyes-Lamothe, R.; Weber, S. C.
Clusters of bacterial RNA polymerase are biomolecular condensates 
that assemble through liquid–liquidphase separation.
{\it Proc. Natl. Acad. Sci. U.S.A.} {\bf 2020}, {\it 117}, 18540--18549.

\bibitem{chong2016}
Chong, P. A.; Forman-Kay, J. D.
Liquid-liquid phase separation in cellular signaling systems.
{\it Curr. Opin. Struct. Biol.} {\bf 2016}, {\it 41}, 180--186.

\bibitem{babu2018}
Li, X.-H.; Chavali, P. L.; Pancsa, R.; Chavali, S.; Babu, M. M.
Function and regulation of phase-separated biological condensates.
{\it Biochemistry} {\bf 2018}, {\it 57}, 2452--2461.

\bibitem{mingjie2020}
Chen, X.; Wu, X.; Wu, H.; Zhang, M. Phase separation at the synapse.
{\it Nat. Neurosci.} {\bf 2020}, {\it 23}, 301--310.

\bibitem{roland2020}
Cinar, H.; Oliva, R.; Lin, Y.-H.; Chen, X.; Zhang, M.; Chan, H. S.;
Winter, R. Pressure sensitivity of SynGAP/PSD-95 condensates as a model for 
postsynaptic densities and its biophysical and neurological ramifications. 
{\it Chem. Eur. J.} {\bf 2020}, {\it 26}, 11024--11031. 

\bibitem{shorter2019}
Gomes, E., and Shorter, J. The molecular language of membraneless
organelles. {\it J. Biol. Chem.} {\bf 2019}, {\it 294}, 7115--7127.

\bibitem{mekhail2020}
Laflamme, G.; Mekhail, K.
Biomolecular condensates as arbiters of biochemical reactions inside 
the nucleus. {\it Commun. Biol.} {\bf 2020}, {\it 3}, 773.

\bibitem{Schaffert_etal}
Schaffert, N.; Hossbach, M.; Heintzmann, R.; Achsel, T.; L\"uhrmann, R.
RNAi knockdown of hPrp31 leads to an accumulation of U4/U6 di‐snRNPs 
in Cajal bodies. {\it EMBO J.} {\bf 2004}, {\it 23}, 3000--3009.

\bibitem{Nott16}
Nott, T. J.; Craggs, T. D.; Baldwin, A. J. 
Membraneless organelles can melt nucleic acid duplexes and act 
as biomolecular filters. {\it Nat. Chem.} {\bf 2016}, {\it 8}, 569--575.

\bibitem{Lu_etal2020}
Lu, Y.; Wu, T.; Gutman, O.; Lu, H.; Zhou, Q.; Henis, Y. I.; Luo, K.
Phase separation of TAZ compartmentalizes the transcription machinery 
to promote gene expression.
{\it Nat. Cell Biol.} {\bf 2020}, {\it 22}, 453--464.

\bibitem{VeronikaRev2016}
Csizmok, V.; Follis, A. V.; Kriwacki, R. W.; Forman-Kay, J. D.
Dynamic protein interaction networks and new structural paradigms in signaling.
{\it Chem. Rev.} {\bf 2016}, {\it 116}, 6424--6462.

\bibitem{Fawzi2017}
Monahan, Z.; Ryan, V. H.; Janke, A. M.; Burke, K. A.; Rhoads, S. N.; 
Zerze, G. H.; O'Meally, R.; Dignon, G. L.; Conicella, A. E.; Zheng, W.;
Best, R. B.; Cole, R. N.; Mittal, J.; Shewmaker, F.; Fawzi N. L.
Phosphorylation of the FUS low‐complexity domain disrupts phase 
separation, aggregation, and toxicity.
{\it EMBO J.} {\bf 2017}, {\it 36}, 2951--2967.

\bibitem{cliff2015}
Elbaum-Garfinkle, S.; Kim, Y.; Szczepaniak, K.; Chen, C. C.-H.; Eckmann, C. R.;
Myong, S.; Brangwynne, C. P.
The disordered P granule protein LAF-1 drives phase separation 
into droplets with tunable viscosity and dynamics.
{\it Proc. Natl. Acad. Sci. U.S.A.} {\bf 2015}, {\it 112}, 7189--7194.

\bibitem{BabuRev2014}
van der Lee, B.; Buljan, M.; Lang, B.; Weatheritt, R. J.; Daughdrill, G. W.;
Dunker, A. K.; Fuxreiter, M.; Gough, J.; Gsponer, J.; Jones, D. T.; Kim, P. M.;
Kriwacki, R. W.; Oldfield, C. J.; Pappu, R. V.; Tompa, P.; Uversky, V. N.;
Wright, P. E.; Babu, M. M.
Classification of intrinsically disordered regions and proteins.
{\it Chem. Rev.} {\bf 2014}, {\it 114}, 6589–-6631.

\bibitem{linJML}
Lin, Y.-H.; Song, J.; Forman-Kay, J. D.; Chan, H. S.
Random-phase-approximation theory for sequence-dependent, biologically 
functional liquid-liquid phase separation of intrinsically disordered proteins. 
{\it J. Mol. Liq.}, 2017, {\bf 228}, 176--193.

\bibitem{HXZhouRev2018}
Zhou, H.-X.; Pang, X.
Electrostatic interactions in protein structure, folding, binding, and 
condensation. {\it Chem. Rev.} {\bf 2018}, {\it 118}, 1691--1741.

\bibitem{robert}
Vernon, R. M.; Chong, P. A.; Tsang, B.; Kim, T. H.; Bah, A.; Farber, P.;
Lin, H.; Forman-Kay, J. D. Pi-Pi contacts are an overlooked protein feature
relevant to phase separation. {\it eLife} {\bf 2018}, {\it 7}, e31486.

\bibitem{moleculargrammar}
Wang, J.; Choi, J. M.; Holehouse, A. S.; Lee, H. O.; Zhang, X.; Jahnel,
M.; Maharana, S.; Lemaitre, R.; Pozniakovsky, A.; Drechsel, D.; Poser, I.;
Pappu, R. V.; Alberti, S.; Hyman, A. A. A molecular grammar governing the
driving forces for phase separation of prion-like RNA binding proteins. 
{\it Cell} {\bf 2018}, {\it 174}, 688--699.

\bibitem{julieRev}
Vernon, R. M.; Forman-Kay, J. D.
First-generation predictors of biological protein phase separation.
{\it Curr. Opin. Struct. Biol.} {\bf 2019}, {\it 58}, 88--96.

\bibitem{TanjaScience2020}
Martin, E. W.; Holehouse, A. S.; Peran, I.; Farag, M.; Incicco, J. J.; 
Bremer, A.; Grace, C. R.; Soranno, A.; Pappu, R. V.; Mittag, T.
Valence and patterning of aromatic residues determine the phase 
behavior of prion-like domains.
{\it Science} {\bf 2020}, {\it 367}, 694--699.

\bibitem{SumanPNAS}
Das, S.; Lin, Y.-H.; Vernon, R. M.; Forman-Kay, J. D.; Chan, H. S.
Comparative roles of charge, $\pi$, and hydrophobic interactions in 
sequence-dependent phase separation of intrinsically disordered proteins. 
{\it Proc. Natl. Acad. Sci. U.S.A.} {\bf 2020}, {\it 117}, 28795--28805.

\bibitem{keeley2018}
Muiznieks, L. D.; Sharpe, S.; Pom\`es, R.; Keeley, F. W.
Role of liquid–liquid phase separation in assembly of elastin and other 
extracellular matrix proteins.
{\it J. Mol. Biol.} {\bf 2018}, {\it 430}, 4741--4753.

\bibitem{biochemrev}
Lin, Y.-H.; Forman-Kay, J. D.; Chan, H. S.
Theories for sequence-dependent phase behaviors of biomolecular condensates. 
{\it Biochemistry} {\bf 2018}, {\it 57}, 2499--2508.

\bibitem{roland18}
Cinar, H.; Cinar, S.; Chan, H. S.; Winter, R.
Pressure-induced dissolution and reentrant formation of condensed, 
liquid-liquid phase-separated elastomeric $\alpha$-elastin.
{\it Chem. Eur. J.} {\bf 2018} {\it 24}, 8286--8291.

\bibitem{jeetainACS}
Dignon, G. L.; Zheng, W.; Kim, Y. C.; Mittal, J.
Temperature-controlled liquid-liquid phase separation of disordered proteins.
{\it ACS Cent. Sci.} {\bf 2019}, {\it 5}, 821--830.

\bibitem{Roland2019}
Cinar, H.; Fetahaj, Z.; Cinar, S.; Vernon, R. M.; Chan, H. S.; Winter, R.
Temperature, hydrostatic pressure, and osmolyte effects on liquid-liquid 
phase separation in protein condensates: Physical chemistry and biological 
implications. {\it Chem. Eur. J.} {\bf 2019}, {\it 57}, 13049--13069.

\bibitem{Brocca2020}
Bianchi, G.; Longhi, S.; Grandori, R.; Brocca, S.
Relevance of electrostatic charges in compactness, aggregation, 
and phase separation of intrinsically disordered proteins.
{\it Int. J. Mol. Sci.} {\bf 2020}, {\it 21}, 6208.

\bibitem{obermeyer2020}
Yeong, V.; Werth, E. G.; Brown, L. M.; Obermeyer, A. C.
Formation of biomolecular condensates in bacteria by tuning protein 
electrostatics.
{\it ACS Cent. Sci.} {\bf 2020}, {\it 6}, 2301--2310.

\bibitem{BrianTsang2019}
Tsang, B.; Arsenault, J.; Vernon, R. M.; Lin, H.; Sonenberg, N.; 
Wang, L. Y.; Bah, A.; Forman-Kay, J. D.
Phosphoregulated FMRP phase separation models activity-dependent 
translation through bidirectional control of mRNA granule formation.
{\it Proc. Natl. Acad. Sci. U.S.A.} {\bf 2019}, {\it 116}, 4218--4227.

\bibitem{Schuster2020}
Schuster, B. S.; Dignon, G. L.; Tang, W. S.; Kelley, F. M.; Ranganath, A. K.;
Jahnke, C. N.; Simplins, A. G.; Regy, R. M.; Hammer, D. A.; Good, M. C.;
Mittal, J.
Identifying sequence perturbations to an intrinsically disordered protein 
that determine its phase separation behavior.
{\it Proc. Natl. Acad. Sci. U.S.A.} {\bf 2020}, {\it 117}, 11421--11431.

\bibitem{rohit2013}
Das, R. K.; Pappu, R. V.
Conformations of intrinsically disordered proteins are influenced by 
linear sequence distributions of oppositely charged residues.
{\it Proc. Natl. Acad. Sci. U.S.A.} {\bf 2013}, {\it 110}, 13392--13397.

\bibitem{kings2015}
Sawle, L.; Ghosh, K.
A theoretical method to compute sequence dependent configurational 
properties in charged polymers and proteins.
{\it J. Chem. Phys.} {\bf 2015}, {\it 143}, 085101.

\bibitem{ZhaoetalJCP2015}
Zhao, M.; Zhou, J.; Su, C.; Niu, L.; Liang, D.; Li, B.
Complexation behavior of oppositely charged polyelectrolytes: Effect of 
charge distribution.
{\it J. Chem. Phys.} {\bf 2015}, {\it 142}, 204902.

\bibitem{lin2017}
Lin, Y.-H.; Chan, H. S.
Phase separation and single-chain compactness of charged disordered proteins 
are strongly correlated.
{\it Biophys, J.} {\bf 2017}, {\it 112}, 2043--2046.

\bibitem{jacob2017}
Brady, J. P.; Farber, P. J.; Sekhar, A.; Lin, Y.-H.; Huang, R.; Bah, A.;
Nott, T. J.; Chan, H. S.;  Baldwin, A. J.; Forman-Kay, J. D.; Kay, L. E.
Structural and hydrodynamic properties of an intrinsically disordered region of
a germ cell-specific protein on phase separation. {\it Proc. Natl. Acad. Sci. 
U.S.A.} {\bf 2017}, {\it 114}, E8194--E8203.

\bibitem{Alan}
Amin, A. N.; Lin, Y.-H.; Das, S.; Chan, H. S.
Analytical theory for sequence-specific binary fuzzy complexes of charged 
intrinsically disordered proteins. 
{\it J. Phys. Chem. B} {\bf 2020}, {\it 124}, 6709--6720.

\bibitem{njp2017}
Lin, Y. -H.; Brady, J. P.; Forman-Kay J. D.; Chan, H. S. Charge pattern
matching as a ‘fuzzy’ mode of molecular recognition for the functional phase
separations of intrinsically disordered proteins. 
{\it New J. Phys.} {\bf 2017}, {\it 19}, 115003. 

\bibitem{singperry2017}
Chang, L.-W.; Lytle, T. K.; Radhakrishna, M.; Madinya, J. J.; V\'elez, J.; 
Sing, C. E.; Perry, S. L.
Sequence and entropy-based control of complex coacervates.
{\it Nat. Comm.} {\bf 2017}, {\it 8}, 1273.

\bibitem{kings2020}
Lin, Y.-H.; Brady, J. P.; Chan, H. S.; Ghosh, K.
A unified analytical theory of heteropolymers for sequence-specific 
phase behaviors of polyelectrolytes and polyampholytes.
{\it J. Chem. Phys.} {\bf 2020}, {\it 152}, 045102.

\bibitem{dignon18}
Dignon, G. L.; Zheng, W.; Kim, Y. C.; Best, R. B.; Mittal, J.
Sequence determinants of protein phase behavior from a coarse-grained model.
{\it PLoS Comput. Biol.} {\bf 2018}, {\it 14}, e1005941.

\bibitem{suman1}
Das, S.; Eisen, A.; Lin, Y.-H.; Chan, H. S.
A lattice model of charge-pattern-dependent polyampholyte phase separation.
{\it J. Phys. Chem. B} {\bf 2018}, {\it 122}, 5418--5431.

\bibitem{jeetainPNAS}
Dignon, G. L.; Zheng, W.; Best, R. B.; Kim, Y. C.; Mittal, J.
Relation between single-molecule properties and phase behavior of 
intrinsically disordered proteins.
{\it Proc. Natl. Acad. Sci. U.S.A.} {\bf 2018}, {\it 115}, 9929--9934.

\bibitem{suman2}
Das, S.; Amin, A. N.; Lin, Y.-H.; Chan, H. S.
Coarse-grained residue-based models of disordered protein condensates: 
Utility and limitations of simple charge pattern parameters.
{\it Phys. Chem. Chem. Phys.} {\bf 2018}, {\it 20}, 28558--28574.

\bibitem{koby2020}
Hazra, M. K.; Levy, Y.
Charge pattern affects the structure and dynamics of polyampholyte condensates.
{\it Phys. Chem. Chem. Phys.} {\bf 2020}, {\it 22}, 19368--19375. 

\bibitem{stefan2019}
Robichaud,N. A. S.; Saika-Voivod, I.; Wallin, S.
Phase behavior of blocky charge lattice polymers: Crystals, liquids, sheets, 
filaments, and clusters. {\it Phys. Rev. E.} {\bf 2019}, {\it 100}, 052404.

\bibitem{joanElife}
Lin, Y.; McCarty, J.; Rauch, J. N.; Delaney, K. T.;
Kosik, K. S.; Fredrickson, G. H.; Shea, J.-E.; Han, S.
Narrow equilibrium window for complex coacervation of tau and RNA under
cellular conditions 
{\it eLife} {\bf 2019}, {\it 8}, e42571.

\bibitem{joanJPCL}
McCarty, J.; Delaney, K. T.; Danielsen, S. P. O.; Fredrickson, G. H.; 
Shea, J.-E.
Complete phase diagram for liquid-liquid phase separation of intrinsically 
disordered proteins.
{\it J. Phys. Chem. Lett.} {\bf 2019}, {\it 10}, 1644--1652.

\bibitem{joanPNAS}
Danielsen, S. P. O.; McCarty, J.; Shea, J.-E.; Delaney, K. T.; 
Fredrickson, G. H.
Molecular design of self-coacervation phenomena in block polyampholytes.
{\it Proc. Natl. Acad. Sci. U.S.A.} {\bf 2019}, {\it 116}, 8224--8232.

\bibitem{joanJCP}
Danielsen, S. P. O.; McCarty, J.; Shea, J.-E.; Delaney, K. T.; 
Fredrickson, G. H.
Small ion effects on self-coacervation phenomena in block polyampholytes.
{\it J. Chem. Phys.} {\bf 2019}, {\it 151}, 034904.

\bibitem{Pal2021}
Pal, T.; Wess\'en, J.; Das, S.; Chan, H. S. 
Subcompartmentalization of polyampholyte species in organelle-like 
condensates is promoted by charge-pattern mismatch and strong excluded-volume 
interaction. {\it Phys. Rev. E} {\bf 2021}, in press (accepted for 
publication); 
preprint available on arXiv: {\tt https://arxiv.org/abs/2006.12776}.


\bibitem{honig}
Gilson, M. K.; Honig, B. H.
The dielectric constant of a folded protein.
{\it Biopolymers} {\bf 1986}, {\it 25}, 2097--2119.

\bibitem{bertrand0}
Garc{\'\i}a-Moreno, B.; Dwyer, J.; Gittis, A. G.; Lattman, E. E.; 
Spencer, D. S.; Stites, W. E.
Experimental measurement of the effective dielectric in the hydrophobic 
core of a protein.
{\it Biophys. Chem.} {\bf 1997}, {\it 64}, 211--224.

\bibitem{bertrand}
Dwyer, J. J.; Gittis, A. G.; Karp, D. A.; Lattman, E. E.; Spencer, D. S.;
Stites, W. E.; Garc{\'\i}a-Moreno E, B.
High apparent dielectric constants in the interior of a protein 
reflect water penetration.
{\it Biophys. J.} {\bf 2000}, {\it 79}, 1610--1620.

\bibitem{vanGunsteren2001}
Pitera, J. W.; Falta, M.; van Gunsteren, W. F. Dielectric properties of
proteins from simulation: The effects of solvent, ligands, pH, and temperature.
{\it Biophys. J}. {\bf 2001}, {\it 80}, 2546-2555.

\bibitem{warshel2001}
Schutz, C. N.; Warshel, A.
What are the dielectric ``constants'' of proteins and how to validate
electrostatic models?
{\it Proteins} {\bf 2001}, {\it 44}, 400--417.

\bibitem{plotkinPCCP}
Guest, W. C.;   Cashman, N. R.; Plotkin, S. S.
A theory for the anisotropic and inhomogeneous dielectric 
properties of proteins.
{\it Phys. Chem. Chem. Phys.} {\bf 2011}, {\it 13}, 6286--6295.

\bibitem{kings2017}
Sawle, L.; Huihui, J.; Ghosh, K.
All-atom simulations reveal protein charge decoration in the folded and 
unfolded ensemble is key in thermophilic adaptation.
{\it J. Chem. Theory Comput.} {\bf 2017}, {\it 13}, 5065--5075.

\bibitem{DNA}
Cuervo, A.; Dans, P. D.; Carrascosa, J. L.; Orozco, M.; Gomila, G.;
Fumagalli, L.
Direct measurement of the dielectric polarization properties of DNA.
{\it Proc. Natl. Acad. Sci. U.S.A.} {\bf 2014}, {\it 111}, E3624--E3630.

\bibitem{Geim}
Fumagalli, L.; Esfandiar, A.; Fabregas, R.; Hu1, S.; Ares, P.; Janardanan, A.;
Yang, Q.; Radha1, B.; Taniguchi, T.; Watanabe, K. Gomila, G.; Novoselov, K. S.;
Geim, A. K.  
Anomalously low dielectric constant of confined water.
{\it Science} {\bf 2018}, {\it 360}, 1339--1342.

\bibitem{bragg}
Bragg, W. L.; Pippard, A. B. The form birefringence of macromolecules.
{\it Acta Cryst}. {\bf 1953}, {\it 6}, 865–-867.

\bibitem{jackson}
Jackson, J. D. 
{\it Classical Electrodynamics} (2nd edition); Wiley, :New York, 1975; 
pp 154--155.

\bibitem{markel2016}
Markel, V. A. Introduction to the Maxwell Garnett approximation: Tutorial.
{\it J. Opt. Soc. Am. A} {\bf 2016} {\it 33} 1244--1256.

\bibitem{uversky2016}
Ferreira, L. A.; Loureiro, J. A.; Gomes, J.; Uversky, V. N.; Madeira, P. P.;
Zaslavsky, B. Y.  
Why physicochemical properties of aqueous solutions of various compounds 
are linearly interrelated.
{\it J. Mol. Liq.} {\bf 2016}, {\it 221}, 116--123.

\bibitem{mann2020}
Moreau, N. G.; Martin, N.; Gobbo, P.; Tang, T.-Y. D.; Mann, S.  
Spontaneous membrane-less multi-compartmentalization via aqueous two-phase 
separation in complex coacervate micro-droplets.
{\it Chem. Commun.} {\bf 2020}, {\it 56}, 12717.

\bibitem{feig2014}
Yildirim, A.; Sharma, M.; Varner, B. M.; Fang, L.; Feig, M.
Conformational preferences of DNA in reduced dielectric environments.
{\it J. Phys. Chem. B} {\bf 2014}, {\it 118}, 10874--10881.

\bibitem{bloomfield1995} 
Arscott, P. G.; Ma, C.; Wenner, J. R.; Bloomfield, V. A.  
DNA condensation by cobalt hexaammine(III) in alcohol-water mixtures: 
Dielectric constant and other solvent effects.
{\it Biopolymers}, {\bf 1995}, {\it 36}, 345--364. 

\bibitem{spruijt2019}
Nakashima, K. K.; Vibhute, M. A.; Spruijt, E.
Biomolecular chemistry in liquid phase separated compartments.
{\it Frontiers Mol. Biosci.} {\bf 2019}, {\it 6}, 21. 

\bibitem{deniz2018}
Milin, A. N.; Deniz, A. A.
Reentrant phase transitions and non-equilibrium dynamics in 
membraneless organelles.
{\it Biochemistry} {\bf 2018}, {\it 57}, 2470--2477.

\bibitem{uversky2018}
Zaslavsky, B. Y.; Uversky, V. N.  
In aqua veritas: The indispensable yet mostly ignored role of water 
in phase separation and membrane-less organelles.
{\it Biochemistry} {\bf 2018}, {\it 57}, 2437--2451.

\bibitem{uversky2018_1}
Zaslavsky, B. Y.; Ferreira, L. A.; Darling, A. L.; Uversky, V. N.  
The solvent side of proteinaceous membrane-less organelles in light of 
aqueous two-phase systems.
{\it Int. J. Biol. Macromol.} {\bf 2018}, {\it 117}, 1224--1251.

\bibitem{Prasad2019}
Pliss, A.;  Levchenko, S. M.; Liu, L.; Peng, X.; Ohulchanskyy, T. Y.; Roy, I.;
Kuzmin, A. N.; Qu, J.; Prasad, P. N.  
Cycles of protein condensation and discharge in nuclear organelles studied 
by fluorescence lifetime imaging.
{\it Nat. Comm.} {\bf 2019}, {\it 10}, 455.

\bibitem{JeetainAtom}
Zheng, W.; Dignon, G. L.; Jovic, N.; Xu, X.; Regy, R. M.; Fawzi, N. L.; Kim, Y. C.; Best, R. B.; Mittal, J.
Molecular details of protein condensates probed by microsecond long 
atomistic simulations.
{\it J. Phys. Chem. B} {\bf 2020}, {\it 124}, 11671--11679.

\bibitem{gerhard2020} 
Benayad, Z.; von B\"ulow, S.; Stelzl, L. S.; Hummer, G.
Simulation of FUS protein condensates with an adapted coarse-grained model.
{\it J. Chem. Theory Comput.} {\bf 2021}, {\it 17}, 525--537.

\bibitem{martini2013}
de Jong, D. H.; Singh, G.; Bennett, W. F. D.; Arnarez, C.; Wassenaar, T. A.; 
Sch\"afer, L. V.; Periole, X.; Tieleman, D. P.; Marrink, S. J. 
Improved parameters for the Martini coarse-grained protein force field.
{\it J. Chem. Theory Comput.} {\bf 2013}, {\it 9}, 687--697.

\bibitem{Smiatek2017}
Michalowsky, J.; Sch\"afer, L. V.; Holm, C.; Smiatek, J.
A refined polarizable water model for the coarse-grained MARTINI force 
field with long-range electrostatic interactions.
{\it J. Chem. Phys}. {\bf 2017}, {\it 146}, 054501.

\bibitem{gerhard2009}
K\"ofinger, J.; Hummer, G.; Dellago, C.
A one-dimensional dipole lattice model for water in narrow nanopores.
{\it J. Chem. Phys.} {\bf  2009},  {\it 130}, 154110.

\bibitem{aluru2018}
Motevaselian, M. H.; Mashayak, S. Y.; Aluru, N. R.
Extended coarse-grained dipole model for polar liquids: Application to 
bulk and confined water.
{\it Phys. Rev. E} {\bf 2018}, {\it 98}, 052135.

\bibitem{orlandPRL2012}
Levy, A.; Andelman, D.; Orland, H.
Dielectric constant of ionic solutions: A field-theory approach.
{\it Phys. Rev. Lett.} {\bf 2012}, {\it 108}, 227801.

\bibitem{orlandJCP2013}
Levy, A.; Andelman, D.; Orland, H.
Dipolar Poisson-Boltzmann approach to ionic solutions: A mean field and 
loop expansion analysis.
{\it J. Chem. Phys.} {\bf 2013}, {\it 139}, 164909.

\bibitem{ZGWangJCP2018}
Zhuang, B.; Wang, Z.-G.
Statistical field theory for polar fluids.
{\it J. Chem. Phys.} {\bf 2018}, {\it 149}, 

\bibitem{GHFJCP2016}
Martin, J. M.; Li, W.; Delaney, K. T.; Fredrickson, G. H.
Statistical field theory description of inhomogeneous polarizable soft matter.
{\it J. Chem. Phys.} {\bf 2016}, {\it 145}, 154104.

\bibitem{FredricksonPRL2019}
Grzetic, D. J.; Delaney, K. T.; Fredrickson, G. H.
Constrasting dielectric properties of electrolyte solutions with
polar and polarizable solvents.
{\it Phys. Rev. Lett.} {\bf 2019}, {\it 122}, 128007.

\bibitem{GHFJCP2020}
Martin, J. M.; Delaney, J. T.; Fredrickson, G. H.
Effect of an electric field on the stability of binary 
dielectric fluid mixtures.
{\it J. Chem. Phys.} {\bf 2020}, {\it 152}, 234901.

\bibitem{panag2017}
Silmore, K. S.; Howard, M. P.; Panagiotopoulos, A. Z.
Vapor-liquid equilibrium and surface tension of fully flexible Lennard-Jones
chains. {\it Mol. Phys.} {\bf 2017}, {\it 115}, 320--327.

\bibitem{TraPPE1}
Mundy, C. J.; Siepmann, J. I.; Klein, M. L. 
Calculation of the shear viscosity of decane using a reversible multiple 
timestep algorithm. 
{\it J. Chem. Phys.} {\bf 1995}, {\it 102}, 3376--3380.

\bibitem{TraPPE2}
Martin, M. G.; Siepmann, J. 
I. Transferable potentials for phase equilibria. 1. United-atom description
of n-alkanes. 
{\it J. Phys. Chem. B} {\bf 1998}, {\it 102}, 2569--2577.

\bibitem{TraPPE3}
Nicolas, J. P.; Smit, B. 
Molecular dynamics simulations of the surface tension of n-hexane, n-decane
and n-hexadecane. 
{\it Mol. Phys.} {\bf 2002}, {\it 100}, 2471--2475.

\bibitem{TraPPE4}
P\`amies, J. C.; McCabe, C.; Cummings, P. T.; Vega, L. F. 
Coexistence densities of methane and propane by canonical molecular 
dynamics and gibbs ensemble Monte Carlo simulations. 
{\it Mol. Simul.} {\bf 2003}, {\it 29}, 463--470.

\bibitem{WCA}
Weeks, J. D.; Chandler, D.; Andersen, H. C. 
Role of repulsive forces in determining the equilibrium
structure of simple liquids. {\it J. Chem. Phys.} {\bf 1971}, 
{\it 54}, 5237--5247.

\bibitem{Anderson}
Anderson, J. A.; Lorenz, C. D.; Travesset, A. 
General purpose molecular dynamics simulations fully
implemented on graphics processing units. 
{\it J. Comput. Phys.} {\bf 2008}, {\it 227}, 5342--5359.

\bibitem{Glaser}
Glaser, J.; Nguyen, T. D.; Anderson, J. A.; Lui, P.; Spiga, F.; 
Millan, J. A.; Morse, D. C.; Glotzer, S. C.
Strong scaling of general-purpose molecular dynamics simulations on GPUs. 
{\it Comput. Phys. Comm.} {\bf 2015}, {\it 192}, 97--107.

\bibitem{LeBard}
LeBard, D. N.; Levine, B. G.; Mertmann, P.; Barr, S. A.; Jusufi, A.; 
Sanders, S.; Klein, M. L.; Panagiotopoulos, A. Z. 
Self-assembly of coarse-grained ionic surfactants accelerated by graphics
processing units. 
{\it Soft Matter} {\bf 2012}, {\it 8}, 2385--2397.

\bibitem{Martinez}
Mart{\'\i}nez, L.; Andrade, R.; Birgin, E. G.; 
Mart{\'\i}nez, J. M. PACKMOL: A package for building initial
configurations for molecular dynamics simulations. 
{\it J. Comput. Chem.} {\bf 2009}, {\it 30}, 2157--2164.

\bibitem{orlandPRL2007}
Abrashkin, A.; Andelman, D.; Orland, H.
Dipolar Poisson-Boltzmann equation: Ions and dipoles close to charged 
interfaces.
{\it Phys. Rev. Lett.} {\bf 2007}, {\it 99}, 077801.

\bibitem{HansenJCP2005}
Ballenegger, V.; Hansen, J.-P. 
Dielectric permittivity profiles of confined polar fluids. 
{\it J.  Chem. Phys.} {\bf 2005}, {\it 122}, 114711.

\bibitem{SternFeller2003}
Stern, H. A.; Feller, S. E.
Calculation of the dielectric permittivity profile for a nonuniform 
system: Application to a lipid bilayer simulation.
{\it J. Chem. Phys.} {\bf 2003}, {\it 118}, 3401--3412.

\bibitem{Zhu_etal2020}
Zhu, H.; Yang, F.; Zhu, Y.; Li, A.; He, W.; Huang, J.; Li, G. 
Investigation of dielectric constants of water in a nano-confined pore. 
{\it RSC Adv.}, {\bf 2020}, {\it 10}, 8628--8635.

\bibitem{ChanDill97rev}
Chan, H. S.; Dill, K. A.
Solvation: How to obtain microscopic energies from partitioning and solvation 
experiments. 
{\it Annu. Rev. Biophys. Biomol. Struct.} {\bf 1997}, {\it 26}, 425--459.

\bibitem{ray1971}
Ray, A.
Solvophobic interactions and micelle formation in structure forming 
nonaqueous solvents.
{\it Nature} {\bf 1971}, {\it 231}, 313--315.

\bibitem{waterdielectric}
Malmberg, C. G.; Maryott, A. A.
Dielectric constant of water from $0^\circ$ to $100^\circ$C.
{\it J. Res. Natl. Bur. Stand. (U. S.)} {\bf 1956}, {\it 56}, 1--8.

\bibitem{Wang2010}
Wang, Z.-G.
Fluctuation in electrolyte solutions: The self energy.
{\it Phys. Rev. E} {\bf 2010}, {\it 81}, 021501.

\bibitem{Chandler1977}
Chandler, D. The dielectric constant and related equilibrium properties 
of molecular fluids: Interaction site cluster theory analysis. 
{\it J. Chem. Phys.} {\bf 1977}, {\it 67}, 1113--1124.

\bibitem{Edwards1965}
Edwards, S. F. 
The statistical mechanics of polymers with excluded volume. 
{\it Proc. Phys. Soc.} {\bf 1965}, {\it 85}, 613--624.

\bibitem{Fredrickson2006}
Fredrickson, G. H. 
{\it The Equilibrium Theory Of Inhomogeneous Polymers}; Oxford
    University Press Inc.:New York, 2006.

\bibitem{FredricksonGanesanDrolet2002}
Fredrickson, G. H.; Ganesan, V.; Drolet, F. Field-theoretic computer 
simulation methods for polymers and complex fluids. 
{\it Macromolecules} {\bf 2002}, {\it 35}, 16--39.

\bibitem{Parisi1983}
Parisi, G. 
On complex probabilities. 
{\it Phys. Lett. B} {\bf 1983}, {\it 131}, 393--395.

\bibitem{Klauder1983}
Klauder, J. R. 
A Langevin approach to fermion and quantum spin correlation functions. 
{\it J. Phys. A: Math. Gen.} {\bf 1983}, {\it 16}, L317--L319.

\bibitem{ParisiWu1981}
Parisi, G.; Wu, Y.-S. Perturbation theory without gauge fixing. 
{\it Scientia Sinica} {\bf 1981} {\it 24}, 483--496. 

\bibitem{HSCMartyGhost}
Chan, H. S.; Halpern, M. B.
New ghost-free infrared-soft gauges. 
{\it Phys. Rev. D} {\bf 1986}, {\it 33}, 540--547. 

\bibitem{HSCMartyGravity}
Chan, H. S.; Halpern, M. B.
Continuum-regularized quantum gravity. 
{\it Zeitschrift F\"ur Physik C} {\bf 1987}, {\it 36}, 669--693.

\bibitem{Lennon_etal_2008}
Lennon, E. M.; Mohler, G. O.; Ceniceros, H. D.; Garca-Cervera, C. J.; 
Fredrickson, G. H. 
Numerical solutions of the complex Langevin equations in polymer
field theory. 
{\it Multiscale Modeling \& Simulation} {\bf 2008}, {\it 6}, 1347--1370.

\bibitem{GHFJCPchi2018}
Grzetic, D. J.; Delaney, K. T.; Fredrickson, G. H.
The effective $\chi$ parameter in polarizable polymeric systems: One-loop 
perturbation theory and field-theoretic simulations. 
{\it J. Chem. Phys.} {\bf 2018}, {\it 148}, 204903.

\bibitem{Delaney2017}
Delaney, K. T.; Fredrickson, G. H. 
Theory of polyelectrolyte complexation---Complex
coacervates are self-coacervates. 
{\it J. Chem. Phys.} {\bf 2017}, {\it 146}, 224902.

\bibitem{Riggleman2010}
Riggleman, R. A.; Fredrickson, G. H. 
Field-theoretic simulations in the gibbs ensemble. 
{\it J. Chem. Phys.} {\bf 2010}, {\it 132}, 024104.

\bibitem{harvey1972}
Harvey, S. C.; Hoekstra, P.
Dielectric relaxation spectra of water adsorbed on lysozyme.
{\it J. Phys. Chem.} {\bf 1972}, {\it 76}, 2987--2994.

\bibitem{hole1976}
Tredgold, R. H.; Hole, P. N.
Dielectric behaviour of dry synthetic polypeptides.
{\it Biochim. Biophys. Acta} {\bf 1976}, {\it 443}, 137--142.

\bibitem{bone1982}
Bone, S.; Pethig, R.
Dielectric studies of the binding of water to lysozyme.
{\it J. Mol. Biol.} {\bf 1982}, {\it 157}, 571--575.

\bibitem{banerjee2021}
Alshareedah, I.; Kaur, T.; Banerjee, P. R.
Methods for characterizing the material properties of biomolecular condensates.
{\it Methods Enzymol.} {\bf 2021}, {\it 646}, 143--183.

\end{thebibliography}
\end{document}